\documentclass[journal, letterpaper]{IEEEtran}
\usepackage{acronym}

\ifCLASSOPTIONcompsoc
  \usepackage[nocompress]{cite}
\else
  \usepackage{cite}
\fi
\usepackage{multirow}

\usepackage{times,amsmath,amsthm}
\usepackage{epsfig}
\usepackage[latin1]{inputenc}
\usepackage{amsfonts}
\usepackage{color}
\usepackage{orcidlink}
\usepackage{etoolbox}
\usepackage{graphicx}
\usepackage{epstopdf}
\usepackage{bm}
\usepackage{subfig}
\usepackage{paralist}
\usepackage{url}
\usepackage{setspace}

\usepackage{pgfplots}
\usepgfplotslibrary{groupplots,dateplot}
\usetikzlibrary{patterns,shapes.arrows}
\pgfplotsset{compat=newest}
\usepackage{algorithm}
\usepackage{algorithmicx}
\usepackage{caption}
\usepackage{stfloats}
\usepackage{algpseudocode}
\usepackage{booktabs}

\usepackage{mdframed}
\mdfdefinestyle{ProblemFrame}{%
    linecolor=black,
    outerlinewidth=2pt,
    innertopmargin=2pt,
    innerbottommargin=4pt,
    innerrightmargin=4pt,
    innerleftmargin=4pt,
        leftmargin = 4pt,
        rightmargin = 4pt
    }
\theoremstyle{plain}

\theoremstyle{plain}
\newtheorem{proposition}{Proposition}
\theoremstyle{remark}
\newtheorem{remark}{Remark}
\DeclareMathOperator*{\argmin}{\arg\!\min}
\DeclareMathOperator*{\argmax}{\arg\!\max}

\title{EASE: Energy-Aware Job Scheduling for Vehicular Edge Networks With Renewable Energy Resources}

\author{Giovanni~Perin\IEEEauthorrefmark{1}\,\orcidlink{0000-0002-7333-3004},~\IEEEmembership{Graduate Student Member,~IEEE,}
        Francesca~Meneghello\IEEEauthorrefmark{1}\,\orcidlink{0000-0002-9905-0360},~\IEEEmembership{Member,~IEEE,}
        Ruggero~Carli\,\orcidlink{0000-0002-6506-5898},~\IEEEmembership{Member,~IEEE,}
        Luca~Schenato\,\orcidlink{0000-0003-2544-2553},~\IEEEmembership{Fellow,~IEEE,}\\
        and~Michele~Rossi\,\orcidlink{0000-0003-1121-324X},~\IEEEmembership{Senior Member,~IEEE} %
\thanks{\scriptsize{This work has been supported, in part, by the Italian Ministry of Education, University and Research (MIUR) through the PRIN project no. 2017NS9FEY entitled ``Realtime Control of 5G Wireless Networks: Taming the Complexity of Future Transmission and Computation Challenges", and by MIUR through the initiative ``Departments of Excellence" (Law 232/2016). The views and opinions expressed in this work are those of the authors and do not necessarily reflect those of the funding institutions.}}
\thanks{\scriptsize{All authors are with the Department of Information Engineering, University of Padova, via Gradenigo 6/b, 35131, Padova, Italy.}}
\thanks{\scriptsize{Michele Rossi is also with the Department of Mathematics ``Tullio Levi-Civita'', University of Padova, via Trieste 63, 35121, Padova, Italy.}}
\thanks{\scriptsize{\IEEEauthorrefmark{1}Corresponding authors, emails: \{name.surname\}@dei.unipd.it.}}}

\IEEEpubid{\begin{minipage}{\textwidth}\ \\[12pt]
  This work has been submitted to the IEEE for possible publication.\\ 
  Copyright may be transferred without notice, after which this version may no longer be accessible.
\end{minipage}} 

\begin{document}
\acrodef{BS}{base station}
\acrodef{5G}{\mbox{fifth-generation}}
\acrodef{ETSI}{European Telecommunications Standards Institute}
\acrodef{CPU}{central processing unit}
\acrodef{MEC}{multi-access edge computing} 
\acrodef{MEH}{mobile edge host} 
\acrodef{MPC}{model predictive control} 
\acrodef{QP}{quadratic programming}  
\acrodef{EH}{energy harvesting}
\acrodef{UPS}{uninterrupted power supply}
\acrodef{IoT}{Internet of Things}
\acrodef{MCC}{\textit{mobile cloud computing}}
\acrodef{VM}{virtual machine}
\acrodef{eNB}{evolved node B}
\acrodef{UE}{user equipment}
\acrodef{QoS}{quality of service}
\acrodef{QoE}{quality of experience}
\acrodef{MIP}{mixed integer programming}
\acrodef{IoV}{internet of vehicles}
\acrodef{PV}{photovoltaic panel}
\acrodef{M2M}{\mbox{machine-to-machine}}
\acrodef{ID}{identifier}
\acrodef{EASE}{Energy-Aware job Scheduling at the Edge}

\IEEEtitleabstractindextext{%
\begin{abstract}
%Serving the users as close as possible in mobile and vehicular networks is of paramount importance for ensuring a high \ac{QoS} and \ac{QoE}. This paper tackles the problem of time-sensitive job handovers for the \ac{IoV} in a resources-constrained \ac{MEC} platform, \mbox{co-powered} by renewables. The working pipeline is composed of the alternation of a i) centralized optimization step, solved through \ac{MPC}, to manage the local resources and estimate the future availability, and ii) a distributed consensus step, obtained via dual ascent in closed form, to agree on service migrations. The approach is compared with common state-of-the-art strategies, namely, always and never migrating the tasks. The results show that the proposed approach outperforms other strategies, by always considering local power constraints and minimizing the carbon footprint while keeping the jobs close to the end users.
The energy sustainability of \ac{MEC} platforms is here addressed by developing \ac{EASE}, a computing resource scheduler for edge servers co-powered by renewable energy resources and the power grid. The scenario under study involves the optimal allocation and migration of \mbox{time-sensitive} computing tasks in a resource-constrained \ac{IoV} context. This is achieved by tackling, as the main objective, the minimization of the carbon footprint of the edge network, whilst delivering adequate \ac{QoS} to the end users (e.g., meeting task execution deadlines). \Ac{EASE} integrates i) a centralized optimization step, solved through \ac{MPC}, to manage the renewable energy that is locally collected at the edge servers and their local computing resources, estimating their future availability, and ii) a distributed consensus step, solved via dual ascent in closed form, to reach agreement on service migrations. \ac{EASE} is compared with four existing migration strategies. Quantitative results demonstrate its greater energy efficiency, which often gets close to complete carbon neutrality, while also improving the \ac{QoS}.
\end{abstract}

\begin{IEEEkeywords}
multi-access edge computing, energy efficiency, green computing networks, mobility management, service migration, distributed scheduling. %\vspace{-0.2cm}
\end{IEEEkeywords}}

\maketitle

\IEEEdisplaynontitleabstractindextext

\IEEEpeerreviewmaketitle

\section{Introduction}\label{sec:Introduction}
%!TEX root = ./paper.tex

% -- the number of Internet users will grow from $50$\% to $66$\% of the world population by 2023

The future of mobile networks is not only concerned with faster and more reliable wireless connections. The rapid digitalization of the society~\cite{cisco2020}  comes with a need to expedite the service provisioning time,  demanding support for \textit{computation-intensive} and \textit{delay-sensitive} users' applications. Often, these applications cannot be executed on the end devices due to memory and energy scarcity, nor on the network cloud due to a consequent surge in the Internet traffic and excessive delays. 
These facts lead to the introduction of the \ac{MEC} paradigm, entailing the \mbox{de-location} of computation services at the mobile network edge, by empowering the \ac{eNB} sites with adequate computing facilities, referred to as \acp{MEH}. With \ac{MEC}, a user can offload intensive computing jobs to a \ac{MEH}, thus considerably reducing the communication delays with respect to cloud services.
Spurred by the high potential of such innovation, the \ac{ETSI} is extensively working on the standardization of interoperable \ac{MEC} architectures~\cite{ETSI003}, along with their integration with \ac{5G} -- and beyond -- mobile networks~\cite{ETSI031}.

In this work, we consider an \ac{IoV} scenario, where the network users are \ac{5G} -- or beyond \ac{5G} -- enabled vehicles requiring communication and computing support~\cite{feng2019mobile, ETSI030}. According to~\cite{cisco2020}, among \ac{M2M} communications, connected cars are the vertical with the highest expected compound annual growth rate ($30\%$) until at least 2023. Moreover, one of the key challenges in an \ac{IoV} context is ensuring computing service continuity as the vehicles move away from their serving \ac{MEH}~\cite{ETSI022}. This requires implementing online policies to decide whether to move the entity executing the service on a \ac{MEH} that is closer to the user or to complete the computation where it started. 
In the former case, the user spends less energy to communicate with the \ac{MEH}, but resources are spent by the network due to the migration process, both in terms of energy and time. As for the latter, standard network procedures~\cite{3GPP2014} ensure that the user remains connected to the serving \ac{MEH}, thus guaranteeing the delivery of the computation result, at the cost of higher latency.

\IEEEpubidadjcol
\noindent\textbf{Article contribution.} We propose \ac{EASE}, a \textit{proactive} approach to select the most suitable allocation of computing resources considering energy, memory and computation constraints. In the envisioned scenario, \acp{eNB} (\acp{MEH}) are connected to the power grid and empowered with \acp{PV}, which provide green energy that can be exploited without additional costs. Vehicle mobility predictions are leveraged to estimate the best sites where the users' computing jobs can be allocated, accounting for network and users' requirements. 
To the best of our knowledge, this is the first attempt to design a complete framework for the \textit{energy efficient} scheduling of computing jobs over \acp{MEH} networks, by exploiting {\it mobility aware} procedures. 
The devised system provides job schedules that minimize the carbon footprint at the network side -- for the computation and communication services -- subject to job latency and mobility constraints. The job scheduling policy consists of two phases, the former is independently and locally executed at the \acp{eNB} (\acp{MEH}), while the latter is implemented as a decentralized consensus process. 
In the first phase, each \ac{MEH} leverages estimates of the renewable (\mbox{cost-free}) energy, the computational power and the memory available within a prediction window to decide upon the optimal local amount of workload to be executed, subject to users' mobility and delay constraints. Each \ac{MEH} also identifies the jobs that should be migrated to neighboring \acp{MEH}, as belonging to vehicles that are approaching the border of their current serving cell. The mobility predictor developed in~\cite{labriji2021mobility} is used to determine the desired workload to transfer to each neighboring \ac{MEH}.
Then, in the second phase, the \acp{MEH} collectively reach an agreement on the amount of workload to exchange to reduce the overall energy expenditure, while guaranteeing adequate \ac{QoS} to the end-users: an approximated integer solution for jobs migration is derived through a consensus algorithm followed by a rounding step, using mobility predictions to make job migration decisions.
\ac{EASE} is evaluated in a real-world scenario emulated through the ``simulation of urban mobility'' (SUMO) software, considering the vehicular mobility traces for the city of Cologne, and dense city-wide deployment of \ac{5G} \ac{eNB}s with \ac{MEC} functionalities.
Numerical results reveal that the developed allocation strategy significantly reduces the carbon footprint of the edge network, with an increasing gain over heuristic strategies when the available green energy is scarce. At the same time, it properly allocates workload to the processing units according to their specific computing power, by delivering better \ac{QoS} to the users with respect to heuristic solutions and meeting delay constraints. When possible, service migrations also follow the \ac{UE} during handovers, i.e., services are migrated to the \ac{MEH} that is closest to the \ac{UE} after the handover event. 

The present work brings the following innovations.
\begin{itemize}
\item The problem of computation service continuity is solved in a holistic way, designing \ac{EASE}, a complete framework for users' job scheduling and migration within the \acp{MEH} of a mobile edge network with distributed renewable energy resources. The main objective is to reduce the carbon footprint of the computing network by using the renewable energy resources to the maximum extent.
\item A two-step approach for job location management and migration is devised, splitting the problem into local and distributed phases. With it, \acp{MEH} take advantage of user mobility information (and forecasting) to reduce the energy expenditure of the edge network. 
\item For the distributed phase, a consensus strategy is designed to make migration decisions, and solved in closed form by exploiting a dual ascent algorithm. Upon reaching consensus, an original strategy is put forward to obtain an approximated solution for workload and memory management at the \acp{MEH}.
\end{itemize}

The related work is analyzed in the next Section~\ref{sec:RelatedWork}, whereas the solution workflow is presented in Section~\ref{sec:Workflow}, where we also detail the remaining sections of the paper.

\section{Related Work}\label{sec:RelatedWork}
%!TEX root = ./paper.tex

The resource allocation problem in a \ac{MEC} scenario with {\it static} users is extensively addressed in the literature. Among the most recent works, in~\cite{kaur2020keids} the authors present a job scheduler for containers management at the \acp{MEH}, to reduce the network carbon footprint. In~\cite{tran2019joint, feng2021joint}, the task offloading is optimized from a user perspective, minimizing the task completion time and the related energy expenditure. However, as these approaches consider static users and are not suitable for \ac{IoV} scenarios. %, where the users -- vehicles -- move within the network area. 
Specifically, for \ac{IoV}, mobility management is a key aspect toward an effective implementation of \ac{MEC} assisted networks~\cite{rejiba2019survey}. 
%In this article, we consider that jobs are generated by vehicles in an urban area, devising
In this article, we devise \ac{EASE}, a scheduling algorithm to guarantee service continuity in \ac{MEC} assisted \ac{IoV} networks, by properly allocating computation services based on the network energy distribution and the mobility of the users. Moreover, \ac{EASE} is specifically designed to reduce the carbon footprint of \ac{MEC} assisted networks, by considering facilities empowered with renewable energy sources in addition to the supply from the power grid. Note that the user's computation task allocation requires both i) to decide the \ac{MEH} where to place the job together with the workload to be executed based on the available resources and ii) to trigger service handovers based on the user mobility and energy availability predictions.
In fact, computation service handovers entail not only the exchange of control messages, but also the migration of the data associated with the specific job under execution. The users' requests are served at a so called serving \ac{MEH} through the instantiation of a virtual entity -- either a \ac{VM} or a container -- empowered with adequate memory and computing resources to satisfy the service requirements~\cite{doan2019containervsvm}. Therefore, when a computing service handover is triggered, the virtual entity must be transferred to the target \ac{MEH} and computation must be restored from the point where the previous serving \ac{MEH} stopped. This poses several issues associated with the job latency constraints and the network energy migration costs.
A paper addressing the latency challenge, and proposing strategies to reduce the migration time is~\cite{ramanathan2021live}. The main focus is on \textit{how} to migrate the virtual entity, by defining protocols to transfer the container/\ac{VM} from the current location to the target one. Machen et al.~\cite{machen2018live} propose a layered framework to migrate applications encapsulated either in \acp{VM} or containers, showing a reduction in the service downtime. The authors of ~\cite{ma2019efficient} leverage the layered nature of the storage system to reduce the overhead in the container file system synchronization between the serving and the target \acp{MEH}. However, these approaches are \textit{reactive}, i.e., the service migration is performed {\it after} the user has moved to the new MEH site. This results in an unavoidable processing delay due to the time required for the virtual entity re-instantiation at the new \ac{MEH}~\cite{rejiba2019survey}. 
\Ac{EASE} is instead \textit{proactive}, as the virtual entity is migrated {\it before} the handover event occurs, thus reducing the service interruption time. A quantitative evaluation of the difference in the service downtime between the two approaches can be found in, e.g., \cite{Saurez2016incremental, frangoudis2018service, Farris2018providing}, where the authors show that proactive approaches are desirable for time-sensitive applications.

\begin{table*}[tb]
\small
\centering
\resizebox{1\textwidth}{!}{%
\begin{tabular}{c|c|c|c|c|c|c|c|c|c}
\hline
 & \multirow{2}{*}{\textbf{Objective}} & \multicolumn{3}{c|}{\textbf{Network energy}} & \multirow{2}{*}{\textbf{Network computing resources}} & \multicolumn{4}{|c}{\textbf{Users mobility}} \\
 & & \textbf{Migration cost} & \textbf{Computing cost} & \textbf{Carbon footprint} & & \textbf{Paths planning} & \textbf{Users distribution} & \textbf{Previously visited cells} & \textbf{Velocity or trajectory} \\
\hline
\hline
\cite{yuan2020joint} & energy & \checkmark & & & \checkmark & \checkmark & & & \\
\cite{campolo2019support} & latency & & & & & \checkmark & & & \\
\cite{aissioui2018enabling} & energy & \checkmark & & & & & & & \checkmark \\
\cite{Wang2019} & latency & & & & \checkmark & & & \checkmark &  \\
\cite{dalgkitsis2021data} & latency & & & & \checkmark & & & \checkmark &  \\
\cite{Farris2017} & energy & & & & \checkmark & & & \checkmark &  \\
\cite{labriji2021mobility} & energy & \checkmark & & & \checkmark & & & \checkmark & \checkmark  \\
\cite{rago2021anticipatory} & latency & & & & \checkmark & & \checkmark & &  \\
\ac{EASE} & energy & \checkmark & \checkmark & \checkmark  & \checkmark & & & \checkmark & \checkmark  \\
\hline
\end{tabular}%
}
\setlength\belowcaptionskip{-.5cm}
\caption{Summary of the minimization objective quantities and the \ac{MEH} system aspects considered by \ac{EASE} and the proactive computing service migration approaches in the literature.}
\label{tab:literature_comparison}
\end{table*}

Proactive methods require the \ac{MEC} orchestrator to know the user's next point of attachment to trigger the migration process in advance. Some recent works in the literature show the effectiveness of this strategy, but i) they fail to provide a complete framework to properly allocate the computing jobs within the network entities while jointly considering the users' mobility and the energy, memory and computing power constraints, and ii) they rely on a centralized orchestrator that computes the best policy to adopt knowing the state of {\it all} the network entities. 
Among them, in~\cite{yuan2020joint}, the \ac{MEC} service migration process and the physical route for the user to get to the destination are jointly optimized. The problem is solved through a multi-agent deep reinforcement learning approach to meet the job delay requirements with minimum migration cost and travel time. While this work forces the vehicle to follow a specific physical path, \Ac{EASE} leaves the decision on the physical route to the user and leverages mobility predictions to place the jobs.
In~\cite{campolo2019support}, Campolo et al. exploit pre-planned vehicle routes to proactively migrate the \ac{MEH} container so as to follow the user's movements. In~\cite{aissioui2018enabling}, the authors leverage the vehicle velocity and its direction to decide if and where, i.e., to which target \ac{MEH} the virtual entity should be migrated to reduce the cost of multiple successive service migrations while meeting the jobs' delay constraints. This is obtained through a tradeoff between the energy consumed for migrations and the energy needed to eventually transmit the information through the backhaul links that connect the \acp{MEH} for service continuity. However, the strategies in \cite{campolo2019support, aissioui2018enabling} do not consider the constraints on the \acp{MEH} computing power, making the solutions not directly applicable in real-world scenarios.
In~\cite{Wang2019}, the authors design a policy to decide whether to migrate the virtual entity to a target \ac{MEH} -- estimated through a mobility predictor based on Markov chains -- or to keep the job execution on the serving \ac{MEH} where it was initiated, reallocating the service in case the \ac{MEH} capacity is exceeded. In~\cite{dalgkitsis2021data}, the authors use mobility estimates, obtained using a convolutional neural network, to migrate the computation services through a recursive procedure based on genetic algorithms. However, the mobility predictor developed in~\cite{Wang2019, dalgkitsis2021data} only considers the sequence of the user's previously visited cells without leveraging the mobility pattern followed by the user within the current radio cell: this fails to precisely capture real-world mobility patterns, as shown in \cite{labriji2021mobility}. Moreover, these articles are concerned with minimizing the computing service latency, i.e., energy aspects are not considered.
A different approach is presented in~\cite{Farris2017}, where the user's virtual entity is replicated to multiple neighboring \acp{MEH} before the handover event occurs, considering the \acp{MEH} capacity. The authors suggest using mobility estimates to place the replicas, but leave this for future study. Again, the energy aspect is not considered.
These issues are addressed in~\cite{labriji2021mobility}, where the authors integrate accurate predictions -- based on the actual trajectory of the user within the \ac{eNB} coverage area -- into a \ac{VM} replication strategy, to reduce the network energy consumption. However, while the authors show the impact of the \ac{MEH} computing power on the risk of service discontinuity, they do not introduce a strategy to address this problem. %\ac{EASE} leverages the mobility prediction algorithm developed in~\cite{labriji2021mobility} to obtain estimates on the next point of attachment, while the job scheduler is new.}

The above-referenced methods are not concerned with finding the proper allocation of computing jobs when they are offloaded from the user to the network (the service is first placed on the closest \ac{MEH}). In this respect, Rago et al.~\cite{rago2021anticipatory} use predictions on the distribution of the number of users attached to the different \acp{eNB} and estimates of the task requests to proactively allocate jobs on the available \acp{MEH} considering computing power constraints. The proposed strategy does not address service migrations and is mainly concerned with minimizing the latency while the energy consumption is not considered.  

We emphasize that~\cite{yuan2020joint, campolo2019support, aissioui2018enabling, Wang2019, Farris2017, dalgkitsis2021data, labriji2021mobility, rago2021anticipatory} assume that all \acp{MEH} are attached to the power grid for continuous energy provisioning. This makes these approaches not suitable for the scenario considered in the present work, where we target the reduction of the network carbon footprint in the presence of renewable energy.
This aspect was considered in~\cite{perin2021towards}, where the authors study the problem of managing the energy coming from renewable sources to minimize the energy drained from the power grid. In~\cite{perin2021towards}, \ac{MPC} is used to jointly allocate the local resources and to obtain offloading decisions toward other servers. Instead, \ac{EASE} uses \ac{MPC} to control the local processing only and to obtain an average estimate of future resource availability. In this way, \ac{EASE} allows reducing the complexity of the solution with respect to the distributed approach in~\cite{perin2021towards} as discussed in Section~\ref{subsec:additiona_considerations}. Moreover, unlike what we do with \ac{EASE}, user mobility was not considered~\cite{perin2021towards}. 
Table~\ref{tab:literature_comparison} summarizes the key aspects considered in the previous literature.

In the present work, we propose \ac{EASE}, an energy- and mobility-aware, distributed and proactive scheduling framework for computing jobs allocation and virtual entity migration, with the objective of minimizing the carbon footprint of the \ac{MEH} network. 
\ac{EASE} is the first approach that jointly considers all these aspects in addressing the complex problem of efficiently managing \ac{MEC} empowered \ac{IoV} networks.
This is achieved by combining local policies with a decentralized consensus algorithm, thus obviating the need for an orchestrator.
To show the impact of \ac{EASE} on the network carbon footprint, we compare the obtained results with the service migration approach in~\cite{labriji2021mobility} as, using the same mobility predictor, allows revealing the advantages of \ac{EASE}. Moreover, we implemented three heuristic schemes to approach service migration as presented in~\cite{li2019service}, i.e.,
\begin{inparaenum}[i)]
\item never migrate the service (``keep''),
\item always migrate the service when a handover occurs (``migrate''), and
\item define a threshold on a performance metric to decide whether to migrate or not the service (``threshold'').
\end{inparaenum}

\section{High level system overview}\label{sec:Workflow}
%!TEX root = ./paper.tex

The network setup consists of an urban environment covered by a set $\mathcal{N}$ of \acp{eNB}, each co-located with a \ac{MEH}. $\mathcal{V}$ represents the set of vehicles moving within the city, which are constantly connected to the nearest \ac{eNB} node (providing communication support). Vehicle $v \in \mathcal{V}$ sends computing job requests to the closest \ac{MEH}, which can locally execute the required workload or offload it, either partially or in full, to neighboring \acp{MEH}. Also, each vehicle can have a single outstanding job instance (being processed) and can generate a single job request at any time slot only if the previous request has been either fully processed or dropped by the serving \ac{MEH}. For this reason, in the following analysis, we will interchangeably identify a vehicle with the associated outstanding job to be computed. The set of neighboring \acp{eNB} to \ac{eNB} $i$ is denoted by $\mathcal{N}_i$. Jobs are executed through the instantiation of containers, which reserve the required computing and memory resources. Here, containers are favored over \acp{VM} due to their lower memory footprint, which permits a faster migration process -- a desirable feature in the considered scenario~\cite{ramanathan2021live}. Jobs that are being executed on one \ac{MEH} but associated with vehicles that are about to leave the \ac{eNB}/\ac{MEH} coverage area are assessed by the migration controller. The latter decides whether to migrate their execution to another (target) \ac{MEH} or to finish it locally and send the processing result to the vehicle in a multi-hop fashion (from the old to the new serving \ac{eNB}).
\acp{eNB} are equipped with energy harvesting \ac{PV} devices, whose collected energy is managed by the system. We assume that \acp{eNB} are also connected to the power grid as relying only upon harvested energy would be risky due to its intermittent nature; so energy can be drained from the grid when the incoming green energy is scarce or surplus energy can be injected into the grid. \acp{MEH} are batteryless, as batteries are often expensive and need periodic replacement -- \ac{EASE} aims at reducing the carbon footprint of such batteryless \ac{eNB}/\ac{MEH} system while meeting memory, processing constraints and accounting for the user mobility.

\begin{figure*}[tb]
	\begin{center}
	\includegraphics[width=\textwidth]{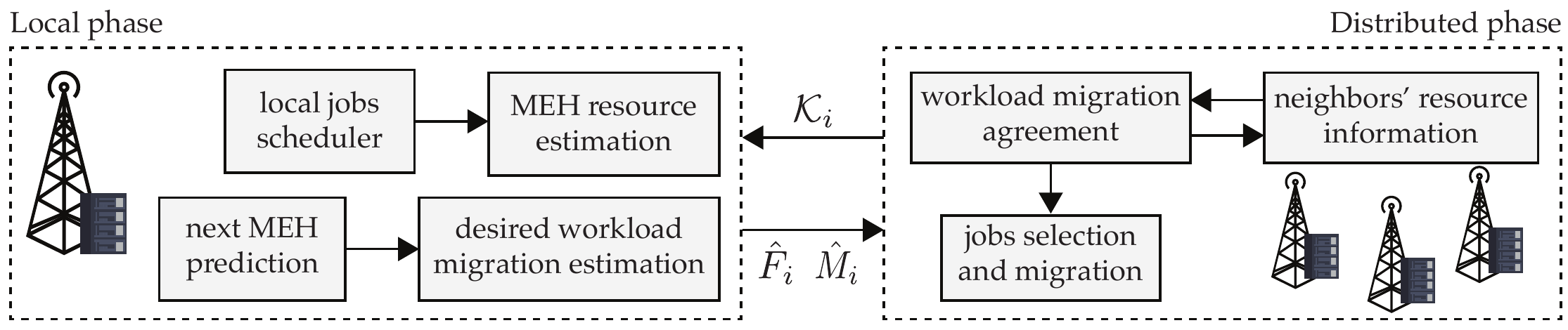}
	\setlength\abovecaptionskip{-.4cm}
	\setlength\belowcaptionskip{-.4cm}
	\caption{\textbf{High level diagram of \ac{EASE}.} The local steps (left) provide the resource and the desired workload migration estimates for each \ac{MEH} in isolation. The distributed algorithm (right) allows \acp{MEH} to reach a consensus on the jobs allocation and trigger their migration.}
	\label{fig:block_diagram}
	\end{center}
\end{figure*}

\begin{figure}[tb]
	\begin{center}
	\includegraphics[width=.9\columnwidth]{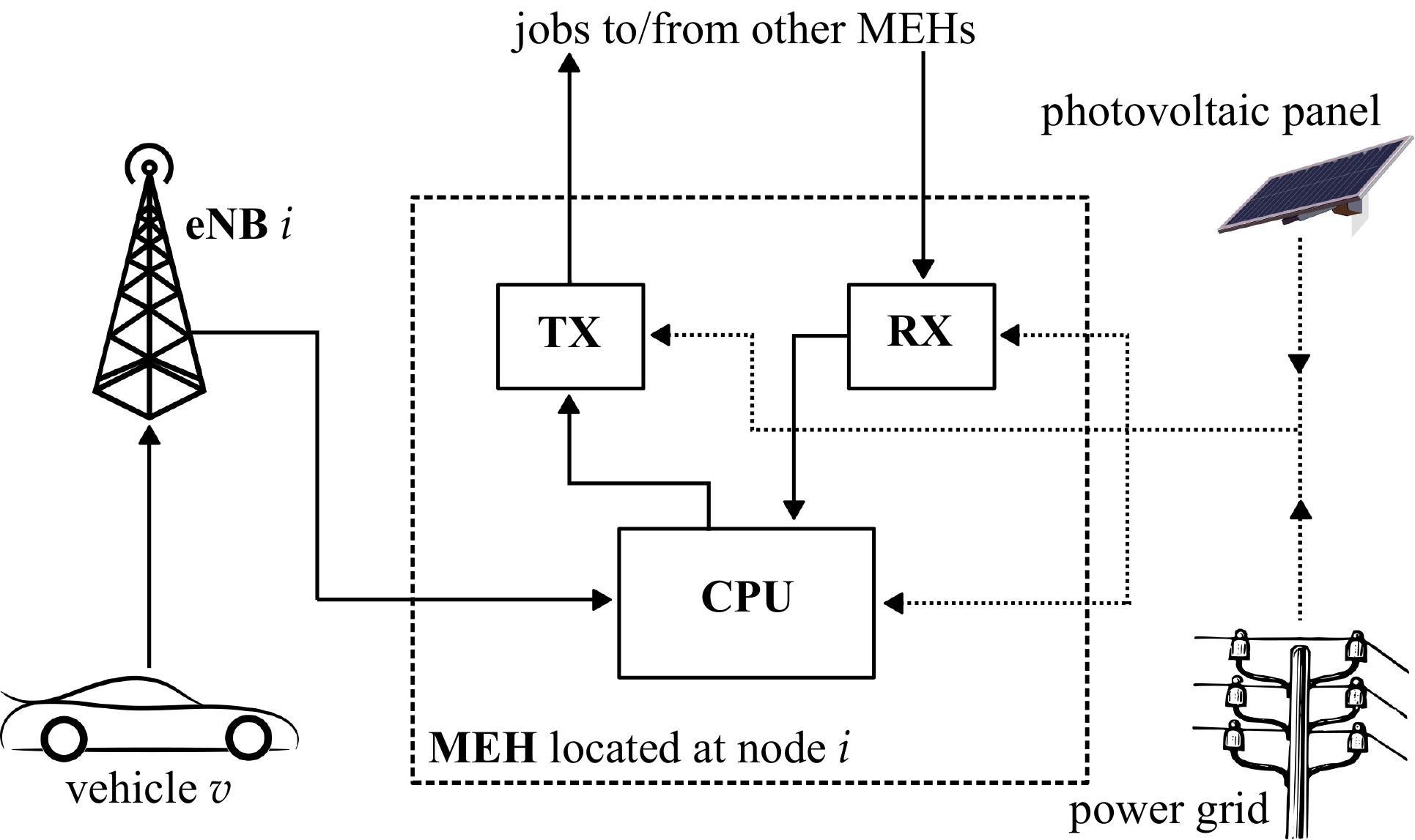}
	\setlength\abovecaptionskip{-.4cm}
	\setlength\belowcaptionskip{-1cm}
	\end{center}
	\caption{\textbf{\ac{eNB}/\ac{MEH} node}. Job requests arrive from connected vehicles $v$ moving within the \ac{eNB} coverage area. Containers handling the execution of the jobs are created at the serving \ac{MEH}, and possibly migrated to other \acp{MEH} in case the associated vehicles exit the \ac{eNB} coverage area. %Nodes are equipped with energy harvesting devices and connected to the power grid.
	}\label{fig:access_node}
\end{figure}

The diagram of an \ac{eNB}/\ac{MEH} node is shown in Fig.~\ref{fig:access_node}, while a high level diagram of \ac{EASE} is presented in \figurename~\ref{fig:block_diagram}. The scheduler operates according to two optimization phases:
\begin{inparaenum}
\item a local phase (left of the diagram): a predictive control phase, performed locally at each \ac{MEH} node, and
\label{step1}
\item a distributed phase (right): a collaborative optimization based on distributed consensus (solved via message passing).
\label{step2} 
\end{inparaenum}
In phase \ref{step1}, the \acp{MEH} locally control the ongoing computations, estimating the local processing capacity and energy availability within a given prediction horizon. At the same time, the local algorithm assesses the amount of workload that should be migrated (``desired workload migration estimation'') to the neighboring \ac{MEH} nodes, predicts the availability of local resources (``\ac{MEH} resource estimation''), and accounts for mobility estimates (``next \ac{MEH} prediction''), i.e., the vehicle that generated the job request is about to hand over to a neighboring radio cell.

With phase \ref{step2}, taking the desired workload to be migrated from phase \ref{step1} as input (``neighbors' resource information''), the \acp{MEH} collectively reach an agreement (``workload migration agreement'') about \emph{how many} and \emph{which} jobs are to be actually migrated, as well as about the target \ac{MEH} for their migration (``job selection and migration'').

After phase~\ref{step2}), each node updates its local state equations with the new jobs generated by the vehicles under coverage and those received from the neighbors, and goes back to phase~\ref{step1}).

In the remainder, the system model is presented in Section~\ref{sec:SystemModel}. The problem formulation for the optimal scheduling is detailed in Section~\ref{sec:ProblemFormulation}. The final scheduling solution, composed of the two phases (local and distributed) is presented in Section~\ref{sec:ProblemSolution}. The performance assessment is reported in Section~\ref{sec:NumericalResuls} and final remarks are provided in Section~\ref{sec:Conclusions}.

%\revision{GP: It is not necessary that the procedure is synchronized. It may be that step 1 is done repeatedly and an iteration of step 2 only when an update from the neighbors is available. In my opinion, step 1, which is centralized and with lower complexity, should be done at a high rate. Step 2 can be an asynchronous algorithm to reach consensus on the migrations, performed continuously when updates are available.} 

\section{System Model}\label{sec:SystemModel}
%!TEX root = ./paper.tex

Next, we detail the mathematical models for computing and communication services, along with the statistical processes involved in the envisioned scenario and the system constraints. Time $t$ is discrete and evolves according to slots of fixed duration $\tau$, i.e., $t=0,\tau, 2\tau, \dots$. The mathematical notation is summarized in Table~\ref{tab:system_symbols}.

%\begin{figure}[tb]
%	\begin{center}
%	\unitlength=.284mm
%	\begin{picture}(250,156)(0,0) 
%	\put(-15,0){\includegraphics[width=.9\columnwidth]{./figures/access_node.eps}}
%	\put(91,97){\footnotesize \textbf{TX}}
%	\put(161,97){\footnotesize \textbf{RX}}
%	\put(121.5,42){\small \textbf{CPU}}
%	\put(24,110){\small \textbf{\ac{eNB}} $i$}
%	\put(-13,3){\footnotesize vehicle $v \in \mathcal{V}$}
%	\put(75,15){\footnotesize \textbf{\ac{MEH}} located at node $i \in\mathcal{N}$}
%	\put(222,146){\footnotesize solar panel}
%	\put(222,-12){\footnotesize power grid}
%	\end{picture}
%	\end{center}
%	\vspace{2mm}
%	\caption{\textbf{\ac{eNB}/\ac{MEH} node}. Job requests arrive from connected vehicles $v$ moving within the \ac{eNB} coverage area. Containers handling the execution of the jobs are created at the serving \ac{MEH}, and possibly migrated to other \acp{MEH} in case the associated vehicles exit the \ac{eNB} coverage area. %Nodes are equipped with energy harvesting devices and connected to the power grid.
%	}\label{fig:access_node}
%\end{figure}

\begin{table}[tb]
\small
\centering
\resizebox{\columnwidth}{!}{%
\begin{tabular}{l|l|l}
\textbf{Symbol} & \textbf{Meaning}  & \textbf{Unit} \\
\hline
\hline
$v \in \mathcal{V}$ & vehicle \ac{ID} and set of vehicles & - \\
%$v$ & vehicle \ac{ID} & $\in \mathcal{V}$\\
$i \in \mathcal{N}$ & \ac{eNB}/\ac{MEH} \ac{ID} and set of \acp{eNB}/\acp{MEH} & - \\
%$i$ & \ac{eNB}/\ac{MEH} \ac{ID} & $\in \mathcal{N}$ \\
$\mathcal{N}_i$ and $N_i$ & set of neighboring nodes of node $i$ and its cardinality $|\mathcal{N}_i|$& - \\
\multirow{2}{*}{$k \in \mathcal{K}_i(t)$ and $K_i(t)$} & job \ac{ID}, set of jobs in execution at \ac{MEH} $i$ at slot $t$, & \multirow{2}{*}{-}\\
& and its cardinality $|\mathcal{K}_i(t)|$ &  \\
\multirow{2}{*}{$\hat{\mathcal{K}}_{ij}$ and $\hat{K}_{ij}$} & set of jobs running on \ac{MEH} $i$ with probable next \ac{MEH} $j$ &\multirow{2}{*}{-} \\
& and its cardinality & \\
% $N_i$ & $|\mathcal{N}_i|$, i.e., cardinality of $\mathcal{N}_i$ & - \\
$T$ & no. of slots in the prediction horizon & -\\
$t=[0,\ldots,T]$ & scheduling time slot index & -\\
$\tau$ & length of a scheduling slot & s\\
\multirow{2}{*}{$V_i(t)$ and $C_i(t)$} & no. of results to be sent in the coverage area of \ac{eNB} $i$ at & \multirow{2}{*}{-}\\
& slot $t$ and to be routed through the backhaul network &\\
%$K_i(t)$ & $|\mathcal{K}_i(t)|$, i.e., cardinality of $\mathcal{K}_i(t)$ & - \\
%$k$ & job \ac{ID} & \\
%$\ell$ & job type \ac{ID} & -\\
$I_k$ & intensity of job $k$ & cyc.\\
$D_k$ & deadline of job $k$ & s\\
$S_k$ & size of job $k$ & bit\\
%$D^\ell$ & starting deadline of a job of type $\ell$ & s\\
%$S^\ell$ & starting data size of a job of type $\ell$ & bits\\
%$I_{i,k}(t)$ & residual intensity of job $k$ hosted by \ac{MEH} $i$ at slot $t$ & cyc.\\
%$D_{i,k}(t)$ & residual deadline of job $k$ hosted by \ac{MEH} $i$ at slot $t$ & s\\
%$S_{i,k}(t)$ & residual data size of job $k$ hosted by \ac{MEH} $i$ at slot $t$ & bits\\
$p_v$ and $p_\ell$ & job generation probability and probability that it is of type $\ell$ & -\\
%$p_\ell$ & probability that the generated job is of type $\ell$ & -\\
$\bm p_{i,k}(t)$ & handover prob vector for vehicle $v$ (job $k$) at slot $t$ & \\
$w_{i,k}(t)$ & workload of job $k$ processed by \ac{MEH} $i$ in slot $t$ & cyc.\\
%$\bm{w}_i(t)$ & vector collecting $w_{i,k}(t)$ for all jobs at \ac{MEH} $i$, slot $t$ & cyc.\\
%$\bm{I}_i(t)$ & residual intensity vector of jobs hosted by \ac{MEH} $i$ at slot $t$ & cyc.\\
%$\bm{D}_i(t)$ & residual deadline vector of jobs hosted by \ac{MEH} $i$ at slot $t$ & s\\
$L$ & (fixed) size of a container instantiated on a \ac{MEH} & bit\\
$E_{b}^{\rm RAN}$ & energy per bit for \ac{eNB}-vehicle wireless transmissions & J/bit\\
$E_{b}^{\rm wired}$ & energy per bit for \ac{eNB}-\ac{eNB} wired transmission & J/bit\\
$\sigma_s$ and $\sigma_d$  & energy per bit for migration at the source (destination) \ac{MEH} & J/bit\\
%$\sigma_d$ & energy per bit for migration at the target \ac{MEH} & J/bit\\
$E_s$ and $E_d$ & (fixed) energy for migration at the source (destination) \ac{MEH} & J\\
%$E_d$ & (fixed) energy for migration at the target \ac{MEH} & J\\
$E^{\rm H}_{i}(t)$ & harvested energy available at slot $t$ & J\\
$P_{i}^{\rm PV}(t)$ & power supplied by the \ac{PV} at node $i$, instant $t$ & W\\
$P_{\rm RAN}$ and $P_{\rm wired}$ & (fixed) power to keep the wireless (wired) unit switched on & W\\
$P_{i}^{\rm idle}$ & (fixed) power to keep the server switched on & W \\
$N_{i}^{\rm inc}(t)$ and $N_{i}^{\rm out}(t)$  & no. of \ac{MEH} incoming (outgoing) jobs at slot $t$ & -\\
%$N_{i,\rm out}(t)$ & number of \ac{MEH} outgoing jobs at slot $t$ & - \\
$F_i$ & maximum computational power of server $i$ & W \\
$M_i$ & maximum amount of RAM available at server $i$ & bit\\
%$C_i(t)$ & number of jobs' results to be routed through the backhaul & -\\
%$\mathcal{W}_i$ & stacks of vectors $\bm{w}_i(t)$ for $T$ slots & cyc.\\
%$\mathcal{I}_i$ & stacks of vectors $\bm{I}_i(t)$ for $T$ slots & cyc.\\
%$\mathcal{D}_i$ & stacks of vectors $\bm{D}_i(t)$ for $T$ slots & s\\
%$\bm{V}_i$ & vector collecting $V_i(t)$ over $T$ slots & -\\
%$\bm{C}_i$ & vector collecting $C_i(t)$ over $T$ slots & -\\
%$\bm E_{{\rm H}, i}$ & vector collecting $E_{{\rm H}, i}(t)$ over $T$ slots & J\\
%$\delta_{F_i(t)}$ & auxiliary variable for the local controller & - \\
%$\delta_{M_i(t)}$ & auxiliary variable for the local controller & - \\
$\bar{w}_{ij}$ & desired intensity requested by \ac{MEH} $i$ to neighbor $j$ & cyc./s\\
%$\bar{\bm{w}}_i$ & vector collecting $\bar{w}_{ij}$ for all the neighbors & cyc./s\\
%$\tilde{\bm{w}}_i$ & vector collecting $\bar{w}_{ji}$ for all the neighbors & cyc./s\\
$\bar{m}_{ji}$ & memory space requested by \ac{MEH} $i$ to neighbor $j$ & bit \\
$\hat{P}^{\rm H}_{i}$ & residual green power at node $i$ after the local scheduling& W\\
%$\hat{\bm P}_{\rm H}$ & vector collecting $\hat{P}_{{\rm H}, i}$ $\forall i \in \mathcal{N}$ & W\\
$\hat{F}_i$ & residual computing power at node $i$ after local scheduling & W \\
$\hat{M}_i$ & residual RAM memory at node $i$ after the local scheduling & bit\\
$o_{ij}$ & optimal amount of \ac{MEH} $i$ processing load to offload to $j$ & cyc./s\\
%$\bm o_i$ & vector collecting $o_{ij}$ for all the neighbors & cyc.\\
$\tilde{o}_{ji}$ & optimal processing load to be received at \ac{MEH} $i$ from $j$ & cyc./s\\
%$\tilde{\bm o}_i$ & vector collecting $\tilde{o}_{ji}$ for all the neighbors & cyc.\\
%$\rho$ & weight of the quadratic penalty in Eq.\eqref{eq:distr_cost} & \\
%$\xi_{M_i}$ & coefficient for the job migration & cyc./(s $\cdot$ bit) \\
%$\hat{\delta}$ & Eq. 19-20 & \\
%$\hat{c_i}$ & Eq. 20& \\
%$\hat{\delta_i}$ & Eq. 20 & \\
%$\hat{\bm \delta}$ & Eq. 21 & \\
%$\bm d$ & $\{\bm d_i=[\min\{\hat{F}_i, \xi_{M_i} \hat{M}_i\}, \bm 0], i \in\mathcal{N}\}$ & \\
%$\bm q_i$ &linear cost vector $[\bm q^{\textrm{tx}}_i - \bm q^{\textrm{proc}}_i, \bm q^{\textrm{rx}}_i + \bm q^{\textrm{proc}}_i, 0]$\\
%$\bm b_i$ & tracking target vector $[\bar{\bm w}_i, \tilde{\bm w}_i, 0]$ & \\
%$\bm m_i$ & $[\rho/2 \ldots \rho/2, \hat{c}_i]$ & \\
%$Q_i$ & $I_{2N_i+1}\bm m_i$ & \\
\hline
\end{tabular}%
}
\setlength\belowcaptionskip{-.6cm}
\caption{Summary of the symbols used within the paper. ``cyc.'' stands for ``CPU cycles''.}
\label{tab:system_symbols}
\end{table}

\subsection{Computation and communication models}
\noindent\textbf{Computing job parameters.}
At time $t$, each job $k$ served by \ac{MEH} $i$ is characterized by the triplet $(I_{i,k}(t), D_{i,k}(t), S_{i,k}(t))$, where
\begin{inparaenum}[i)]
\item $I_{i,k}(t)$ is the residual job intensity, expressed in CPU cycles,
\item $D_{i,k}(t)$ is the residual (hard) execution deadline, in seconds, i.e., the time still available to execute the job, and
\item $S_{i,k}(t)$ is the remaining data to be processed, in bits.
\end{inparaenum}
As the job is processed by the server, the intensity, deadline, and data size decrease according to %the following state equations 
\begin{align}
\label{eq:intensity}
I_{i,k}(t+\tau) &= I_{i,k}(t) - w_{i,k}(t)\,,\\
\label{eq:deadline}
D_{i,k}(t+\tau) &= D_{i,k}(t) - \tau\,,\\
\label{eq:size}
S_{i,k}(t+\tau) &= S_{i,k}(t) - \frac{S_{i,k}(0)}{I_{i,k}(0)}w_{i,k}(t)\,,
\end{align}
where $w_{i,k}(t)$ is the amount of workload (CPU cycles) belonging to job $k$ and processed by \ac{MEH} $i$ in slot $t$, $S_{i,k}(0)$ represents the initial job size (bits), whereas $I_{i,k}(0)$ is the total number of CPU cycles required to fully process the job. Eq.~\eqref{eq:size} means that the amount of data that is still to be processed decreases linearly with the amount of workload allotted to a job, irrespective of how the workload is distributed in time. Note that \eqref{eq:intensity} makes it possible to rewrite \eqref{eq:size} as
\begin{align}
S_{i,k}(t) &= \frac{S_{i,k}(0)}{I_{i,k}(0)} I_{i,k}(t).
\end{align}
%where the factor $S_{i,k}(0) / I_{i,k}(0)$ regulates the linear correlation between the remaining job intensity $I_{i,k}(t)$ and the data that still has to be processed for job $k$, $S_{i,k}(t)$.

\noindent\textbf{Communication models.}
For the \ac{5G} wireless links between the \acp{eNB} and the vehicles we adopt
\begin{inparaenum}[i)]
\item the massive-MIMO energy consumption model of~\cite{bjornson2017}, and
\item the \mbox{mm-wave} -- $28$~GHz -- urban NLoS channel model of~\cite{Ko2017}.
\end{inparaenum}
Specifically, from~\cite{bjornson2017} the following system parameters are obtained:
\begin{inparaenum}[i)]
\item the power needed to keep the wireless unit switched on (fixed circuit power consumption), $P_{\textrm{RAN}}$,
\item the energy required per transmitted bit via wireless links, $E_{b}^{\rm RAN}$,
\item the fixed wired circuit power consumption, $P_{\textrm{wired}}$,
\item the energy expenditure for the wired backhaul links connecting the \acp{eNB}, $E_{b}^{\rm wired}$.
\end{inparaenum}
Note that the vehicles' energy utilization is not involved in the scheduling and, in turn, only the energy consumption at the \ac{eNB} side is considered.
The model in \cite{Ko2017} is used for the vehicle-\ac{eNB} association.
%According to~\cite{bjornson2017}, a massive-MIMO system with $100$ antennas transmitting with power $P_{\textrm{RF}}=400$~mW, consumes $P_{\textrm{RAN}}=50.2$~W to keep the wireless unit switched on (fixed circuit power consumption). In the downlink (DL) direction, there is an additional term accounting for the transmission power of the \ac{eNB}. Specifically, with a power $P_{\textrm{tx}}=100$ mW and a fixed bitrate of $R_b=100$ Mb/s, the energy required per transmitted bit is $E_{b}^{\rm RAN} = P_{\textrm{tx}} / R_b = 1$~nJ/bit. For the wired backhaul links connecting the \acp{eNB} we consider an energy expenditure of $E_{b}^{\rm wired} = 250$~pJ/bit according to~\cite{bjornson2017}, and a circuitry fixed power $P_{\rm wired}=20$~W. 

\noindent\textbf{Container migration model.} The migration of a container requires the hosting \ac{MEH} to spend energy to freeze the status of the virtual entity and prepare the data to be sent to the target \ac{MEH} for the correct re-instantiation. Hence, the target \ac{MEH} has to spend energy to create the new virtual entity using the received information. The energy expenditure on the two sides consists of~\cite{liu2011performance}: 
\begin{inparaenum}[i)]
\item a contribution proportional to the size of the migration data, through the parameters  $\sigma_s$ and $\sigma_d$ respectively, plus 
\item a fixed energy contribution, equal to $E_s$ for the source \ac{MEH} and $E_d$ for the target one, respectively.
\end{inparaenum}
Additionally, the source spends some energy to transmit the data over the wired channel $E_{b}^{\rm wired}$. 
Overall, it holds
\begin{align}
E^{\textrm{migr}}_ {\textrm{source}}(t)&=\sigma_s L + E_{b}^{\rm wired} S_{k}(t) + E_s, \quad \text{and} \label{eq:e_mig1} \\
E^{\textrm{migr}}_{\textrm{dest}}(t)&=\sigma_d L+E_d,\label{eq:e_mig2}
\end{align}
where $S_{k}(t)$ is the (variable) data size associated with job $k$, and $L$ is the (fixed) container size. 
According to \cite{campolo2019support}, we account for a service downtime of $T_k^{\rm migr}$ when migrating the entities. In turn, \mbox{$T_k^{\rm migr}$}~seconds are additionally removed from the job's deadline $D_{k}(t)$ at every migration occurrence. Note that the delay associated with wired transmissions is negligible as compared to the service downtime.
%Following~\cite{liu2011performance}, we consider \mbox{$\sigma_s=\sigma_d = 500$}~nJ/bit and \mbox{$E_s=E_d=250$}~mJ. We use $L=50$~MB for the containers' size and account for a service downtime of $2$~s when migrating the entities~\cite{campolo2019support}. The delay associated with wired transmissions is negligible as compared to the service downtime. Thus, \mbox{$T_k^{\rm migr}=2$}~s are additionally removed from the job's deadline $D_{k}(t)$ at every migration occurrence.

\subsection{Statistical processes}
\label{sec:processes}
\noindent\textbf{Energy harvesting model.} We refer to $P_{i}^{\rm PV}(t)$ as the power supplied by the \ac{PV} co-located with \ac{eNB}/\ac{MEH} $i$ at instant $t$ and that varies from a minimum of $P^{\rm PV}_{\rm min}$ to a maximum of $P^{\rm PV}_{\rm max}$. 
%Depending on the solar irradiance and the hour of the day, $P_{i}^{\rm PV}(t)$ for a typical modern \ac{PV} varies from a minimum of $P^{\rm PV}_{\rm min}=250$~W to a maximum of $P^{\rm PV}_{\rm max}=400$~W. 
Accounting for the power required to keep the server ($P_{i}^{\rm idle}$) and the communication channels ($P_{\rm RAN}$ and $P_{\rm wired}$) switched on, and the fixed amount of energy required for the container migration, the harvested energy available at \ac{eNB}/\ac{MEH} $i$ for computations and data transmissions at time slot $t$ is
%Not all the amount of generated power, however, will be available for computations and data transmissions. Specifically, the fixed terms to keep the server ($P_{i, \rm idle}$) and the wireless channel ($P_{\rm RAN}$) switched on are to be removed from the available power, as well as the fixed amount of energy required for the container migration. Therefore, the remaining useful harvested energy for slot $t$ is
\begin{equation}
\begin{aligned}
\label{eq:harvested}
&&E^{\rm H}_{i}(t)=&\left(P_{i}^{\rm PV}(t)-P_{\rm RAN}-P_{\rm wired}-P_{i}^{\rm idle}\right) \tau\, + \\
&&& - N_{i}^{\rm inc}(t)\left(\sigma_d \,L + E_d\right) \,+\\
&&& - N_{i}^{\rm out}(t)\left[\left(\sigma_s + E_{b}^{\rm wired} \right)L + E_s\right]
\end{aligned} 
\end{equation}
where $N_{i}^{\rm inc}(t)$ and $N_{i}^{\rm out}(t)$ are the known number of \ac{MEH} incoming and outgoing jobs at \ac{MEH} $i$ and time $t$, which are scheduled at the previous step $t-\tau$. The terms in Eqs.~\eqref{eq:e_mig1}-\eqref{eq:e_mig2} that depend on the data size $S_{i,k}(t)$ are not considered in $E^{\rm H}_{i}(t)$ as they will be integrated in the optimization function (see Eq.~\eqref{eq:distr_cost}).
%$E^{\rm H}_{i}(t)$ is known for the current slot only, while its estimates for future time slots are obtained by considering a Gaussian distribution with an average computed over the previously observed outcomes, and fixed variance. %is kept at each \ac{MEH} %with an average equal to the average value of the previously observed outcomes, and fixed variance.
Note that being $E^{\rm H}_{i}(t)$ a difference between the harvested energy and that required to deliver the services, its value can be negative. $E^{\rm H}_{i}(t)$ is known for the current slot $t$ only. However, the developed \ac{MPC} framework also needs estimates for \mbox{$[E_i^H(t+\tau),\ldots,E_i^H(t+\tau (T-1))]$}, within the time window $t+\tau,\dots,t+\tau (T-1)$, where $T$ is the prediction horizon. These estimates are computed by forecasting the time-dependent quantities in \eqref{eq:harvested}: future values of $P_i^{\rm PV}(t+\cdot)$ are estimated using a Gaussian r.v. with average $P_{\rm PV}$ and standard deviation $\sigma_{\rm PV}$, estimates for the number of incoming \mbox{$N_i^{\rm inc}(t+\cdot)$} and outgoing $N_i^{\rm out}(t+\cdot)$ jobs at \ac{eNB} $i$ in slot $t$ are obtained considering the vehicles in the external annulus of the \ac{eNB}'s coverage area. Finally, $P_{i}^ {\rm idle}$ depends on the specific \ac{MEH} characteristics at \ac{eNB} $i$, as specified in Section~\ref{sec:NumericalResuls}.

\noindent\textbf{Jobs types and arrival model.} Three job types are considered for the numerical results of Section~\ref{sec:NumericalResuls}, having different intensities, deadlines, and data sizes and identified through the index $\ell=\{1,2,3\}$. Every job type is associated with a generation triplet~\mbox{$(I^\ell, D^\ell, S^\ell)$}, and a generation probability $p_\ell$. Each vehicle $v \in\mathcal{V}$ can submit at most one computing job at a time to the network facilities, so that a bijective mapping vehicle-job \ac{ID} can be derived. Once a job is finished or expired, the vehicle submits a new job to the \ac{MEH} with probability $p_v$ at each slot. This parameter is tuned in the simulations. Also in this case, for predictive optimization, an estimate for the future incoming jobs is needed. For this purpose, a circular buffer containing the values of $I_{i,k}/D_{i,k}$ of the newly generated jobs is kept. A fixed estimate of the average of the last $W$ seconds is used to predict the incoming traffic. In~\cite{perin2021towards}, the authors verified that even simple predictors are still effective with \ac{MPC} if $T$ is large enough. 
%In the simulations, we use a circular buffer of length $W=5$ minutes. 

\noindent\textbf{Handover probabilities.} Each job $k$ is associated with a probability vector that depends on the position of the vehicle $v$ requesting the service. Being $i$ the serving  \ac{eNB} for vehicle $v$, we define $\bm p_{i,k}(t)$ as the \mbox{$N_i$-dimensional} vector containing the probabilities that vehicle $v$ will hand over to any of the \mbox{$|\mathcal{N}_i| = N_i$} neighboring radio cells, i.e., \mbox{$\bm p_{i,k}(t)= [p_{i1,k}(t), p_{i1,k}(t), \dots, p_{iN_i,k}(t)]$}, with \mbox{$\sum_j p_{ij,k} = 1$}. Vector $\bm p_{i,k}(t)$ is updated every time a new trajectory sample is available for the associated vehicle $v$, either inside the same cell or in a new cell after performing the handover. 
%\begin{equation}
%	\sum_{b=0}^{N-1} p_{k, n}(t) = 1
%\end{equation}
%Note that the non-zeros entries of the vector $\bm p^i_{k}(t)$ depend on the cell $j$ where the vehicle is located, which may differ from the cell where job $k$ is running as the communication and computation services can be offered by entities in two different cells, i.e., eNB $j$ for the communication, server $i$ for the computation.

\subsection{System constraints}\label{subsec:constraints}
The set $\mathcal{K}_i(t)$, with cardinality$K_i(t) = |\mathcal{K}_i(t)|$, collects the jobs being executed at time slot $t$ at \ac{MEH} $i$. The following systems constraints apply\newline
%\noindent\textbf{\ac{VM}s creation.} The number of \ac{VM}s that can be activated at server $i$ is limited to $\bar{K}_i$, that is
%\begin{equation}
%\label{eq:tot_VM}
%0 \leq K_i \leq \bar{K}_i\,.
%\end{equation}
%\newline
\noindent\textbf{Processing capacity.} Indicating with $F_i$ the maximum computing power of server $i$ -- expressed in CPU cycles per second -- the following inequality on the sum of the workloads holds
\begin{equation}
\label{eq:proc_capacity}
\frac{1}{\tau}\sum_{k=1}^{K_i(t)} w_{i,k}(t) \leq F_i\,.
\end{equation}
%We allow the constraint to be violated by $\delta_F$, penalizing this term in the cost function.
\noindent\textbf{Storage capacity.} Being $M_i$~[bits] the maximum amount of RAM available at server $i$, the sum of the data sizes $S_{i,k}(t)$ of all the active jobs at \ac{MEH} $i$ must obey
\begin{equation}
\label{eq:storage_capacity}
\sum_{k=1}^{K_i(t)} S_{i,k}(t) \leq M_i\,.
\end{equation}
%Also here we allow the constraint to be violated with a penalty term $\delta_M$, added to the cost function. 
%\noindent\textbf{Channel capacity.} We denote with $\mathcal{N}_i$ the set of neighboring nodes of $i$ and $R_{ij}$ the maximum transmission rate of the backhaul link between server $i$ and server $j \in \mathcal{N}_i$. When migrating a \ac{VM} to another server, the relative data must also be transmitted. Therefore, at any time slot $t$,
%\begin{equation}
%\label{eq:max_rate}
%\frac{1}{\tau}\sum_{k=1}^{K_{ij}} S_k \leq R_{ij} \qquad \forall j \in \mathcal{N}_i
%\end{equation}
%must hold, where $K_{ij}$ is the total number of \ac{VM} migrated from server $i$ to $j$. When routing just the result to the \ac{UE}, the transmission overhead introduced is assumed to be negligible.
%\newline
\noindent\textbf{Job execution time.} In case the deadline of job $k$, $D_{i,k}(t)$, expires in the current time slot $t$, the job must be processed entirely and immediately at server $i$ and cannot be further migrated, i.e.,
\begin{equation}
\label{eq:forced}
w_{i,k}(t) = I_{i,k}(t) \qquad \text{if} \; D_{i,k}(t) \leq \tau\,.
\end{equation}
This guarantees the timely delivery of the computation result to the requesting vehicle, avoiding that the outcome becomes useless. 
As Eqs.~\eqref{eq:proc_capacity}-\eqref{eq:forced} may not be jointly satisfied, in the following we will relax Eq.~\eqref{eq:proc_capacity}.

\noindent\textbf{Workload conservation.} Finally, note that, in general, the inequalities
\begin{equation}
\label{eq:conservation}
0 \leq w_{i,k}(t) \leq I_{i,k}(t), \; \forall \; i \in \mathcal{N}, \; \forall \; k \in \mathcal{K}_i(t),  \; \forall \; t
\end{equation}
must always hold, because of the workload conservation principle. 

\section{Problem Formulation}\label{sec:ProblemFormulation}
%!TEX root = ./paper.tex

Here we formulate the optimization problems concerning the 1) local and 2) distributed scheduling phases introduced in Section~\ref{sec:Workflow}. As shown in \figurename~\ref{fig:block_diagram}, the local and distributed schedulings are run in parallel as distinct tasks that exchange information.

\subsection{Local phase: Local controller and resources estimation}
Each \ac{MEH} $i \in \mathcal{N}$ estimates $w_{i,k}(t)$ for every job $k \in \mathcal{K}_i(t)$ to be executed at time $t$: in the analysis, $w_{i,k}(t)$ stands for the optimal fraction of computing intensity $I_{i,k}(t)$ to be locally executed at time slot $t$ for the hosted job $k$. We define vectors $\bm{w}_i(t)$, $\bm I_i(t)$ and $\bm D_i(t)$ respectively collecting $w_{i,k}(t)$, $I_{i,k}(t)$ and $D_{i,k}(t)$ for all $k \in \mathcal{K}_i(t)$. As for the energy spent to transmit the processing results back to the vehicles, \mbox{$V_i(t)\,E_{b}^{\rm RAN}$} is the (per bit) energy cost of sending the results to the $V_i(t)$ vehicles in the wireless coverage area, while \mbox{$C_i(t)\,E_{b}^{\rm wired}$} is the energy cost entailed in routing the $C_i(t)$  jobs that are completed at node $i$ and that have to be routed via the backhaul links to reach the corresponding user (vehicle). $R_k$ is the size of the processing result of job $k$, and $q_i^{\rm proc}$ is the energy cost of processing a unit of workload. 

Given these quantities, we define two local (at node $i$) functions $f_i(\cdot)$ and $g_i(\cdot)$, as follows.
\begin{equation}
\begin{aligned}
\label{eq:res_energy}
&&f_i(\bm w_i; V_i, C_i, E^{\rm H}_i) = &\;q_i^{\rm proc} \bm 1^{\scriptscriptstyle{T}} \bm w_i(t) + V_i(t)\,E_{b}^{\rm RAN}R_k\,+\\
&&&+C_i(t)\,E_{b}^{\rm wired}R_k - E^{\rm H}_{i}(t)\,,
\end{aligned}
\end{equation}
\vspace{-0.3cm}
\begin{equation}
\label{eq:state_cost}
g_i(\bm I_i(t); \bm D_i(t)) = \sum_{k=1}^{K_i(t)} \left(\frac{I_{i,k}(t)}{D_{i,k}(t)} \right)^2.
\end{equation}
% \mbox{$\bm{w}_i(t) = [w_{i,1}(t),\ldots,w_{i,K_i}(t)]$}
$f_i(\cdot)$ quantifies the difference between the total energy expenditure at node $i$ in slot $t$ (due to processing and communications processes) and the energy that is locally harvested at this node. Hence, $-f(\bm w_i; \cdot)$ represents the residual cost-free energy available for the migration process in the distributed phase. Minimizing $f_i(\cdot)$ corresponds to maximizing the local energy available at the node. $g_i(\cdot)$ represents the residual processing cost, which is proportional to $(I_{i,k}/D_{i,k})^2$. Minimizing $g_i(\cdot)$ forces the node to execute the jobs, especially prioritizing those with high intensity and whose deadline is about to expire. Note also that, due to Eq.~\eqref{eq:intensity}, $I_{i,k}(t)$ depends on the optimization variable $w_{i,k}$ at previous time slots.

Considering a forecast optimization window of $T$ slots into the future, and letting $t=0$ be the current time slot, the local cost function at node $i$ over the whole time horizon is formulated by combining $f_i(\cdot)$ and $g_i(\cdot)$, as
\begin{equation}
\begin{aligned}
\label{eq:centr_cost}
&&J_i\left(\mathcal{W}_i, \mathcal{I}_i; \mathcal{D}_i, \bm V_i, \bm C_i, \bm E^{\rm H}_{i}\right) = \,&\gamma \sum_{t=0}^{T-1} \, g_i(\bm I_i(t); \bm D_i(t))\,+\\
&&&+ \sum_{t=0}^{T-1} \max\{f_i(\bm w_i;\cdot), 0\}^2,
\end{aligned}
\end{equation}
where $\mathcal{W}_i$, $\mathcal{I}_i$ and $\mathcal{D}_i$ represent the stacks of vectors $\bm w_i(t)$, $\bm I_i(t)$ and $\bm D_i(t)$ over the considered horizon $T$, respectively, while $\bm V_i$, $\bm C_i$ and $\bm E^{\rm H}_i$ are the vectors collecting $V_i(t)$, $C_i(t)$ and $E^{\rm H}_i(t)$ for $t \in \{0,\tau, \dots, \tau (T-1)\}$. The coefficient $\gamma > 0$ is used to balance the processing state cost term ($g_i(\cdot)$) with respect to the energy cost ($f_i(\cdot)$).
\begin{remark}
From a physical perspective, the processing energy consumption is not necessarily a quadratic function, but it varies based on the specific computing architecture~\cite{specpower}. A quadratic function for $f_i(\cdot)$ was chosen, as it promotes smoothness of the controller in the transitions from one slot to the next one, and has the same curvature order of the processing state cost $g_i(\cdot)$. Also, the $\max\{ \cdot \}$ function is used to make the cost positive only when $f_i(\cdot)>0$, i.e., the renewable energy is fully used and the node has to resort to the power grid.
\end{remark} 
%\begin{remark}
%A positive cost on the left branch of the parabola $f^2(\bm w_i; \cdot)$ is also chosen, because using \emph{now} less energy than what is available may be harmful to the \emph{future} scheduling. Since, however, the primary goal is to avoid the power grid facilities, we set $\xi < 1$, so that the right branch is more costly.
%\end{remark}
Next, the cost function in Eq.~\eqref{eq:centr_cost} is modified through the addition of a penalty term proportional to two non-negative auxiliary variables \mbox{$\bm \delta_i(t) = [\delta_{F_i}(t), \delta_{M_i}(t)]$}, to ensure that the problem does not become infeasible when resources are scarce. Therefore, rewriting the constraints~\eqref{eq:proc_capacity} and~\eqref{eq:storage_capacity}, we define for each \ac{MEH} the following local problem at node $i$,
\begin{equation}
\label{eq:centr_problem}
\begin{aligned}
P^{\rm loc}_i: &&\min_{\mathcal{W}_i, \bm\delta_i} \quad & J_i\left(\mathcal{W}_i, \bm \delta_i; \cdot \right) + \sum_{t=0}^{T-1} \bm c_i^{\scriptscriptstyle{T}}\bm\delta_i(t)\\
&&\text{s.t.} \quad & \eqref{eq:intensity}\,\text{-}\,\eqref{eq:size},\, \eqref{eq:forced},\, \eqref{eq:conservation},\\
&&& \frac{1}{\tau} \sum_{k \in \mathcal{K}_i} w_{i,k}(t) \le F_i + \delta_{F_i}(t) ,\\
&&& \sum_{k \in \mathcal{K}_i} S_{i,k}(t) \le M_i + \delta_{M_i}(t), \\
&&& \delta_{F_i}(t) \geq 0, \, \delta_{M_i}(t) \geq 0,
\end{aligned}
\end{equation}
where $\bm c_i = [c_{F_i}, c_{M_i}]$ is the vector collecting the coefficients weighting the penalty variables, with $c_{F_i}, c_{M_i} > 0$. By solving \eqref{eq:centr_problem}, each \ac{MEH} obtains the optimal control $\bm w_i(0)$ which is implemented in the current time step.
%and an estimate of the available resources over the prediction horizon $T$.

%\revision{GP: processing consumption is a linear function, but with a penalty for switching the server on (about $30\%$ of power just for being on) \cite{barroso2007case, sen2017energy}. Should we add the penalty if $\bm w_i > 0$? Also, maybe it is a good idea to put a cost on the states $I_k$ to control it to $0$ (or even better, weighing it with the ratio $I_k/D_k$, so that if a job is near the deadline, it costs more).}

\subsection{Distributed phase: Workload migration agreement}
\label{sec:migration_agreement}

From \eqref{eq:centr_problem}, each server estimates its future energy and processing resources. Specifically, let $\hat{P}^{\rm H}_{i}$ be the residual available green power, possibly negative if the grid support is sought, $\hat{F}_i$, and $\hat{M}_i$ be the residual computational power, and RAM memory at node $i$, respectively. Note that, since constraints~\eqref{eq:proc_capacity} and~\eqref{eq:storage_capacity} are relaxed in~\eqref{eq:centr_problem}, $\hat{F}_i$ and $\hat{M}_i$ can be negative. These estimates are obtained by averaging the values over the prediction horizon, excluding the current instant $t=0$. Due to this averaging operation, while in \eqref{eq:centr_problem} we deal with energy expenditures, in the following we refer to power quantities. 

The migration task presents itself as a combinatorial \ac{MIP} problem, which is non-convex and is generally difficult to solve in a distributed fashion. Thus, we use heuristics to derive approximated solutions. In this work, the popular \emph{relax and round} method is used, which consists in solving the convex counterpart of the original problem, and rounding the result to a feasible solution afterward. The reason for this choice is that it allows tackling the problem in a distributed fashion via \emph{message passing}, solving the continuous form problem exactly to the optimum. Other approaches would have required a centralized solution or the design of a heuristic inspired by the optimization objective.

Based on the handover probability vector $\bm p_{i,k}$ presented in Section~\ref{sec:processes}, each \ac{MEH} determines the average resource demand requested from its neighbors in the migration process. Specifically, the CPU cycles per second and memory space that are requested from neighbor $j$ are
%\begin{align}
%\label{eq:avg_proc}
%\bar{w}_{ij} &= \sum_{k \in \hat{\mathcal{K}}_{ij}} \frac{I_{i,k}}{D_{i,k}}\,, \qquad\text{and} \\
%\label{eq:avg_mem}
%\bar{m}_{ij} &= \sum_{k \in \hat{\mathcal{K}}_{ij}} S_{i,k}\,,
%\end{align}
\begin{equation}
\label{eq:avg_proc}
\bar{w}_{ij} = \sum_{k \in \hat{\mathcal{K}}_{ij}} \frac{I_{i,k}}{D_{i,k}}\,, \,\,\, \text{and} \,\,\, \bar{m}_{ij} = \sum_{k \in \hat{\mathcal{K}}_{ij}} S_{i,k}\,,
\end{equation}
respectively, where $\hat{\mathcal{K}}_{ij}$ contains the set of jobs that are currently running at server $i$, associated with vehicles that are about to leave the coverage area of the co-located \ac{eNB} $i$ and whose most probable next \ac{eNB} is co-located with \ac{MEH} $j$. With $\bar{\bm w}_i = [\bar{w}_{i1}, \ldots,\bar{w}_{iN_i}]$ we denote the vector collecting the {\it desired} processing intensity per second to be sent to each of the $N_i$ neighbors of \ac{MEH} $i$, computed via \eqref{eq:avg_proc}. We also introduce the new optimization variables \mbox{$\bm o_i = [o_{i1},\ldots,o_{iN_i}]$} and \mbox{$\tilde{\bm o}_i = [\tilde{o}_{1i},\ldots,\tilde{o}_{N_ii}]$} representing the optimal total amount of processing load to be sent to, and to be received from each neighbor, respectively. The deviation from the desired $\bar{\bm w}_i$ to be migrated is penalized with the $l_2$-norm $\Vert \bar{\bm w}_i - \bm o_i\Vert^2$, and the migration cost is defined as
%\revision{IDEA: Based on $\bm p_{k}$, we compute the desired average workload occupation for the next slot, for each server in the network, in the following way. $\forall \;\text{job}\; k$ that we want to migrate to any $j \in \mathcal{N}_i$, the average processing requirements that we are asking to $j$ is $I_k/D_k$, so that $\sum_{k=1}^{K_{ij}} I_k/D_k \leq \hat{F_j}$. In short, we find the processing workload $\bm o_{ij}$ to be sent to $j$ in such a way that the previous constraint is kept (together with rate, nr. \ac{VM}s and memory constraints), to minimize the energy drained from the power grid. However, maybe $j$ is not going to accept everything that $i$ wants to give to it, because it does not have availability or other servers insist on it. Still, we want to be as close as possible to the desired migration place. 
%Therefore, a penalty is added considering a distance $d(\hat{\bm p_i},\bm o_i)$, where $\bm o_i = \lbrace o_{ij} \rbrace$ is the variable which keeps the \emph{total} computational effort in terms of workload to be sent to each server in the network. 
%$\hat{\bm p_i}$ is instead derived from vectors $\bm p_{k}$ for all jobs $k$ running on server $i$, and selects the \emph{desired} total computational effort to be sent to the servers in the network.}
%\revision{Specifically, $\sum_k (\argmax \bm p_{k} == j)$ gives the number of \ac{VM}s we want to migrate to server $j$. Then consider their average computational intensity. We can take $\lambda \Vert \hat{\bm p_i} - \bm o_i\Vert^2$ as a distance penalty.}
\begin{equation}
\begin{aligned}
\label{eq:distr_cost}
&&\Gamma_i\left(\bm o_i, \tilde{\bm o}_i; \bar{\bm w}_i, \hat{P}^{\rm H}_{ i}\right) =& \max\big\{\left(q_i^{\rm tx}-q_i^{\rm proc}\right) \bm 1^{\scriptscriptstyle{T}} \bm o_i\, +\\
&&&\quad+ \left(q_i^{\rm rx} + q_i^{\rm proc}\right) \bm 1^{\scriptscriptstyle{T}} \tilde{\bm o}_i - \hat{P}^{\rm H}_{i}, 0\big\}\,+\\
&&& + \rho \Vert \bm o_i - \bar{\bm w}_i\Vert^2\,,
\end{aligned}
\end{equation}
where $q_i^{\rm proc}$, $q_i^{\rm tx}$ and $q_i^{\rm rx}$ are the processing, transmission and reception costs of server $i$ (expressed as powers), respectively. The $\max\{\cdot\}$ term accounts for the power that would be drained from the power grid to migrate the jobs, whereas the quadratic term encodes the fact that the optimal $\bm o_i $ should be as close as possible to the desired $\bar{\bm w}_i$ -- this corresponds to moving the jobs to the next serving \ac{eNB}. Finally, $\rho > 0$ is a weight balancing the importance of the two cost terms. Note that minimizing Eq.~\eqref{eq:distr_cost} returns a solution $\bm o_i$ that matches vector $\bar{\bm w}_i$ if the residual harvested power is sufficient and the constraints are satisfied. Specifically, as system constraint we consider the following variation of~\eqref{eq:proc_capacity} and~\eqref{eq:storage_capacity}, introducing a variable $\hat{\delta}_i \geq 0$, as follows,
\begin{equation}
\label{eq:distr_proc_constraint}
\sum_{j \in \mathcal{N}_i} \left(\tilde{o}_{ji} - o_{ij}\right) \leq \min\{\hat{F}_i, \xi_{M_i}\hat{M}_i\} + \hat{\delta}_i, \qquad \forall \, i \in \mathcal{N}.
\end{equation}
%\revision{GP: "Philosophical" problem: if $i$ has a lot of energy, it will try to use it to send jobs away with this cost function. But maybe the reverse is better: if $i$ has a lot of energy foreseen, it should \emph{receive} jobs for processing. Should we add a term for this (maybe just adding something to $q_{rx}$)?}
\begin{remark}
The meaning of \eqref{eq:distr_proc_constraint} is that the workload surplus that server $i$ has during the following time steps, i.e., the \emph{incoming} workload minus the \emph{outgoing} one, should satisfy the average (\mbox{long-term}) power ($\hat{F}_i$) and memory ($\hat{M}_i$) availability at node $i$. The coefficient $\xi_{M_i}$ relates the memory availability to the residual computational power. This follows from the assumption of direct proportionality between the data size $S_k$ and the processed workload $w_k$. 
\end{remark}
Since the general goal is to minimize the energy drained network-wide from the power grid, a cost function that represents the global welfare and at that at the same time is amenable to a distributed solution is the sum
\begin{equation}
\label{eq:opt_distr}
\Gamma\left(\bm o, \tilde{\bm o}, \hat{\bm \delta}; \bar{\bm w}, \hat{\bm P}^{\rm H}\right) = \sum_{i \in \mathcal{N}} \left[\Gamma_i\left(\bm o_i, \tilde{\bm o}_i; \bar{\bm w}_i, \hat{P}^{\rm H}_{i}\right)+\hat{c}_i\hat{\delta}^2_i\right],
\end{equation}
where $\hat{c}_i > 0$ is the cost coefficient associated with the penalty term $\hat{\delta}_i^2$. This leads to the constrained optimization problem
\begin{equation}
\label{eq:distr_problem}
\begin{aligned}
P^{\rm glob}: && \min_{\bm o, \tilde{\bm o}, \hat{\bm \delta}} \quad& \Gamma\left(\bm o, \tilde{\bm o}, \hat{\bm \delta}; \bar{\bm w}, \hat{\bm P}^{\rm H}\right)\\
&& \text{s.t.} \quad &\bm o, \tilde{\bm o}, \hat{\bm \delta} \ge 0, \; \eqref{eq:distr_proc_constraint},\\
&&&o_{ij} = \tilde{o}_{ij} \; \forall\, i, j,
\end{aligned}
\end{equation}
with $\bm o$, $\tilde{\bm o}$, $\hat{\bm \delta}$, $\bar{\bm w}$ and $\hat{\bm P}^{\rm H}$ are vectors collecting $\bm {o}_i$, $\tilde{\bm o}_i$, $\hat{\delta}_i$, $\bar{\bm w}_i$ and $\hat{P}_i^{{\rm H}}$ respectively, for all the \acp{MEH} $i \in \mathcal{N}$. The equality $o_{ij} = \tilde{o}_{ij}$ is called \emph{consensus constraint} and ensures that the amount of workload exiting node $i$ and directed to $j$ equals the one that $j$ expects to receive from $i$.

\subsection{On the interaction between local and distributed phases}
\label{sec:local_vs_distributed}

The local problem~\eqref{eq:centr_problem} is used to schedule the amount of workload $w_i$ (CPU cycles) that is to be executed locally at each \ac{MEH} in the current time slot $t$. Since the solution is predictive, it uses future memory availability ($\hat{M}_i$) and residual computational power ($\hat{F}_i$) estimates to set the global problem constraints~\eqref{eq:distr_proc_constraint}. Thanks to the global problem~\eqref{eq:distr_problem} an agreement is reached on which jobs are to be migrated and where. The solution ${\bm o}_i$ of the global problem is utilized to move workload across the MEHs: this entails an update of sets $\mathcal{K}_i(t+1)$ containing the jobs that are assigned to MEH $i$ at the next time slot $t+1$. The optimization keeps iterating between local and distributed phases.

%, i.e., an agreement has to be reached among the servers.
% The satisfaction of constraint \eqref{eq:storage_capacity} is checked afterward, when the solution is rounded to the closest feasible to the one computed by problem~\eqref{eq:distr_problem}.
%where $\bar{K}_i$, $F_i$, and $M_i$ in constraints \eqref{eq:tot_VM}$\,$-$\,$\eqref{eq:storage_capacity} are replaced with the estimated $\hat{K}_i$, $\hat{F}_i$, and $\hat{M}_i$, respectively. The equality constraints in problem~\eqref{eq:distr_problem} are needed to ensure that the amount of outgoing workload that $i$ sends to $j$ is equal to what $j$ receives from $i$.
%\revision{GP: Now we need also a step 3, actually. An algorithm to round the result of \eqref{eq:distr_problem}, preserving the mass and checking that the solution is feasible with the constraints. Also, I don't know how to satisfy \eqref{eq:max_rate}. Should we add a variable for the memory occupation or consider an average association with the existing variables?}

\section{Final Scheduling Solution via Local and Distributed Processes}\label{sec:ProblemSolution}
%!TEX root = ./paper.tex

\subsection{Phase 1: local MPC solution}

At each \ac{MEH}, the local \ac{MPC} problem of \eqref{eq:centr_problem} is solved over the whole horizon $T$~\cite{rawlings2009model}. \ac{MPC} uses the \emph{receding horizon} technique, which consists of solving the given problem within a prediction window of size $T$, applying the optimal computed control only for the current time step $t=0$, moving forward the optimization window by one time slot ($\tau$ seconds) and repeating the procedure. In this way, the controller progressively adapts to new observations and estimates of the exogenous processes. Also, at any given instant, \ac{MEH} $i$ computes the optimal policy throughout the whole horizon of $T$ slots, but only $\bm w_i(0)$ is applied as the control action. The exogenous processes are the future jobs and the harvested energy availability, see Section~\ref{sec:processes}.
%\revision{GP: Must write somewhere between sec. 3 and 4 how the exogenous processes are statistically described. Here also needed some further maths to explain problem transformation in MPC?}

\subsection{Phase 2a: distributed workload migration}
\label{sec:distributed}

In the following, the scheduling slot index $t$ is omitted in the interest of readability. Eq.~\eqref{eq:distr_problem} is a \emph{consensus} problem, i.e., it entails reaching an agreement on the value of some variables among multiple agents in a distributed system. In our context, the \acp{MEH} must agree on the amount of processing load to exchange among each other. A way to solve this problem -- written as the sum of separable convex cost functions -- is via the \emph{dual ascent} algorithm~\cite{bertsekas1999nonlinear}. 
Given a generic cost function $\psi(\bm x)$, its  \emph{Lagrangian} is defined as
\begin{equation}
\label{eq:lagrangian}
\mathcal{L}(\bm x, \bm z) = \psi(\bm x) + \bm z^T (A \bm x - \bm d)\,,
\end{equation}
where $\bm z$ are the Lagrange multipliers associated with the constraints $A\bm x = \bm d$. The dual ascent solves the problem by iteratively 
\begin{inparaenum}[i)]
\item minimizing $\mathcal{L}(\bm x,\bm z)$ with respect to $\bm x$ (primal step), and
\item updating the value of $\bm z$ (dual step).
\end{inparaenum}
To formalize the solution of problem~\eqref{eq:distr_problem} via dual ascent, we split the local cost functions~\eqref{eq:distr_cost} as
\begin{equation}
\begin{aligned}
\label{eq:distr_cost_trans}
\tilde{\Gamma}_i\left(\bm o_i, \tilde{\bm o}_i, \hat{\delta}\right)=&\max\big\{\left(q_i^{\rm tx}-q_i^{\rm proc}\right) \bm 1^{\scriptscriptstyle{T}} \bm o_i \,+\\
&\quad+ \left(q_i^{\rm rx} + q_i^{\rm proc}\right) \bm 1^{\scriptscriptstyle{T}} \tilde{\bm o}_i 
- \hat{P}^{\rm H}_i, 0 \big\}\,+\\
&+ \frac{\rho}{2} \Vert \bm o_i - \bar{\bm w}_i\Vert^2 + \frac{\rho}{2} \Vert \tilde{\bm o}_i - \tilde{\bm w}_i\Vert^2 + \hat{c}_i\hat{\delta}_i^2,
\end{aligned}
\end{equation}
exploiting the fact that \mbox{$o_{ij} = \tilde{o}_{ij}$}, and defining \mbox{$\tilde{\bm w}_i = \{\bar{w}_{ji} \,\vert\, j \in \mathcal{N}_i\}$}. Intuitively, node $i$ is responsible for half of the quadratic cost from its neighbors and for half of its own local cost. For compactness, let \mbox{$\bm x = \{\bm x_i=[\bm o_i, \tilde{\bm o}_i, \hat{\delta}_i], \,\forall\, i\in\mathcal{N}\}$} be the global optimization variable, \mbox{$\bm b_i = [\bar{\bm w}_i, \tilde{\bm w}_i, 0]$} the tracking target vector, and \mbox{$\bm q_i = [\bm q^{\textrm{tx}}_i - \bm q^{\textrm{proc}}_i, \bm q^{\textrm{rx}}_i + \bm q^{\textrm{proc}}_i, 0]$} the linear costs vector. Moreover, we define matrix \mbox{$Q_i=I_{2N_i+1}\bm m_i$}, with \mbox{$\bm m_i=[\frac{\rho}{2}, \ldots \frac{\rho}{2}, \hat{c}_i]$}, and the global block diagonal matrix $Q$, collecting each $Q_i$ on the diagonal.  
%\revision{There is a problem with constraint (16), which may make the problem infeasible. A possible solution is to make (16) a soft constraint as
%\begin{align}
%\min \; & \; \tilde{\Gamma}_i(\cdot)+\sigma_i\max\left(\sum_{j\in\mathcal{N}_i}(\tilde{\bm o}_{ji}-\bm o_{ij})-\hat{F}_i,0\right)^2\\
%\text{s.t.} \; & \; \tilde{\bm o_i} = \bm o_{i} \geq 0 
%\end{align}
%Need to check how the solution changes in this case.
%As an alternative, resort to the same soft constraint formulation of the centralized problem, adding a variable.}
With these definitions, problem~\eqref{eq:distr_problem} can be expressed in the following form
\begin{align}
\label{eq:final_distr_prob}
&&\min_{\bm x} \quad &\sum_{i \in \mathcal{N}} \Big(\Vert\bm x_i - \bm b_i\Vert^2_{Q_i} + \max\left\{\bm q_i^T\bm x_i - \hat{P}^{\rm H}_{i},0\right\}\Big)\\
&&\text{s.t.}\quad& A_1\, \bm x \leq \bm d, \label{eq:ineq_cons} \\
&&& A_2\, \bm x = \bm 0, \label{eq:eq_cons}
\end{align}
where \mbox{$\Vert \bm x \Vert^2_Q = \bm x^{\scriptscriptstyle{T}}Q\bm x$}. The inequalities~\eqref{eq:ineq_cons} collect~\eqref{eq:distr_proc_constraint} and the \mbox{non-negativity} constraints \mbox{$\bm o, \tilde{\bm o}, \hat{\bm\delta} \geq 0$}, while the equalities~\eqref{eq:eq_cons} correspond to the consensus constraints \mbox{$o_{ij} = \tilde{o}_{ij}$}, \mbox{$\forall i \in \mathcal{N}$}, \mbox{$j \in \mathcal{N}_i$}. 
Here, matrices $A_1$ and $A_2$ are used to select the concerned variables, whereas $\bm d = \{\bm d_i=[\min\{\hat{F}_i, \xi_{M_i} \hat{M}_i\}, \bm 0] \mid i \in\mathcal{N}\}$. We can now write the Lagrangian as
\begin{equation}
\label{eq:compact_lagr}
\mathcal{L}\left(\bm x, \bm y, \bm z\right) = \sum_{i \in \mathcal{N}} \tilde{\Gamma}_i\left(\bm x_i; \bm b_i, \hat{P}^{\rm H}_i\right)+\bm y^T\left(A_1 \bm x - \bm d\right) + \bm z^T A_2 \bm x\,,
\end{equation}
where \mbox{$\bm y = \{\bm y_i=[\lambda_i, \bm \gamma_i, \tilde{\bm \gamma}_i, \hat{\varphi}_i] \, \vert\, i \in \mathcal{N}\}$} are the Lagrange multipliers associated with the inequality constraints~\eqref{eq:ineq_cons}, and \mbox{$\bm z =\{\bm z_i=\bm \mu_i\,\vert\, i \in \mathcal{N}\}$} are the multipliers associated with equalities~\eqref{eq:eq_cons}. Specifically, the Lagrange multipliers $\lambda_i$ refer to constraints~\eqref{eq:distr_proc_constraint}, \mbox{$\bm \gamma_i = \{\gamma_{ij}\}$}, \mbox{$\tilde{\bm \gamma}_i = \{\tilde{\gamma}_{ji}\}$} and $\hat{\varphi}_i$ to \mbox{$\bm o_i \geq 0$}, \mbox{$\tilde{\bm o}_i \geq 0$}, and \mbox{$\hat{\delta}_i \geq 0$}, respectively, and \mbox{$\bm \mu_i = \{\mu_{ij}\}$} to \mbox{$o_{ij} = \tilde{o}_{ij}$}, for every server \mbox{$i \in \mathcal{N}$}, and \mbox{$j \in \mathcal{N}_i$}. 
Using the $^+$ sign to denote the update at the following iteration, we detail in Algorithm~\ref{alg:dual_ascent} the dual ascent procedure that solves the problem
\begin{equation}
\label{eq:lagrangian_problem}
\inf_{\bm x}\;\sup_{\bm y \ge 0, \bm z}\; \mathcal{L}\left(\bm x, \bm y, \bm z\right).
\end{equation}

\begin{algorithm}[t]
\caption{Dual ascent algorithm solving problem~\eqref{eq:distr_problem}}
\label{alg:dual_ascent}
\begin{algorithmic}[1]
\State $\bm x^+ = \argmin_{\bm x}\;\mathcal{L}\left(\bm x; \bm y, \bm z\right)$
\Comment	primal
\State $\bm y^+ =\max\left\{\bm y + \bm \alpha_{\bm y}\left(A_1\,\bm x^+ - \bm d\right),0\right\}$
\Comment	dual (ineq.)
\State $\bm z^+ = \bm z + \bm \alpha_{\bm z} A_2\,\bm x^+$
\Comment	dual (eq.)
\end{algorithmic}
\end{algorithm}

The dual update requires in this case two different forms, depending on whether the constraint is an equality or an inequality one. Inequality constraints may actually be inactive, and the associated Lagrange multipliers would be null in this case. 
The parameters $\bm \alpha_{\bm y}$ and $\bm \alpha_{\bm z}$ in the algorithm tune the stability and the convergence speed.
The presented compact version of the dual ascent translates into the following local procedure, from a server perspective. 
Defining vectors \mbox{$\tilde{\bm \mu}_i = \{\mu_{ji}\}$} and \mbox{$\bar{\bm o}_i = \{\tilde{o}_{ij}\}$} to collect those variables that are kept in memory by the neighborhoods of $i$, the local Lagrangian at node $i$ is
\begin{equation}
\begin{aligned}
\label{eq:local_lagrangian}
\mathcal{L}_i\left(\bm x_i; \bar{\bm w}_i, \tilde{\bm w}_i, \hat{P}^{\rm H}_{i}, \bm \nu_i \right) = &\;\tilde{\Gamma}_i\left(\bm o_i, \tilde{\bm o}_i; \bar{\bm w}_i, \tilde{\bm w}_i, \hat{P}^{\rm H}_{i}\right)\,+\\
&+ \lambda_i \left[\bm 1^T(\tilde{\bm o}_i-\bm o_i)-\hat{\delta}_i\right]+\\
&-\bm\gamma_i^T\bm o_i -\tilde{\bm \gamma}_i^T\tilde{\bm o}_i+\bm \mu_i^T\bm o_i\,+\\
&-\tilde{\bm \mu}_i^T\tilde{\bm o}_i-\hat{\varphi}_i\hat{\delta}_i,
\end{aligned}
\end{equation}
with $\bm x_i = [\bm o_i^+, \tilde{\bm o}_i^+, \hat{\delta}_i]$ and $\bm \nu_i =  [\lambda_i\bm 1, \varphi_i, \bm \gamma_i, \tilde{\bm \gamma}_i, \bm\mu_i, \tilde{\bm\mu}_i]$ to collect the Lagrange multipliers. The local procedure is presented in Algorithm~\ref{alg:dual_ascent_server}, where a fixed step size $\alpha$ is assumed.

\begin{algorithm}[t]
\caption{Dual ascent from a server perspective}
\label{alg:dual_ascent_server}
\begin{algorithmic}[1]
\State		receive $\tilde{\bm \mu}_i = \{\mu_{ji}\}$ from the neighbors
\State		$[\bm o_i^+, \tilde{\bm o}_i^+, \hat{\delta}_i] = \argmin_{\bm x_i}\; \mathcal{L}_i\left(\bm x_i; \bar{\bm w}_i, \tilde{\bm w}_i, \hat{P}^{\rm H}_{i}, \bm \nu_i \right)$
\label{eq:primal}
\State		send $\tilde{o}_{ji}^+$ to the corresponding neighbor $j$
\State		$\lambda_i^+ = \max\left\{\lambda_i+\alpha\left(\sum_{j \in \mathcal{N}_i}\left(\tilde{o}_{ji}^+-o_{ij}^+\right)-\hat{F}_i\right),0\right\}$
\label{eq:dual_lambda}
\State		$\hat{\varphi}_i^+ = \max\left\{\hat{\varphi}_i-\alpha\,\hat{\delta}_i^+,0\right\}$
\label{eq:dual_phi}
\State		$\bm\gamma_i^+=\max\left\{\bm\gamma_i-\alpha\,\bm o_i^+,0\right\}$
\label{eq:dual_gamma}
\State		$\tilde{\bm \gamma}_i^+ = \max\left\{\tilde{\bm \gamma}_i-\alpha\,\tilde{\bm o}_i^+,0\right\}$
\label{eq:dual_gamma_tilde}
\State		receive $\bar{\bm o}_i^+ = \{\tilde{o}_{ij}^+\}$ from the neighbors
\State		$\bm \mu_i^+ = \bm \mu_i + \alpha  \left(\bm o_i^+-\bar{\bm o}_i^+\right)$
\label{eq:dual_mu}
\State		send $\mu_{ij}^+$ to the corresponding neighbor $j$
\end{algorithmic}
\end{algorithm}

Note that, to minimize the Lagrangian in the primal step at line~\ref{eq:primal}, server $i$ not only needs its own Lagrange multipliers, but also the introduced $\tilde{\bm \mu}_i$, which collects the $\mu_{ji}$ of neighbors \mbox{$j \in \mathcal{N}_i$}. Therefore, node $i$ must first receive these multipliers from the neighborhood. Also, while updating $\bm \mu_i$ in the dual step at line~\ref{eq:dual_mu}, $\bar{\bm o}_i^+$ is needed, which collects the $\tilde{o}_{ij}^+$ variables kept by the neighborhood of $i$, and which are to be received after the computation of $j$'s primal step (\mbox{$\forall j \in \mathcal{N}_i$}). Hence, this amounts to two communication rounds among neighbors per dual ascent iteration. The dual updates are computationally inexpensive, whereas the primal step requires solving a local convex subproblem, which is complicated by the $\max\{\cdot\}$ operator in the cost function~\eqref{eq:distr_cost_trans}. Eventually, note that an additional communication is required at the beginning of the procedure, to inform the neighborhood about the values of $\tilde{\bm w}_i$. 

\noindent\textbf{Solution to the primal step (line~\ref{eq:primal}).} The solution of the local primal subproblems is computed in closed form, distinguishing three cases. We consider the local primal subproblems in compact form with variables $\bm x_i$, and collect the Lagrange multipliers of~\eqref{eq:local_lagrangian} in \mbox{$\bm \nu_i = [\lambda_i\bm 1, \varphi_i, \bm \gamma_i, \tilde{\bm \gamma}_i, \bm\mu_i, \tilde{\bm\mu}_i]$}, with associated variables selection matrix $A_i$. 
We split \mbox{$\mathcal{L}_i\left(\bm x_i; \cdot\right)=u_i\left(\bm x_i\right)+h_i\left(\bm x_i\right)$}, so that
\begin{align}
u_i\left(\bm x_i\right)&=\Vert\bm x_i-\bm b_i\Vert^2_{Q_i}+\bm \nu_i^TA_i \bm x_i, \\
h_i\left(\bm x_i\right) &= \bm q_i^T \bm x_i - \hat{P}^{\rm H}_i.
\end{align}
\begin{proposition}
The solution of the primal step of problem~\eqref{eq:distr_problem} is computed as one of the mutually exclusive cases
\begin{itemize}
\item[i)] \mbox{$\bm x_i^+ = \argmin_{\bm x_i} u_i\left(\bm x_i\right)$}, if \mbox{$h_i\left(\bm x_i^+\right) \leq 0$}, or
\item[ii)] \mbox{$\bm x_i^+ = \argmin_{\bm x_i} u_i\left(\bm x_i\right) + h_i\left(\bm x_i\right)$}, if \mbox{$h_i\left(\bm x_i^+\right) > 0$}, or
\item[iii)] \mbox{$\bm x_i^+ = \argmin_{\bm x_i} u_i\left(\bm x_i\right)$}, s.t. \mbox{$h_i\left(\bm x_i\right) = 0$}.
\end{itemize}
\label{prop:3cases}
\end{proposition}
\noindent The solutions for each of the cases of Proposition~\ref{prop:3cases} are now given in the following result.
\begin{proposition}
Consider the three cases of Proposition~\ref{prop:3cases}. Their closed form optimal solutions are expressed as
\begin{itemize}
\item[i)] $\bm x_i^+ = \bm b_i - \frac{1}{2}\,Q_i^{-1}A_i^{\scriptscriptstyle{T}}\bm\nu_i$
%\frac{1}{\rho}\,A_i^T\bm y_i$}
\item[ii)] $\bm x_i^+ = \bm b_i - \frac{1}{2}\,Q_i^{-1}\left(A_i^{\scriptscriptstyle{T}}\bm\nu_i+\bm q_i\right)$
%\frac{1}{\rho}\left(A_i^T\bm y_i+\bm q_i\right)$}
\item[iii)] \resizebox{.92\hsize}{!}{$\bm x_i^+ = \bm b_i - \frac{1}{2}\,Q_i^{-1}\left(A_i^{\scriptscriptstyle{T}}\bm \nu_i + \bm q_i^{\scriptscriptstyle{T}}\frac{ 2\,Q_i \left(\bm b_i - \frac{\hat{P}^{\rm H}_{i}}{\Vert \bm q_i \Vert ^2}\bm q_i\right) - A_i^{\scriptscriptstyle{T}} \bm\nu_i }{\Vert \bm q_i \Vert ^2}\bm q_i \right)$}
%\frac{1}{\rho}A_i^T\bm y_i+\left[\hat{P}_{{\rm H}, i}-\bm q_i^T\left(\bm b_i-\frac{1}{\rho}A_i^T\bm y_i\right)\right] \frac{\bm q_i}{\Vert\bm q_i\Vert^2}$}
%\left(1 - \frac{1}{\rho^2}\right)\bm b_i+\frac{1}{\rho^2}
\end{itemize}

\label{prop:primal_sol}
\end{proposition}
\noindent The proofs of Propositions \ref{prop:3cases} and \ref{prop:primal_sol} are given in Appendix~\ref{sec:appendix}, together with the theoretical upper bound for the step size $\alpha$ that guarantees convergence.

\subsection{Phase 2b: rounding to a feasible discrete solution}
In this section, we show how to compute the actual discrete allocation of jobs by obtaining new variables $\bm o_i^r$, which are the {\it rounded} versions of the $\bm o_i$ that were previously computed through consensus (see Section~\ref{sec:distributed}). In particular, $\bm o_i$ contains the optimal continuous amount of workload that each \ac{MEH} would like to send to its neighbors. Instead, its rounded version $\bm o_i^r$ contains a feasible allocation accounting for the fact that the number of jobs and the possible ways of allocating them are discrete. 

%The following step of the procedure is to identify \emph{which} jobs have to be migrated, and \emph{where}. 
To compute the new $\bm o_i^r$, as an initial solution, we select the jobs from set $\hat{\mathcal{K}}_{ij}$, whose associated vehicle is about to migrate from \ac{eNB} site $i$ to $j$. The rounded $\bm o_i^r$ is thus initially set to $\bar{\bm w}_i$, assuming that the minimizer of the objective function~\eqref{eq:distr_cost} is the vector that minimizes the quadratic term. Then, the difference between this guess and the actual optimum obtained from the proposed dual ascent algorithm is computed, $\bm o_i^{\rm diff}$. For every neighbor $j$ it is now clear whether more workload is to be added to (in case $o_{ij}^{\rm diff} < 0$) or removed from ($o_{ij}^{\rm diff} > 0$) the initial guess $o_{ij}^r$. The jobs that were initially scheduled for migration to node $j$ but that are eventually retained for computation at node $i$ are those minimizing $\Vert \bm o_i^{\rm diff} \Vert_{_1}$. Instead, new jobs are added to the migration list using the prediction vectors $\bm p_{ij}$. In detail, the added jobs are those for which the handover probabilities towards $j$ are maximized. A threshold $\epsilon_P$ is used to approximate the rounded solution, as the continuous optimum $\bm o_i$ will likely not coincide with any possible discrete approximation. The procedure is detailed in Algorithm~\ref{alg:association}.

\begin{algorithm}[t]
\caption{Job-neighbor association}
\label{alg:association}
\begin{algorithmic}[1]
\State \textbf{Input:} mobility pattern predictions matrix $P_i$; optimal outgoing workload amount $\bm o_i$; set of the jobs $\mathcal{J}_i$ in execution at \ac{MEH} $i$; tolerance threshold $\epsilon_P$.
\State \textbf{Output:} job-neighbor association sets $\mathcal{Z}_{ij} \,\forall \, j\in\mathcal{N}_i$; rounded $\bm o_i^r$.
\State remove jobs $\{k \mid I_{i,k} < \epsilon \vee D_{i,k} < 2\}$ from $\mathcal{J}_i$
\State $\bm o_{i}^r \gets \bar{\bm w}_i$
\State $\mathcal{Z}_{ij} \gets \hat{\mathcal{K}}_{ij}$
\State $\mathcal{J}_i \gets \mathcal{J}_i \setminus \bigcup_{j\in\mathcal{N}_i} \hat{\mathcal{K}}_{ij}$
\State $\bm o_i^{\rm diff} \gets \bm o_i^r - \bm o_i$
\Comment workload to be adjusted
\ForAll{neighbors $j$ in $\mathcal{N}_i$}
	\While{$o_{ij}^{\rm diff} > \epsilon_P$}
		\State $k \gets$ job of $\mathcal{Z}_{ij}$ minimizing $\left|\,o_{ij}^{\rm diff}\right|$
		\State remove job $k$ from $\mathcal{Z}_{ij}$
		\State $o_{ij}^r \gets o_{ij}^r - I_{i,k}/D_{i,k}$
		\State $o_{ij}^{\rm diff} \gets o_{ij}^{\rm diff} - I_{i,k}/D_{i,k}$
		\State add job $k$ to $\mathcal{J}_i$
		\Comment make it available for neighbors
	\EndWhile
	\While{$o_{ij}^{\rm diff} < -\epsilon_P$}
			\State take $k \in \mathcal{J}_i \mid k \in \argmax \bm p_{ij}$
			\Comment most prob. $i \rightarrow j$
			\State add job $k$ to $\mathcal{Z}_{ij}$
			\State $o_{ij}^r \gets o_{ij}^r + I_{i,k}/D_{i,k}$
			\State $o_{ij}^{\rm diff} \gets o_{ij}^{\rm diff} + I_{i,k}/D_{i,k}$
			\State mask entry $p_{ij,k}$
			\Comment s.t. $k$ is not selected again
	\EndWhile
\EndFor
\end{algorithmic}
\end{algorithm}

%\begin{algorithm}[tb]
%\caption{Job-neighbor association}
%\label{alg:association}
%\begin{algorithmic}[1]
%\State \textbf{Input:} mobility pattern predictions matrix $P_i$; optimal outgoing workload amount $\bm o_i$; list of the jobs $J_i$; pre-computed value of $\bm o_i^r$; tolerance threshold $\epsilon$.
%\State \textbf{Output:} job-neighbor association list $Z$.
%\State \textbf{Initialize:} Remove jobs $\{k \mid I_{i,k} < \epsilon \vee D_{i,k} < 2\}$;
%\ForAll{jobs in $J_i$}
%\ForAll{neighbors $n$ in $\mathcal{N}_i$}
%	\State $\bm p^{\rm max} \gets \max_j p_{ij, \cdot}$ \Comment max probability value
%	\State $\bm p^{\rm max}_{in} \gets \bm p^{\rm max}_{j=n}$
%	\Comment remove jobs $\mid p^{\rm max}_{ij} \neq p^{\rm max}_{in}$
%	\If{$\bm p^{\rm max}_{in}$ is empty}
%		\State continue
%		\Comment $n$ not the best neighbor
%	\EndIf
%	\State $k \gets \argmax \bm p^{\rm max}_{in}$
%	\Comment most probable job to $n$
%	\If{$o_{i,n}^r + I_{i,k}/D_{i,k} < o_{i,n} + \epsilon/\tau$}
%		\State $o_{i,n}^r \gets o_{i,n}^r + I_{i,k}/D_{i,k}$
%		\Comment update rounded
%		\State append job $k$ to $Z_n$
%		\Comment assign job $k$ to $n$
%		\State mask row $k$ from $P_i$
%		\Comment association done
%	\Else
%		\State mask entry $p_{k,n}$
%		\Comment low power available at $n$
%	\EndIf
%\EndFor
%\EndFor
%\end{algorithmic}
%\end{algorithm}

\subsection{Additional considerations}\label{subsec:additiona_considerations}

\noindent\textbf{Handling pathological cases:} Since system constraints are made soft to avoid primal infeasibility, three pathological cases may arise, namely,
\begin{inparaenum}
\item \label{pato-one} the optimal processed workload at the current instant exceeds the computational capacity; or
\item \label{pato-two} the data size for the currently running jobs do not fit the RAM memory; or
\item \label{pato-three} the deadline expires during the current slot, but the residual intensity is greater than zero.
\end{inparaenum}
A greedy algorithm is developed to handle all of them. For the first two, the \ac{MEH} ranks the active jobs through a double ordering criterion, considering as the first ranking criterion the time slot when they expire, and as the second their intensity (or data size). Next, it momentarily pauses the execution of the services starting from the last one in the ordered list, until the resources suffice to proceed. In case \ref{pato-one}, when pausing a job $m$, the amount of processed workload becomes \mbox{$\sum_{k\in\mathcal{K}_i}w_{i,k} - w_{i,m}$}, while in case \ref{pato-two}, the data relative to suspended jobs is deleted from the RAM. The number of suspended jobs is the minimum such that the requirements are satisfied. Moreover, in case \ref{pato-one}, it is likely that, when a job is suspended, additional computational power becomes available. In such a case, the new computational resources are assigned to the jobs that are closest to their deadline. Case \ref{pato-three} is managed considering the amount of residual intensity $I_{i,k}$. If $I_{i,k}$ is smaller than a threshold $\epsilon$, then the deadline is extended by a small amount, so that the controller will privilege the execution of the corresponding job in the next slot. In this way, jobs are allowed to finish with a little additional delay (within one slot). If, however, the amount of residual intensity is larger than $\epsilon$, the job is dropped, i.e., in this case the algorithm failed to provide an acceptable solution.

\noindent\textbf{Algorithm complexity:} The decomposition approach adopted in EASE makes the overall algorithm feasible and lightweight to be run even in a complex and highly variable scenario such as the vehicular one here considered. Previous work~\cite{perin2021towards} uses \ac{MPC} to obtain an optimal decision on both the amount of workload to process locally and to offload to other \acp{MEH} in a fully decentralized fashion. This amounts to having a number of shared variables to be optimized via message passing $\mathcal{O}(VT)$, where $V$ is the number of edges in the network graph and $T$ is the prediction window. EASE, instead, by performing a preliminary local optimization phase, estimates the future \emph{on average}, having thus a number of shared variables $\mathcal{O}(V)$. The local phase amounts to solving a constrained convex problem numerically every $\tau$ seconds, while the distributed phase requires broadcasting to the neighborhood (a part of) the primal and dual information of problem~\eqref{eq:distr_problem}, plus inexpensive closed form updates. Note that this information amounts to a few bits only, which can be easily piggybacked on control packets that the \acp{MEH} normally exchange for other reasons. The empirical convergence rate is studied in Section~\ref{sec:convergence}. 

\noindent\textbf{Predictions inaccuracies:} With the adopted approach, a residual migration suboptimality is still possible also due to prediction errors on the mobility of the users, the service request, and the available local resources. Concerning the mobility prediction, the performance is extensively studied in~\cite{labriji2021mobility}, where the authors compare the mobility predictor also used by EASE with a simpler and less accurate approach based on Markov chains, showing the improvement brought by considering the information on the actual trajectory followed by the users. However, we recall that the main objective of EASE is to reduce the carbon footprint and, in turn, even in the case of precise mobility predictions, the scheduler can decide to place the service in a MEH that is far away from the vehicle, it this leads to better use of the energy resources. For this reason, the \acp{eNB}/\acp{MEH} are connected via backhaul links that always ensure that the result is sent back to the user. Regarding instead the statistical processes that control the energy availability and the job requests, with EASE we only assume to know the average income over a prediction horizon of some seconds (e.g., with $T=5$, $15$~s). Previous work~\cite{perin2021towards} assessed the impact of average versus estimated (via Markov chains) or exact knowledge into the future (i.e., a ``genie predictor''), showing that \ac{MPC} is highly effective even when simple predictors are used.

%\begin{algorithm}[tb]
%\small
%\caption{Job-neighbor association}
%\label{alg:association}
%\begin{algorithmic}[1]
%\State \textbf{Input:} mobility pattern predictions matrix $P_i$; optimal outgoing workload amount $\bm o_i$; list of jobs $J_i$; tolerance threshold $\epsilon_P$.
%\State \textbf{Output:} job-neighbor association list $Z$.
%\State remove jobs $\{k \mid I_{i,k} < \epsilon \vee D_{i,k} < 2\}$ from $J_i$
%\State set $\bm o_{i}^r=\bar{\bm w}_i$ and $Z$ as a consequence
%\State $\bm o_i^{\rm diff} \gets \bm o_i^r - \bm o_i$
%\Comment workload to be adjusted
%\ForAll{neighbors $j$ in $\mathcal{N}_i$}
%	\If{$o_{ij}^{\rm diff} > \epsilon_P$}
%		\State remove from $Z_n$ the job minimizing $|\,o_{ij}^{\rm diff}|$
%		\State update $o_{ij}^r$ accordingly
%	\ElsIf{$o_{ij}^{\rm diff} < -\epsilon_P$}
%		\ForAll{jobs in $J_i$}
%			\State $k \gets \argmax \bm p_{ij}$
%			\Comment most probable job to $j$
%			\If{$o_{ij}^{\rm diff} + I_k / D_k < \epsilon_P$}
%				\State add $k$ to $Z_n$
%				\State update $o_{ij}^r$ accordingly
%			\EndIf
%			\State mask entry $p_{ij}^k$
%			\Comment s.t. $k$ is not selected again
%		\EndFor
%	\EndIf
%\EndFor
%\end{algorithmic}
%\end{algorithm}

\section{Numerical Results}\label{sec:NumericalResuls}
%!TEX root = ./paper.tex

EASE is assessed in an emulated environment featuring \mbox{5G-enabled} vehicles moving within an urban scenario. Mobility traces are obtained with SUMO~\cite{SUMO2012}, an \mbox{open-source} traffic simulator to obtain mobility traces around a predefined city road map. For this, we use the ``TAPAS Cologne'' scenario, which mimics the vehicular traffic within the city of Cologne for a whole day based on the traveling habits of the city dwellers~\cite{tapasCologne}.
%The vehicle density changes during the day presenting two peaks, in the time intervals between $8$-$9$ a.m. and $4$-$6$ p.m., reflecting the mobility due to getting to and leaving work.
The mobile network is composed of $8$ \acp{eNB} endowed with \ac{MEH} functionalities, wired connected through optical links. 
The mobility area is covered with hexagonal cells with an \ac{eNB} in the center, and with an inter-distance among nodes of $400$~m.
We generated and collected $24$h long SUMO mobility traces with $25$~ms granularity, for each of the $8$ \acp{eNB} in the deployment. The first $15$ hours were used to train and validate the mobility prediction algorithm, which is taken from~\cite{labriji2021mobility}, whereas the remaining ones to assess the performance of \ac{EASE}. For the evaluation, we considered vehicles approaching the edge of the serving \ac{eNB} coverage area, i.e., that are about to hand over to a new \ac{eNB}/\ac{MEH}. With the considered setup, this occurs, on average, when a user is less than $40$~meters apart from the radio cell's border.
The energy consumption of the \acp{MEH} is computed based on the SPECpower benchmark~\cite{specpower}. We selected two different edge computing platforms, namely, an HP ProLiant DL 110 Gen 10 Plus and a Nettrix R620 G40, obtaining two clusters of edge servers with different energy consumption, processing speed and memory, see Table~\ref{tab:params_server}. In Table~\ref{tab:jobs_params}, we report the jobs intensities, deadlines, data sizes, and generation probabilities, according to the system model of Section~\ref{sec:processes}. The other system parameters are listed in Table~\ref{tab:sim_params}.

\begin{table}[tb]
\centering
\resizebox{0.9\columnwidth}{!}{%
\begin{tabular}{l|c|c}
&   \textbf{HP ProLiant DL 110}    &   \textbf{Nettrix R620 G40}       \\
\hline
\hline
%\vspace{0.1cm}
idle power $P_{i}^{\rm idle}$         & $94$~W               & $110$~W   \\
max load power $P_i^{\rm max}$        & $299$~W              & $468$~W   \\
computational power $F_i$             & $3.3$~Gflops         & $7.6$~Gflops  \\     
RAM memory $M_i$                      & $64$~GB              & $256$~GB  \\
%\vspace{0.1cm}  
\hline                                                                                                         
\end{tabular}%
}
\caption{Servers specifications~\cite{specpower}.}
\label{tab:params_server}
\end{table}
\begin{table}[tb]
\centering
\begin{tabular}{l|c|c|c|c}
&$I^{\ell}$~[Gflop]  &$D^{\ell}$~[s]  &$S^{\ell}$~[GB]  & $p_{\ell}$ \\
\hline
\hline
type 1 & $10$ & $20$ & $2$ & $0.4$ \\
type 2 & $16$ & $30$ & $10$ & $0.2$ \\
type 3 & $12$ & $40$ & $0.1$ & $0.4$ \\
\hline
\end{tabular}%
\caption{Jobs parameters for the simulations.}
\setlength\belowcaptionskip{-1cm}
\label{tab:jobs_params}
\end{table}

\begin{table}[t]
\centering
\resizebox{0.9\columnwidth}{!}{%
\begin{tabular}{l|c}
\textbf{Parameter} & \textbf{Value} \\
\hline
\hline
number of nodes $|\mathcal{N}|$ & $8$ \\
fixed wireless circuit power consumption $P_{\textrm{RAN}}$ & $50.2$~W \\
fixed wired circuit power consumption $P_{\textrm{wired}}$ & $20$~W \\
energy per transmitted bit via wireless link $E_{b}^{\rm RAN}$ & $1$~nJ/bit\\
energy per transmitted bit via wired link $E_{b}^{\rm wired}$ & $250$~pJ/bit\\
\ac{PV} panel minimum power $P^{\rm PV}_{\rm min}$ & $250$~W\\
\ac{PV} panel maximum power $P^{\rm PV}_{\rm max}$ & $400$~W\\
\ac{PV} panel average power $P_{\rm PV}$ & $370$~W \\
\ac{PV} panel power std $\sigma_{\rm PV}$ & $10$~W \\
containers' size $L$ & $50$~MB \\
weight parameters for $L$ in~\eqref{eq:e_mig2} $\sigma_s$, $\sigma_d$ & $500$~nJ/bit \\
fixed container migration energy expenditure $E_s$, $E_d$ & $250$~mJ \\
delay associated with wired transmissions $T_k^{\rm migr}$ & $2$~s \\
window size to predict incoming traffic $W$ & $5$ minutes \\
scheduler time slot $\tau$ &  $3$~s \\
\ac{MPC} horizon $T$ & $\{2, 5, 20\}$ \\
job generation probability $p$ & $0.25$ \\
weight of the soft constraint penalty $c_i$ of~\eqref{eq:centr_problem} & $500$ \\
weight of the soft constraint penalty $\hat{c}_i$ of~\eqref{eq:distr_problem} & $10$ \\ 
state cost $\gamma$ for Eq.~\eqref{eq:centr_cost} & $100$ \\
weight of the quadratic term of~\eqref{eq:distr_problem} $\rho$ & $2.5$ \\
%maximum number of dual ascent iterations & $100$ \\
\hline
\end{tabular}%
}
\setlength\belowcaptionskip{-.3cm}
\caption{Summary of simulation parameters.}
\label{tab:sim_params}
\end{table}

In the following analysis, the edge energy consumption is evaluated through
\begin{inparaenum}[i)]
\item the processing and migration power, averaged across all the \acp{MEH},
\item the energy efficiency, defined as \mbox{$\eta=E_h/E_{\rm tot}$}, i.e., the fraction of harvested (green) energy used over the total energy drained (green plus grid energy),
\item the fraction of executed and finished jobs, and
\item the fraction of jobs finishing in the \ac{MEH} that is co-located with the \ac{eNB} serving the vehicle. 
%fraction of finished jobs at \ac{MEH} $i$ whose associated vehicle $v$ is under \ac{eNB} $i$ coverage.
\end{inparaenum}
First, we assess the impact of the prediction window size $T$ on the performance of EASE, then we compare it with the three heuristic migration strategies proposed in~\cite{li2019service} (i.e., ``keep'', ``migrate'' and ``threshold'') and the solution of~\cite{labriji2021mobility}, based on Lyapunov optimization and termed thus ``lyapunov'' (see Section~\ref{sec:RelatedWork} for details). 
The migrations in the ``threshold'' strategy are triggered whenever the current serving \ac{MEH} starts to have a positive carbon footprint, according to equation~\eqref{eq:res_energy}. Note that, for a fair comparison, the approaches we compare our strategy with are all based on the local resource allocation algorithm we devise in this paper. Hence, their differing performance only depends on the adopted migration policy.

\subsection{EASE performance varying the resources prediction window}
\begin{figure*}[tb]
  \centering
  \subfloat[\label{fig:proc_power}$P_{\rm PV}=370$~W]{\resizebox{0.25\textwidth}{!}{% This file was created with tikzplotlib v0.9.12.
\begin{tikzpicture}

\definecolor{color0}{rgb}{0.4,0.76078431372549,0.647058823529412}
\definecolor{color1}{rgb}{0.988235294117647,0.552941176470588,0.384313725490196}
\definecolor{color2}{rgb}{0.552941176470588,0.627450980392157,0.796078431372549}
\definecolor{color3}{rgb}{0.905882352941176,0.541176470588235,0.764705882352941}
\definecolor{color4}{rgb}{0.650980392156863,0.847058823529412,0.329411764705882}
\definecolor{color5}{rgb}{1,0.85098039215,0.18431372549}
\definecolor{color6}{rgb}{0.89803921568,0.76862745098,0.58039215686}

\begin{axis}[
legend cell align={left},
legend style={
  fill opacity=0.8,
  draw opacity=1,
  text opacity=1,
  at={(0.97,0.03)},
  anchor=south east,
  draw=white!80!black,
  font=\footnotesize
},
label style={font=\small},
ymajorgrids=true,
xmajorgrids=true,
tick align=outside,
tick pos=left,
x grid style={white!69.0196078431373!black},
xlabel={Job generation probability $p$},
xmin=0.05, xmax=0.5,
xtick style={color=black},
y grid style={white!69.0196078431373!black},
ylabel={Average processing power [W]},
ymin=24.5986903640945, ymax=136.407123010561,
ytick style={color=black}
]
\path [draw=color0, fill=color0, opacity=0.2]
(axis cs:0.05,36.4355902025473)
--(axis cs:0.05,34.1985061814893)
--(axis cs:0.1,57.3892708689603)
--(axis cs:0.15,74.9090327004432)
--(axis cs:0.2,86.932816527307)
--(axis cs:0.25,96.6680573364578)
--(axis cs:0.3,104.470307351999)
--(axis cs:0.35,111.135781604577)
--(axis cs:0.4,116.003467312024)
--(axis cs:0.45,120.503993807054)
--(axis cs:0.5,125.315483453744)
--(axis cs:0.5,129.33250937337)
--(axis cs:0.5,129.33250937337)
--(axis cs:0.45,124.485466911187)
--(axis cs:0.4,119.998779734072)
--(axis cs:0.35,114.934306359635)
--(axis cs:0.3,108.295866724918)
--(axis cs:0.25,100.680861518212)
--(axis cs:0.2,90.2546038773709)
--(axis cs:0.15,78.1291547498607)
--(axis cs:0.1,60.3152721252105)
--(axis cs:0.05,36.4355902025473)
--cycle;

\path [draw=color1, fill=color1, opacity=0.2]
(axis cs:0.05,35.3851144723565)
--(axis cs:0.05,33.0445282116611)
--(axis cs:0.1,55.4123484332974)
--(axis cs:0.15,71.0249775282413)
--(axis cs:0.2,81.8302253280146)
--(axis cs:0.25,91.838932546879)
--(axis cs:0.3,97.9795917693383)
--(axis cs:0.35,104.361248100292)
--(axis cs:0.4,109.247504671865)
--(axis cs:0.45,112.142902165281)
--(axis cs:0.5,116.359116280994)
--(axis cs:0.5,120.073085144384)
--(axis cs:0.5,120.073085144384)
--(axis cs:0.45,116.011293243105)
--(axis cs:0.4,113.087579078297)
--(axis cs:0.35,108.100535951871)
--(axis cs:0.3,101.580240427471)
--(axis cs:0.25,95.4493726195057)
--(axis cs:0.2,85.3207474462326)
--(axis cs:0.15,74.2995016690673)
--(axis cs:0.1,58.2501575114806)
--(axis cs:0.05,35.3851144723565)
--cycle;

\path [draw=color2, fill=color2, opacity=0.2]
(axis cs:0.05,36.462206408467)
--(axis cs:0.05,34.1474925432077)
--(axis cs:0.1,55.5605870893777)
--(axis cs:0.15,71.847820109758)
--(axis cs:0.2,83.4120618181449)
--(axis cs:0.25,92.1872820192308)
--(axis cs:0.3,99.0888181098288)
--(axis cs:0.35,104.444541463237)
--(axis cs:0.4,109.094414720325)
--(axis cs:0.45,113.420783116645)
--(axis cs:0.5,117.385112526853)
--(axis cs:0.5,121.166045635275)
--(axis cs:0.5,121.166045635275)
--(axis cs:0.45,117.275049387757)
--(axis cs:0.4,112.80277519414)
--(axis cs:0.35,108.328838806129)
--(axis cs:0.3,102.546798621902)
--(axis cs:0.25,95.7579655411998)
--(axis cs:0.2,86.8752643451907)
--(axis cs:0.15,74.9072410844162)
--(axis cs:0.1,58.5183319192092)
--(axis cs:0.05,36.462206408467)
--cycle;

\addplot [thick, color0, mark=triangle*, mark size=4.5, mark options={solid,draw=white}]
table {%
0.05 35.3501600931158
0.1 58.9218191251759
0.15 76.5103425572618
0.2 88.6321473117824
0.25 98.656740659963
0.3 106.276963680343
0.35 112.927802907976
0.4 117.91467288047
0.45 122.534156470564
0.5 127.31857835813
};
\addlegendentry{$T = 2$}
\addplot [thick, color1, mark=square*, mark size=3.5, mark options={solid,draw=white}]
table {%
0.05 34.2293065287903
0.1 56.7906845282257
0.15 72.7103917836207
0.2 83.5863265393843
0.25 93.6717024597484
0.3 99.8211126065799
0.35 106.204846531075
0.4 111.089339697973
0.45 114.08806805061
0.5 118.234601400376
};
\addlegendentry{$T = 5$}
\addplot [thick, color2, mark=diamond*, mark size=3.5, mark options={solid,draw=white}]
table {%
0.05 35.2874568859268
0.1 57.0660865092009
0.15 73.3511502198666
0.2 85.0697378508968
0.25 93.9838159185195
0.3 100.836847347931
0.35 106.3692511061
0.4 110.857483354515
0.45 115.342299240066
0.5 119.141181522164
};
\addlegendentry{$T = 20$}
\end{axis}

\end{tikzpicture}}}
  \subfloat[\label{fig:migr_power}$P_{\rm PV}=370$~W]{\resizebox{0.25\textwidth}{!}{% This file was created with tikzplotlib v0.9.12.
\begin{tikzpicture}

\definecolor{color0}{rgb}{0.4,0.76078431372549,0.647058823529412}
\definecolor{color1}{rgb}{0.988235294117647,0.552941176470588,0.384313725490196}
\definecolor{color2}{rgb}{0.552941176470588,0.627450980392157,0.796078431372549}
\definecolor{color3}{rgb}{0.905882352941176,0.541176470588235,0.764705882352941}
\definecolor{color4}{rgb}{0.650980392156863,0.847058823529412,0.329411764705882}
\definecolor{color5}{rgb}{1,0.85098039215,0.18431372549}
\definecolor{color6}{rgb}{0.89803921568,0.76862745098,0.58039215686}

\begin{axis}[
legend cell align={left},
legend style={
  fill opacity=0.8,
  draw opacity=1,
  text opacity=1,
  at={(0.97,0.03)},
  anchor=south east,
  draw=white!80!black,
  font=\footnotesize
},
label style={font=\small},
ymajorgrids=true,
xmajorgrids=true,
tick align=outside,
tick pos=left,
x grid style={white!69.0196078431373!black},
xlabel={Job generation probability $p$},
xmin=0.05, xmax=0.5,
xtick style={color=black},
y grid style={white!69.0196078431373!black},
ylabel={Average migration power [W]},
ymin=0.25, ymax=1.71276415660682,
ytick style={color=black}
]
\path [draw=color0, fill=color0, opacity=0.2]
(axis cs:0.05,0.484945373675596)
--(axis cs:0.05,0.391077188952454)
--(axis cs:0.1,0.643445930294363)
--(axis cs:0.15,0.896865514677708)
--(axis cs:0.2,1.04689454573621)
--(axis cs:0.25,1.05097718893829)
--(axis cs:0.3,1.16156814066122)
--(axis cs:0.35,1.20428913080619)
--(axis cs:0.4,1.25036425792688)
--(axis cs:0.45,1.34383193311837)
--(axis cs:0.5,1.43042612738722)
--(axis cs:0.5,1.59696852029244)
--(axis cs:0.5,1.59696852029244)
--(axis cs:0.45,1.51326655823515)
--(axis cs:0.4,1.41021349013295)
--(axis cs:0.35,1.35910769498627)
--(axis cs:0.3,1.31562916094041)
--(axis cs:0.25,1.19851189662255)
--(axis cs:0.2,1.20733459764607)
--(axis cs:0.15,1.03620043212166)
--(axis cs:0.1,0.766428374230289)
--(axis cs:0.05,0.484945373675596)
--cycle;

\path [draw=color1, fill=color1, opacity=0.2]
(axis cs:0.05,0.453502716199124)
--(axis cs:0.05,0.358653521304881)
--(axis cs:0.1,0.505514966777185)
--(axis cs:0.15,0.723673278910016)
--(axis cs:0.2,0.752179495774984)
--(axis cs:0.25,0.896131163082572)
--(axis cs:0.3,1.05704623182218)
--(axis cs:0.35,1.11643718437477)
--(axis cs:0.4,1.19808044294263)
--(axis cs:0.45,1.27991577659775)
--(axis cs:0.5,1.28414266871811)
--(axis cs:0.5,1.44131662482308)
--(axis cs:0.5,1.44131662482308)
--(axis cs:0.45,1.44357144445182)
--(axis cs:0.4,1.35076838073526)
--(axis cs:0.35,1.26213135405852)
--(axis cs:0.3,1.21346460645247)
--(axis cs:0.25,1.03612607364665)
--(axis cs:0.2,0.880294487415522)
--(axis cs:0.15,0.851644215780462)
--(axis cs:0.1,0.616928586048259)
--(axis cs:0.05,0.453502716199124)
--cycle;

\path [draw=color2, fill=color2, opacity=0.2]
(axis cs:0.05,0.492704548276627)
--(axis cs:0.05,0.392844447612494)
--(axis cs:0.1,0.633797289251687)
--(axis cs:0.15,0.852534164301359)
--(axis cs:0.2,0.902753027690178)
--(axis cs:0.25,1.0093280076328)
--(axis cs:0.3,1.04896699053749)
--(axis cs:0.35,1.13941457472331)
--(axis cs:0.4,1.17957843225611)
--(axis cs:0.45,1.29995691617048)
--(axis cs:0.5,1.35866185337078)
--(axis cs:0.5,1.52764492107372)
--(axis cs:0.5,1.52764492107372)
--(axis cs:0.45,1.46425536004085)
--(axis cs:0.4,1.34452635814988)
--(axis cs:0.35,1.30284683747361)
--(axis cs:0.3,1.19352429373247)
--(axis cs:0.25,1.16144944337266)
--(axis cs:0.2,1.03749059208501)
--(axis cs:0.15,0.996667661175832)
--(axis cs:0.1,0.75785001112338)
--(axis cs:0.05,0.492704548276627)
--cycle;

\addplot [thick, color0, mark=triangle*, mark size=4.5, mark options={solid,draw=white}]
table {%
0.05 0.436263032870894
0.1 0.706204344514933
0.15 0.968896383637523
0.2 1.12378716388673
0.25 1.12495991762129
0.3 1.23622898328226
0.35 1.2845089071427
0.4 1.32991472114876
0.45 1.42685741089548
0.5 1.5179782334597
};
\addlegendentry{$T = 2$}
\addplot [thick, color1, mark=square*, mark size=3.5, mark options={solid,draw=white}]
table {%
0.05 0.404837643720749
0.1 0.560782499672877
0.15 0.787455505298869
0.2 0.816880663897559
0.25 0.964660678674002
0.3 1.13200332232296
0.35 1.18978797945869
0.4 1.27494809405528
0.45 1.36232720359111
0.5 1.36227729301731
};
\addlegendentry{$T = 5$}
\addplot [thick, color2, mark=diamond*, mark size=3.5, mark options={solid,draw=white}]
table {%
0.05 0.442494885655523
0.1 0.696500878363271
0.15 0.923420183409749
0.2 0.967005347858766
0.25 1.08226121313699
0.3 1.1182985593979
0.35 1.21911659987859
0.4 1.25953095112369
0.45 1.38113545816095
0.5 1.44056589758819
};
\addlegendentry{$T = 20$}
\end{axis}

\end{tikzpicture}}}
  \subfloat[\label{fig:eff_prob}$P_{\rm PV}=370$~W]{\resizebox{0.25\textwidth}{!}{% This file was created with tikzplotlib v0.9.12.
\begin{tikzpicture}

\definecolor{color0}{rgb}{0.4,0.76078431372549,0.647058823529412}
\definecolor{color1}{rgb}{0.988235294117647,0.552941176470588,0.384313725490196}
\definecolor{color2}{rgb}{0.552941176470588,0.627450980392157,0.796078431372549}
\definecolor{color3}{rgb}{0.905882352941176,0.541176470588235,0.764705882352941}
\definecolor{color4}{rgb}{0.650980392156863,0.847058823529412,0.329411764705882}
\definecolor{color5}{rgb}{1,0.85098039215,0.18431372549}
\definecolor{color6}{rgb}{0.89803921568,0.76862745098,0.58039215686}

\begin{axis}[
legend cell align={left},
legend style={
  fill opacity=0.8,
  draw opacity=1,
  text opacity=1,
  at={(0.03,0.03)},
  anchor=south west,
  draw=white!80!black,
  font=\footnotesize
},
label style={font=\small},
ymajorgrids=true,
xmajorgrids=true,
tick align=outside,
tick pos=left,
x grid style={white!69.0196078431373!black},
xlabel={Job generation probability $p$},
xmin=0.05, xmax=0.5,
xtick style={color=black},
y grid style={white!69.0196078431373!black},
ylabel={Energy efficiency $\eta$},
ymin=0.965, ymax=1,
ytick style={color=black}
]
\path [draw=color0, fill=color0, opacity=0.2]
(axis cs:0.05,0.997579395425118)
--(axis cs:0.05,0.996770338200894)
--(axis cs:0.1,0.992109047651556)
--(axis cs:0.15,0.987871600641454)
--(axis cs:0.2,0.983425493430127)
--(axis cs:0.25,0.980034480081897)
--(axis cs:0.3,0.976995255865608)
--(axis cs:0.35,0.974257095853882)
--(axis cs:0.4,0.97154871781955)
--(axis cs:0.45,0.968552535439143)
--(axis cs:0.5,0.966421188858502)
--(axis cs:0.5,0.969415043136705)
--(axis cs:0.5,0.969415043136705)
--(axis cs:0.45,0.971474689320069)
--(axis cs:0.4,0.974356574828082)
--(axis cs:0.35,0.976941505874812)
--(axis cs:0.3,0.979418722998503)
--(axis cs:0.25,0.982392802643411)
--(axis cs:0.2,0.985527224794071)
--(axis cs:0.15,0.989653069896206)
--(axis cs:0.1,0.993565486453658)
--(axis cs:0.05,0.997579395425118)
--cycle;

\path [draw=color1, fill=color1, opacity=0.2]
(axis cs:0.05,0.997986922601434)
--(axis cs:0.05,0.997275450081671)
--(axis cs:0.1,0.993300412180068)
--(axis cs:0.15,0.989258001590407)
--(axis cs:0.2,0.986007014294612)
--(axis cs:0.25,0.982734924109936)
--(axis cs:0.3,0.980322844166562)
--(axis cs:0.35,0.978025356162146)
--(axis cs:0.4,0.9746033777012)
--(axis cs:0.45,0.973248907646123)
--(axis cs:0.5,0.971416025342057)
--(axis cs:0.5,0.974146280299017)
--(axis cs:0.5,0.974146280299017)
--(axis cs:0.45,0.975971861591322)
--(axis cs:0.4,0.977288835807338)
--(axis cs:0.35,0.980563514304181)
--(axis cs:0.3,0.982579057120073)
--(axis cs:0.25,0.984867206407062)
--(axis cs:0.2,0.987846433771587)
--(axis cs:0.15,0.990936007333606)
--(axis cs:0.1,0.994527789303836)
--(axis cs:0.05,0.997986922601434)
--cycle;

\path [draw=color2, fill=color2, opacity=0.2]
(axis cs:0.05,0.997849358554518)
--(axis cs:0.05,0.997079975900657)
--(axis cs:0.1,0.992916307908557)
--(axis cs:0.15,0.988186062928087)
--(axis cs:0.2,0.985396633785878)
--(axis cs:0.25,0.982649547840748)
--(axis cs:0.3,0.979616823058052)
--(axis cs:0.35,0.977402578451831)
--(axis cs:0.4,0.974664609484548)
--(axis cs:0.45,0.973434002245888)
--(axis cs:0.5,0.971161598747889)
--(axis cs:0.5,0.973988880852969)
--(axis cs:0.5,0.973988880852969)
--(axis cs:0.45,0.976071959704272)
--(axis cs:0.4,0.97721916396593)
--(axis cs:0.35,0.979948377895797)
--(axis cs:0.3,0.982107233611451)
--(axis cs:0.25,0.984798735772611)
--(axis cs:0.2,0.987414694846149)
--(axis cs:0.15,0.990052333884161)
--(axis cs:0.1,0.994265860719332)
--(axis cs:0.05,0.997849358554518)
--cycle;

\addplot [thick, color0, mark=triangle*, mark size=4.5, mark options={solid,draw=white}]
table {%
0.05 0.99717125936622
0.1 0.992838975422426
0.15 0.988749754821345
0.2 0.984484883509223
0.25 0.981215379655485
0.3 0.97820667512502
0.35 0.97553754578813
0.4 0.972982463198209
0.45 0.970008837944254
0.5 0.967879953662463
};
\addlegendentry{$T=2$}
\addplot [thick, color1, mark=square*, mark size=3.5, mark options={solid,draw=white}]
table {%
0.05 0.997646356042138
0.1 0.993907965095324
0.15 0.99009882982774
0.2 0.986925864642315
0.25 0.983772537958589
0.3 0.98141853886887
0.35 0.979291124564484
0.4 0.975941764718911
0.45 0.97466425325156
0.5 0.972741900203779
};
\addlegendentry{$T=5$}
\addplot [thick, color2, mark=diamond*, mark size=3.5, mark options={solid,draw=white}]
table {%
0.05 0.997477955800888
0.1 0.993578731505355
0.15 0.989125912451564
0.2 0.986489054098028
0.25 0.983784138662234
0.3 0.980857832499064
0.35 0.978588757961614
0.4 0.975984816656828
0.45 0.97472307181048
0.5 0.972605793676458
};
\addlegendentry{$T=20$}
\end{axis}

\end{tikzpicture}}}
  \subfloat[\label{fig:eff_power}$p=0.25$]{\resizebox{0.25\textwidth}{!}{% This file was created with tikzplotlib v0.9.12.
\begin{tikzpicture}

\definecolor{color0}{rgb}{0.4,0.76078431372549,0.647058823529412}
\definecolor{color1}{rgb}{0.988235294117647,0.552941176470588,0.384313725490196}
\definecolor{color2}{rgb}{0.552941176470588,0.627450980392157,0.796078431372549}
\definecolor{color3}{rgb}{0.905882352941176,0.541176470588235,0.764705882352941}
\definecolor{color4}{rgb}{0.650980392156863,0.847058823529412,0.329411764705882}
\definecolor{color5}{rgb}{1,0.85098039215,0.18431372549}
\definecolor{color6}{rgb}{0.89803921568,0.76862745098,0.58039215686}

\begin{axis}[
legend cell align={left},
legend style={
  fill opacity=0.8,
  draw opacity=1,
  text opacity=1,
  at={(0.97,0.03)},
  anchor=south east,
  draw=white!80!black,
  font=\footnotesize
},
label style={font=\small},
ymajorgrids=true,
xmajorgrids=true,
tick align=outside,
tick pos=left,
x grid style={white!69.0196078431373!black},
xlabel={Average harvested power $P_{\rm PV}$ [W]},
xmin=250, xmax=400,
xtick style={color=black},
xtick={250, 275, 300, 325, 350, 375, 400},
xticklabels={250, 275, 300, 325, 350, 375, 400},
y grid style={white!69.0196078431373!black},
ylabel={Energy efficiency $\eta$},
ymin=0.854414948331404, ymax=1,
ytick style={color=black}
]
\path [draw=color0, fill=color0, opacity=0.2]
(axis cs:250,0.870755067156184)
--(axis cs:250,0.864318465384808)
--(axis cs:265,0.886545363067094)
--(axis cs:280,0.905630904533131)
--(axis cs:295,0.922800203684418)
--(axis cs:310,0.938361264320758)
--(axis cs:325,0.949139103007354)
--(axis cs:340,0.96061289706432)
--(axis cs:355,0.97015556265914)
--(axis cs:370,0.978049853248438)
--(axis cs:385,0.984032104399957)
--(axis cs:400,0.987327250190048)
--(axis cs:400,0.989305223043919)
--(axis cs:400,0.989305223043919)
--(axis cs:385,0.986196687872296)
--(axis cs:370,0.980489533676649)
--(axis cs:355,0.972951877986364)
--(axis cs:340,0.963879287434478)
--(axis cs:325,0.952922962565697)
--(axis cs:310,0.942488548509144)
--(axis cs:295,0.927531411665044)
--(axis cs:280,0.911038756802973)
--(axis cs:265,0.892432139181934)
--(axis cs:250,0.870755067156184)
--cycle;

\path [draw=color1, fill=color1, opacity=0.2]
(axis cs:250,0.883440605880725)
--(axis cs:250,0.87705457891587)
--(axis cs:265,0.897302566459446)
--(axis cs:280,0.913996365060719)
--(axis cs:295,0.930783656078225)
--(axis cs:310,0.94442485064018)
--(axis cs:325,0.955229456448946)
--(axis cs:340,0.964903019808405)
--(axis cs:355,0.973786478869367)
--(axis cs:370,0.980972021758519)
--(axis cs:385,0.986107966939378)
--(axis cs:400,0.989050567316494)
--(axis cs:400,0.990811789779283)
--(axis cs:400,0.990811789779283)
--(axis cs:385,0.98803180509386)
--(axis cs:370,0.983207685273372)
--(axis cs:355,0.97625662238871)
--(axis cs:340,0.968042414796029)
--(axis cs:325,0.958773702015014)
--(axis cs:310,0.948305070402537)
--(axis cs:295,0.935338274296459)
--(axis cs:280,0.919392104710676)
--(axis cs:265,0.902786566842077)
--(axis cs:250,0.883440605880725)
--cycle;

\path [draw=color2, fill=color2, opacity=0.2]
(axis cs:250,0.88156407284586)
--(axis cs:250,0.875258358600564)
--(axis cs:265,0.894527199155822)
--(axis cs:280,0.912951197500598)
--(axis cs:295,0.92875501205927)
--(axis cs:310,0.942721663639985)
--(axis cs:325,0.954767311849208)
--(axis cs:340,0.964576372576441)
--(axis cs:355,0.973543183071339)
--(axis cs:370,0.980939599627116)
--(axis cs:385,0.985804784815575)
--(axis cs:400,0.988982405426614)
--(axis cs:400,0.990639624538853)
--(axis cs:400,0.990639624538853)
--(axis cs:385,0.98779543442634)
--(axis cs:370,0.983168093181788)
--(axis cs:355,0.976083364894765)
--(axis cs:340,0.967663378114649)
--(axis cs:325,0.958254840171146)
--(axis cs:310,0.946813650536763)
--(axis cs:295,0.933440093267651)
--(axis cs:280,0.918156360889884)
--(axis cs:265,0.900393737820278)
--(axis cs:250,0.88156407284586)
--cycle;

\addplot [thick, color0, mark=triangle*, mark size=4.5, mark options={solid,draw=white}]
table {%
250 0.867649003174279
265 0.889585025474341
280 0.908184851474679
295 0.925116055229022
310 0.940312400033136
325 0.951037427042535
340 0.962230522600925
355 0.971597498555802
370 0.979268360116113
385 0.985121775148812
400 0.988290818286301
};
\addlegendentry{$T=2$}
\addplot [thick, color1, mark=square*, mark size=3.5, mark options={solid,draw=white}]
table {%
250 0.880230113118895
265 0.897677071875937
280 0.916872608373147
295 0.933091067483757
310 0.946356505751051
325 0.957058032807868
340 0.966518909792586
355 0.97505130545154
370 0.982025925976004
385 0.987084733079489
400 0.989951330499653
};
\addlegendentry{$T=5$}
\addplot [thick, color2, mark=diamond*, mark size=3.5, mark options={solid,draw=white}]
table {%
250 0.878403790195734
265 0.897677071875937
280 0.915561908014054
295 0.93106492883042
310 0.94482182354695
325 0.956453806484698
340 0.966169404541974
355 0.974784793219731
370 0.982038021502641
385 0.986873478901994
400 0.98982957794602
};
\addlegendentry{$T=20$}
\end{axis}

\end{tikzpicture}}}
 \setlength\belowcaptionskip{-.4cm}
  \caption{\textbf{Results of using EASE with different prediction windows for the local phase.} Average processing (\ref{fig:proc_power}) and migration (\ref{fig:migr_power}) power dissipation of the edge servers. Energy efficiency with respect to the generation probability (\ref{fig:eff_prob}) and to the power generated by the \ac{PV} (\ref{fig:eff_power}) cells.}
  \label{fig:ease}
\end{figure*}
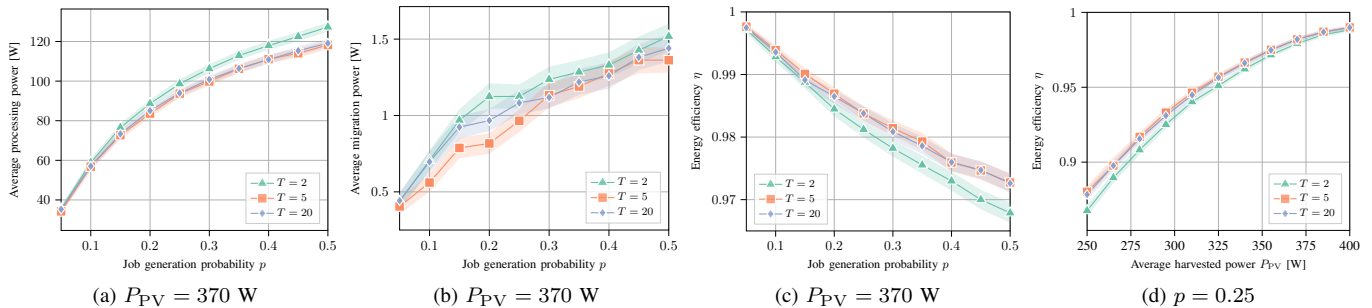

EASE is evaluated by varying the local optimization window size $T$ of \ac{MPC}. By increasing it the controller is likely to find a better solution for the local management of resources and better estimates, which can be used in the migration process of phase 2. Fig.~\ref{fig:ease} shows the results of the aforementioned metrics for $T\in\{2,5,20\}$ time slots. Specifically, in Fig.~\ref{fig:proc_power} the processing power is shown as a function of the job generation probability $p$. While the curves for $T=5$ and $T=20$ substantially overlap, there is a slight increase in the energy consumption using $T=2$ (of about $5\%$). For the migration power (Fig.~\ref{fig:migr_power}), the configuration that drains more energy is still $T=2$, due to a poor prediction of future resources. However, setting $T=5$ leads to a better migration efficiency than $T=20$, but in the latter case the algorithm better captures the future system evolution, thus migrating the jobs to the next serving \ac{eNB} at a slightly higher rate (see Table~\ref{tab:results}). The overall energy efficiency is depicted in Figs.~\ref{fig:eff_prob} and~\ref{fig:eff_power}, showing that EASE is resilient to the prediction window size, as $T=2$ loses at most $0.5\%$ in efficiency when compared to the other two policies. In what follows, we select $T=5$ to be compared with other existing strategies, as it provides the best tradeoff between performance and complexity.

\vspace{-2mm}   

\subsection{EASE vs other migration methods from the literature}

\begin{figure*}[tb]
  \centering
  \subfloat[\label{fig:proc_power_comp}$P_{\rm PV}=370$~W]{\resizebox{0.25\textwidth}{!}{% This file was created with tikzplotlib v0.9.12.
\begin{tikzpicture}

\definecolor{color0}{rgb}{0.4,0.76078431372549,0.647058823529412}
\definecolor{color1}{rgb}{0.988235294117647,0.552941176470588,0.384313725490196}
\definecolor{color2}{rgb}{0.552941176470588,0.627450980392157,0.796078431372549}
\definecolor{color3}{rgb}{0.905882352941176,0.541176470588235,0.764705882352941}
\definecolor{color4}{rgb}{0.650980392156863,0.847058823529412,0.329411764705882}
\definecolor{color5}{rgb}{1,0.85098039215,0.18431372549}
\definecolor{color6}{rgb}{0.89803921568,0.76862745098,0.58039215686}

\begin{axis}[
legend cell align={left},
legend style={
  fill opacity=0.8,
  draw opacity=1,
  text opacity=1,
  at={(0.03,0.97)},
  anchor=north west,
  draw=white!80!black,
  font=\footnotesize
},
label style={font=\small},
ymajorgrids=true,
xmajorgrids=true,
tick align=outside,
tick pos=left,
x grid style={white!69.0196078431373!black},
xlabel={Job generation probability $p$},
xmin=0.05, xmax=0.5,
xtick style={color=black},
y grid style={white!69.0196078431373!black},
ylabel={Average processing power [W]},
ymin=24.5986903640945, ymax=210.407123010561,
ytick style={color=black}
]

\path [draw=color1, fill=color1, opacity=0.2]
(axis cs:0.05,35.3851144723565)
--(axis cs:0.05,33.0445282116611)
--(axis cs:0.1,55.4123484332974)
--(axis cs:0.15,71.0249775282413)
--(axis cs:0.2,81.8302253280146)
--(axis cs:0.25,91.838932546879)
--(axis cs:0.3,97.9795917693383)
--(axis cs:0.35,104.361248100292)
--(axis cs:0.4,109.247504671865)
--(axis cs:0.45,112.142902165281)
--(axis cs:0.5,116.359116280994)
--(axis cs:0.5,120.073085144384)
--(axis cs:0.5,120.073085144384)
--(axis cs:0.45,116.011293243105)
--(axis cs:0.4,113.087579078297)
--(axis cs:0.35,108.100535951871)
--(axis cs:0.3,101.580240427471)
--(axis cs:0.25,95.4493726195057)
--(axis cs:0.2,85.3207474462326)
--(axis cs:0.15,74.2995016690673)
--(axis cs:0.1,58.2501575114806)
--(axis cs:0.05,35.3851144723565)
--cycle;

\path [draw=color3, fill=color3, opacity=0.2]
(axis cs:0.05,41.4712602796053)
--(axis cs:0.05,38.1132115757354)
--(axis cs:0.1,74.0255293269832)
--(axis cs:0.15,103.171798623956)
--(axis cs:0.2,125.326023319778)
--(axis cs:0.25,144.402405212209)
--(axis cs:0.3,159.574683286313)
--(axis cs:0.35,172.139238404186)
--(axis cs:0.4,182.929245631689)
--(axis cs:0.45,190.228478703771)
--(axis cs:0.5,196.737929391041)
--(axis cs:0.5,201.961285162995)
--(axis cs:0.5,201.961285162995)
--(axis cs:0.45,195.1517720611)
--(axis cs:0.4,187.893530855087)
--(axis cs:0.35,177.037385297765)
--(axis cs:0.3,164.672896252656)
--(axis cs:0.25,149.47480617048)
--(axis cs:0.2,130.269754362878)
--(axis cs:0.15,107.995169995668)
--(axis cs:0.1,78.3690661919679)
--(axis cs:0.05,41.4712602796053)
--cycle;

\path [draw=color4, fill=color4, opacity=0.2]
(axis cs:0.05,40.8104584324272)
--(axis cs:0.05,37.555209079044)
--(axis cs:0.1,73.2765144397839)
--(axis cs:0.15,101.363502031186)
--(axis cs:0.2,124.12442978647)
--(axis cs:0.25,143.493482643237)
--(axis cs:0.3,155.691347197786)
--(axis cs:0.35,168.604063015499)
--(axis cs:0.4,178.920231897344)
--(axis cs:0.45,187.164703491194)
--(axis cs:0.5,193.202724684579)
--(axis cs:0.5,198.206222374303)
--(axis cs:0.5,198.206222374303)
--(axis cs:0.45,192.19591856633)
--(axis cs:0.4,184.109621698064)
--(axis cs:0.35,173.686913237208)
--(axis cs:0.3,160.643463323874)
--(axis cs:0.25,148.840650367203)
--(axis cs:0.2,129.271848843732)
--(axis cs:0.15,105.86433834748)
--(axis cs:0.1,77.530871554734)
--(axis cs:0.05,40.8104584324272)
--cycle;

\path [draw=color5, fill=color5, opacity=0.2]
(axis cs:0.05,39.3025782435573)
--(axis cs:0.05,36.1966271025375)
--(axis cs:0.1,69.908285112487)
--(axis cs:0.15,95.9633230627061)
--(axis cs:0.2,117.590346546052)
--(axis cs:0.25,135.846919877861)
--(axis cs:0.3,150.028538322356)
--(axis cs:0.35,161.736493289761)
--(axis cs:0.4,170.210210410389)
--(axis cs:0.45,177.756993239093)
--(axis cs:0.5,184.766734702951)
--(axis cs:0.5,190.085901836696)
--(axis cs:0.5,190.085901836696)
--(axis cs:0.45,183.255226172103)
--(axis cs:0.4,175.906025177063)
--(axis cs:0.35,166.970346257766)
--(axis cs:0.3,155.508143376974)
--(axis cs:0.25,141.225094760068)
--(axis cs:0.2,122.567622642279)
--(axis cs:0.15,100.497942137011)
--(axis cs:0.1,74.0783663661574)
--(axis cs:0.05,39.3025782435573)
--cycle;

\path [draw=color6, fill=color6, opacity=0.2]
(axis cs:0.05,41.8652889157068)
--(axis cs:0.05,38.4835159719225)
--(axis cs:0.1,73.7491324814323)
--(axis cs:0.15,100.193894811613)
--(axis cs:0.2,123.751441488532)
--(axis cs:0.25,141.723101623438)
--(axis cs:0.3,156.561574890482)
--(axis cs:0.35,167.874042768607)
--(axis cs:0.4,176.700150062006)
--(axis cs:0.45,183.717093837289)
--(axis cs:0.5,190.202724684579)
--(axis cs:0.5,195.206222374303)
--(axis cs:0.45,188.784163078419)
--(axis cs:0.4,181.772239266987)
--(axis cs:0.35,173.144325947897)
--(axis cs:0.3,161.877116009494)
--(axis cs:0.25,146.653596677809)
--(axis cs:0.2,128.646963689068)
--(axis cs:0.15,104.959426288801)
--(axis cs:0.1,78.0850531574266)
--(axis cs:0.05,41.8652889157068)
--cycle;

\addplot [thick, color1, mark=square*, mark size=3.5, mark options={solid,draw=white}]
table {%
0.05 34.2293065287903
0.1 56.7906845282257
0.15 72.7103917836207
0.2 83.5863265393843
0.25 93.6717024597484
0.3 99.8211126065799
0.35 106.204846531075
0.4 111.089339697973
0.45 114.08806805061
0.5 118.234601400376
};
\addlegendentry{EASE}
\addplot [thick, color3, mark=*, mark size=3.5, mark options={solid,draw=white}]
table {%
0.05 39.8137761939572
0.1 76.2889232229345
0.15 105.567110509984
0.2 127.849908132113
0.25 147.075695252821
0.3 162.022168313054
0.35 174.597383930493
0.4 185.303786067386
0.45 192.694201197415
0.5 199.318225963783
};
\addlegendentry{keep}
\addplot [thick, color4, mark=pentagon*, mark size=3, mark options={solid,draw=white}]
table {%
0.05 39.1834303531653
0.1 75.331616026297
0.15 103.550567102317
0.2 126.867321414063
0.25 146.114186469894
0.3 158.163038700489
0.35 171.198229656948
0.4 181.53011326514
0.45 189.730717385126
0.5 195.703413128336
};
\addlegendentry{migrate}
\addplot [thick, color5, mark=otimes*, mark size=3, mark options={solid,draw=white}]
table {%
0.05 37.73513873105
0.1 71.9384598287254
0.15 98.2697448551636
0.2 120.065504463915
0.25 138.486578023382
0.3 152.74177623245
0.35 164.300514771888
0.4 173.09335093016
0.45 180.60679120858
0.5 187.551734808088
};
\addlegendentry{threshold}
\addplot [thick, color6, mark=oplus*, mark size=3, mark options={solid,draw=white}]
table {%
0.05 40.1504763922512
0.1 75.866110813922
0.15 102.67721535256
0.2 126.113014412658
0.25 144.308807787077
0.3 159.158437188889
0.35 170.526463360385
0.4 179.293786254994
0.45 186.195828509993
0.5 192.895828509993
};
\addlegendentry{lyapunov}
\end{axis}

\end{tikzpicture}}}
  \subfloat[\label{fig:migr_power_comp}$P_{\rm PV}=370$~W]{\resizebox{0.24\textwidth}{!}{% This file was created with tikzplotlib v0.9.12.
\begin{tikzpicture}

\definecolor{color0}{rgb}{0.4,0.76078431372549,0.647058823529412}
\definecolor{color1}{rgb}{0.988235294117647,0.552941176470588,0.384313725490196}
\definecolor{color2}{rgb}{0.552941176470588,0.627450980392157,0.796078431372549}
\definecolor{color3}{rgb}{0.905882352941176,0.541176470588235,0.764705882352941}
\definecolor{color4}{rgb}{0.650980392156863,0.847058823529412,0.329411764705882}
\definecolor{color5}{rgb}{1,0.85098039215,0.18431372549}
\definecolor{color6}{rgb}{0.89803921568,0.76862745098,0.58039215686}

\begin{axis}[
legend cell align={left},
legend style={
  fill opacity=0.8,
  draw opacity=1,
  text opacity=1,
  at={(0.97,0.97)},
  anchor=north east,
  draw=white!80!black,
  font=\footnotesize
},
label style={font=\small},
ymajorgrids=true,
xmajorgrids=true,
tick align=outside,
tick pos=left,
x grid style={white!69.0196078431373!black},
xlabel={Job generation probability $p$},
xmin=0.05, xmax=0.5,
xtick style={color=black},
y grid style={white!69.0196078431373!black},
ylabel={Average migration power [W]},
ymin=0, ymax=4.71276415660682,
ytick style={color=black}
]
\path [draw=color1, fill=color1, opacity=0.2]
(axis cs:0.05,0.453502716199124)
--(axis cs:0.05,0.358653521304881)
--(axis cs:0.1,0.505514966777185)
--(axis cs:0.15,0.723673278910016)
--(axis cs:0.2,0.752179495774984)
--(axis cs:0.25,0.896131163082572)
--(axis cs:0.3,1.05704623182218)
--(axis cs:0.35,1.11643718437477)
--(axis cs:0.4,1.19808044294263)
--(axis cs:0.45,1.27991577659775)
--(axis cs:0.5,1.28414266871811)
--(axis cs:0.5,1.44131662482308)
--(axis cs:0.5,1.44131662482308)
--(axis cs:0.45,1.44357144445182)
--(axis cs:0.4,1.35076838073526)
--(axis cs:0.35,1.26213135405852)
--(axis cs:0.3,1.21346460645247)
--(axis cs:0.25,1.03612607364665)
--(axis cs:0.2,0.880294487415522)
--(axis cs:0.15,0.851644215780462)
--(axis cs:0.1,0.616928586048259)
--(axis cs:0.05,0.453502716199124)
--cycle;

\path [draw=color3, fill=color3, opacity=0.2]
(axis cs:0.05,0)
--(axis cs:0.05,0)
--(axis cs:0.1,0)
--(axis cs:0.15,0)
--(axis cs:0.2,0)
--(axis cs:0.25,0)
--(axis cs:0.3,0)
--(axis cs:0.35,0)
--(axis cs:0.4,0)
--(axis cs:0.45,0)
--(axis cs:0.5,0)
--(axis cs:0.5,0)
--(axis cs:0.5,0)
--(axis cs:0.45,0)
--(axis cs:0.4,0)
--(axis cs:0.35,0)
--(axis cs:0.3,0)
--(axis cs:0.25,0)
--(axis cs:0.2,0)
--(axis cs:0.15,0)
--(axis cs:0.1,0)
--(axis cs:0.05,0)
--cycle;

\path [draw=color4, fill=color4, opacity=0.2]
(axis cs:0.05,0.495330288245249)
--(axis cs:0.05,0.396548027873513)
--(axis cs:0.1,0.74796975219293)
--(axis cs:0.15,1.13477286481845)
--(axis cs:0.2,1.38364667453556)
--(axis cs:0.25,1.70766164501885)
--(axis cs:0.3,1.82200389611393)
--(axis cs:0.35,2.02040863032767)
--(axis cs:0.4,2.10747240175141)
--(axis cs:0.45,2.30557828637286)
--(axis cs:0.5,2.35662909316708)
--(axis cs:0.5,2.58136271831281)
--(axis cs:0.5,2.58136271831281)
--(axis cs:0.45,2.52262470346448)
--(axis cs:0.4,2.31500336395928)
--(axis cs:0.35,2.2247748151323)
--(axis cs:0.3,2.03220123503454)
--(axis cs:0.25,1.8950712763392)
--(axis cs:0.2,1.55175653137393)
--(axis cs:0.15,1.295033844607)
--(axis cs:0.1,0.875077279583617)
--(axis cs:0.05,0.495330288245249)
--cycle;

\path [draw=color5, fill=color5, opacity=0.2]
(axis cs:0.05,0.984567488143076)
--(axis cs:0.05,0.853688687439966)
--(axis cs:0.1,1.28013021116607)
--(axis cs:0.15,1.45199217687106)
--(axis cs:0.2,1.65572802785558)
--(axis cs:0.25,1.8474980212811)
--(axis cs:0.3,1.87942858352005)
--(axis cs:0.35,1.93162453087937)
--(axis cs:0.4,2.03611053854614)
--(axis cs:0.45,2.02931729939058)
--(axis cs:0.5,1.89360258729684)
--(axis cs:0.5,2.10816377500749)
--(axis cs:0.5,2.10816377500749)
--(axis cs:0.45,2.2465721963223)
--(axis cs:0.4,2.24653116982429)
--(axis cs:0.35,2.14999025567869)
--(axis cs:0.3,2.08939736334446)
--(axis cs:0.25,2.04873869843656)
--(axis cs:0.2,1.84901128067486)
--(axis cs:0.15,1.62407499582455)
--(axis cs:0.1,1.44445939877008)
--(axis cs:0.05,0.984567488143076)
--cycle;

\path [draw=color6, fill=color6, opacity=0.2]
(axis cs:0.05,1.13825538194404)
--(axis cs:0.05,0.877579563530393)
--(axis cs:0.1,1.73363885269978)
--(axis cs:0.15,2.66659278958303)
--(axis cs:0.2,3.42749203813479)
--(axis cs:0.25,4.03553811726423)
--(axis cs:0.3,4.62419400165767)
--(axis cs:0.35,5.28894084781851)
--(axis cs:0.4,6.04535070707429)
--(axis cs:0.45,6.4122895283836)
--(axis cs:0.45,7.22270383263436)
--(axis cs:0.45,7.22270383263436)
--(axis cs:0.4,6.80214608747308)
--(axis cs:0.35,5.98437493807842)
--(axis cs:0.3,5.3074217831897)
--(axis cs:0.25,4.61814715839873)
--(axis cs:0.2,3.96166614313831)
--(axis cs:0.15,3.09836200086091)
--(axis cs:0.1,2.11283867464203)
--(axis cs:0.05,1.13825538194404)
--cycle;

\addplot [thick, color1, mark=square*, mark size=3.5, mark options={solid,draw=white}]
table {%
0.05 0.404837643720749
0.1 0.560782499672877
0.15 0.787455505298869
0.2 0.816880663897559
0.25 0.964660678674002
0.3 1.13200332232296
0.35 1.18978797945869
0.4 1.27494809405528
0.45 1.36232720359111
0.5 1.36227729301731
};
\addlegendentry{EASE}
\addplot [thick, color3, mark=*, mark size=3.5, mark options={solid,draw=white}]
table {%
0.05 0
0.1 0
0.15 0
0.2 0
0.25 0
0.3 0
0.35 0
0.4 0
0.45 0
0.5 0
};
\addlegendentry{keep}
\addplot [thick, color4, mark=pentagon*, mark size=3, mark options={solid,draw=white}]
table {%
0.05 0.447750671114561
0.1 0.808528138892863
0.15 1.21517414033826
0.2 1.46444463789487
0.25 1.79847697698649
0.3 1.9275230686536
0.35 2.12395906766257
0.4 2.20952246984801
0.45 2.41085742977226
0.5 2.47025866316789
};
\addlegendentry{migrate}
\addplot [thick, color5, mark=otimes*, mark size=3, mark options={solid,draw=white}]
table {%
0.05 0.921212222222031
0.1 1.36514345098392
0.15 1.53626002306412
0.2 1.74835822298736
0.25 1.94658096808006
0.3 1.97663394004659
0.35 2.04590704325284
0.4 2.13974481219513
0.45 2.14321776213854
0.5 2.00292854845628
};
\addlegendentry{threshold}
\addplot [thick, color6, mark=oplus*, mark size=3, mark options={solid,draw=white}]
table {%
0.05 1.00735891927003
0.1 1.92529241356859
0.15 2.88992877821267
0.2 3.68305527842544
0.25 4.31839635305286
0.3 4.95389411802073
0.35 5.63453635227304
0.4 6.41551241686497
0.45 6.82884423309507
};
\addlegendentry{lyapunov}
\end{axis}

\end{tikzpicture}}}
  \subfloat[\label{fig:eff_prob_comp}$P_{\rm PV}=370$~W]{\resizebox{0.25\textwidth}{!}{% This file was created with tikzplotlib v0.9.12.
\begin{tikzpicture}

\definecolor{color0}{rgb}{0.4,0.76078431372549,0.647058823529412}
\definecolor{color1}{rgb}{0.988235294117647,0.552941176470588,0.384313725490196}
\definecolor{color2}{rgb}{0.552941176470588,0.627450980392157,0.796078431372549}
\definecolor{color3}{rgb}{0.905882352941176,0.541176470588235,0.764705882352941}
\definecolor{color4}{rgb}{0.650980392156863,0.847058823529412,0.329411764705882}
\definecolor{color5}{rgb}{1,0.85098039215,0.18431372549}
\definecolor{color6}{rgb}{0.89803921568,0.76862745098,0.58039215686}

\begin{axis}[
legend cell align={left},
legend style={
  fill opacity=0.8,
  draw opacity=1,
  text opacity=1,
  at={(0.03,0.03)},
  anchor=south west,
  draw=white!80!black,
  font=\footnotesize
},
label style={font=\small},
ymajorgrids=true,
xmajorgrids=true,
tick align=outside,
tick pos=left,
x grid style={white!69.0196078431373!black},
xlabel={Job generation probability $p$},
xmin=0.05, xmax=0.5,
xtick style={color=black},
y grid style={white!69.0196078431373!black},
ylabel={Energy efficiency $\eta$},
ymin=0.891315865611761, ymax=1,
ytick style={color=black}
]

\path [draw=color1, fill=color1, opacity=0.2]
(axis cs:0.05,0.997986922601434)
--(axis cs:0.05,0.997275450081671)
--(axis cs:0.1,0.993300412180068)
--(axis cs:0.15,0.989258001590407)
--(axis cs:0.2,0.986007014294612)
--(axis cs:0.25,0.982734924109936)
--(axis cs:0.3,0.980322844166562)
--(axis cs:0.35,0.978025356162146)
--(axis cs:0.4,0.9746033777012)
--(axis cs:0.45,0.973248907646123)
--(axis cs:0.5,0.971416025342057)
--(axis cs:0.5,0.974146280299017)
--(axis cs:0.5,0.974146280299017)
--(axis cs:0.45,0.975971861591322)
--(axis cs:0.4,0.977288835807338)
--(axis cs:0.35,0.980563514304181)
--(axis cs:0.3,0.982579057120073)
--(axis cs:0.25,0.984867206407062)
--(axis cs:0.2,0.987846433771587)
--(axis cs:0.15,0.990936007333606)
--(axis cs:0.1,0.994527789303836)
--(axis cs:0.05,0.997986922601434)
--cycle;

\path [draw=color3, fill=color3, opacity=0.2]
(axis cs:0.05,0.993491525720363)
--(axis cs:0.05,0.99201969659003)
--(axis cs:0.1,0.979890743045225)
--(axis cs:0.15,0.966525894885204)
--(axis cs:0.2,0.952976745928695)
--(axis cs:0.25,0.941109064418946)
--(axis cs:0.3,0.930368877721654)
--(axis cs:0.35,0.921832715916579)
--(axis cs:0.4,0.913389991651456)
--(axis cs:0.45,0.907488868572248)
--(axis cs:0.5,0.901298226011622)
--(axis cs:0.5,0.906371476058492)
--(axis cs:0.5,0.906371476058492)
--(axis cs:0.45,0.912365002127342)
--(axis cs:0.4,0.918439342728893)
--(axis cs:0.35,0.92661612355808)
--(axis cs:0.3,0.934805908434725)
--(axis cs:0.25,0.945283341523053)
--(axis cs:0.2,0.956789975554312)
--(axis cs:0.15,0.969808334735055)
--(axis cs:0.1,0.98239148888791)
--(axis cs:0.05,0.993491525720363)
--cycle;

\path [draw=color4, fill=color4, opacity=0.2]
(axis cs:0.05,0.989449280341491)
--(axis cs:0.05,0.987635730528811)
--(axis cs:0.1,0.973075100411856)
--(axis cs:0.15,0.959347553659257)
--(axis cs:0.2,0.945838986093466)
--(axis cs:0.25,0.933023840141999)
--(axis cs:0.3,0.925063728650439)
--(axis cs:0.35,0.915648763452642)
--(axis cs:0.4,0.907789068660352)
--(axis cs:0.45,0.900523190370671)
--(axis cs:0.5,0.896701578242029)
--(axis cs:0.5,0.902086991701455)
--(axis cs:0.5,0.902086991701455)
--(axis cs:0.45,0.905610191485115)
--(axis cs:0.4,0.912785773584486)
--(axis cs:0.35,0.920289199404107)
--(axis cs:0.3,0.929781092022228)
--(axis cs:0.25,0.937148335258904)
--(axis cs:0.2,0.949968885904827)
--(axis cs:0.15,0.962767493107693)
--(axis cs:0.1,0.97582216713986)
--(axis cs:0.05,0.989449280341491)
--cycle;

\path [draw=color5, fill=color5, opacity=0.2]
(axis cs:0.05,0.990731647709023)
--(axis cs:0.05,0.989196464587716)
--(axis cs:0.1,0.975875834082247)
--(axis cs:0.15,0.962445668186286)
--(axis cs:0.2,0.94781173599256)
--(axis cs:0.25,0.935314335677301)
--(axis cs:0.3,0.924724350320548)
--(axis cs:0.35,0.915905788834799)
--(axis cs:0.4,0.907351337284211)
--(axis cs:0.45,0.902070578891539)
--(axis cs:0.5,0.896042227186661)
--(axis cs:0.5,0.901276100342464)
--(axis cs:0.5,0.901276100342464)
--(axis cs:0.45,0.9071552404411)
--(axis cs:0.4,0.912534387294475)
--(axis cs:0.35,0.920921122808162)
--(axis cs:0.3,0.92934976357344)
--(axis cs:0.25,0.939795007438763)
--(axis cs:0.2,0.951706417530641)
--(axis cs:0.15,0.965914623441223)
--(axis cs:0.1,0.978436560110834)
--(axis cs:0.05,0.990731647709023)
--cycle;

\path [draw=color6, fill=color6, opacity=0.2]
(axis cs:0.05,0.988839296026592)
--(axis cs:0.05,0.986993058355282)
--(axis cs:0.1,0.971952158545236)
--(axis cs:0.15,0.957427238380703)
--(axis cs:0.2,0.943574187618785)
--(axis cs:0.25,0.931413529708807)
--(axis cs:0.3,0.920270637129134)
--(axis cs:0.35,0.911772037898685)
--(axis cs:0.4,0.904424804438537)
--(axis cs:0.45,0.897997927836228)
--(axis cs:0.5,0.893042227186661)
--(axis cs:0.5,0.895042227186661)
--(axis cs:0.45,0.90337929049596)
--(axis cs:0.4,0.909438931392538)
--(axis cs:0.35,0.916737171265214)
--(axis cs:0.3,0.925070523267434)
--(axis cs:0.25,0.935870820893657)
--(axis cs:0.2,0.947658758798757)
--(axis cs:0.15,0.960920651897327)
--(axis cs:0.1,0.974782586171015)
--(axis cs:0.05,0.988839296026592)
--cycle;

\addplot [thick, color1, mark=square*, mark size=3.5, mark options={solid,draw=white}]
table {%
0.05 0.997646356042138
0.1 0.993907965095324
0.15 0.99009882982774
0.2 0.986925864642315
0.25 0.983772537958589
0.3 0.98141853886887
0.35 0.979291124564484
0.4 0.975941764718911
0.45 0.97466425325156
0.5 0.972741900203779
};
\addlegendentry{EASE}
\addplot [thick, color3, mark=*, mark size=3.5, mark options={solid,draw=white}]
table {%
0.05 0.992778096256302
0.1 0.981153197161418
0.15 0.96822832605346
0.2 0.955017902853224
0.25 0.943049323557722
0.3 0.932479409647857
0.35 0.924257935650937
0.4 0.915796964720835
0.45 0.909887016589873
0.5 0.903875175854626
};
\addlegendentry{keep}
\addplot [thick, color4, mark=pentagon*, mark size=3, mark options={solid,draw=white}]
table {%
0.05 0.988574738971287
0.1 0.974509942504905
0.15 0.961058216011709
0.2 0.947836383069027
0.25 0.935099254131601
0.3 0.927504869707245
0.35 0.918038814324047
0.4 0.910288300280177
0.45 0.903150619081651
0.5 0.89924557448514
};
\addlegendentry{migrate}
\addplot [thick, color5, mark=otimes*, mark size=3, mark options={solid,draw=white}]
table {%
0.05 0.990011956548143
0.1 0.977179259106412
0.15 0.964220794311376
0.2 0.949777521862874
0.25 0.937525537510624
0.3 0.926936912056664
0.35 0.918310098984967
0.4 0.909871647469708
0.45 0.904595850116604
0.5 0.898772557883219
};
\addlegendentry{threshold}
\addplot [thick, color6, mark=oplus*, mark size=3, mark options={solid,draw=white}]
table {%
0.05 0.987941984897043
0.1 0.973415032247657
0.15 0.959196415517143
0.2 0.94569346276337
0.25 0.933753512473284
0.3 0.922676223804334
0.35 0.914293181024037
0.4 0.906958709150353
0.45 0.900688307434928
0.5 0.898272557883219
};
\addlegendentry{lyapunov}
\end{axis}

\end{tikzpicture}}}
  \subfloat[\label{fig:eff_power_comp}$p=0.25$]{\resizebox{0.25\textwidth}{!}{% This file was created with tikzplotlib v0.9.12.
\begin{tikzpicture}

\definecolor{color0}{rgb}{0.4,0.76078431372549,0.647058823529412}
\definecolor{color1}{rgb}{0.988235294117647,0.552941176470588,0.384313725490196}
\definecolor{color2}{rgb}{0.552941176470588,0.627450980392157,0.796078431372549}
\definecolor{color3}{rgb}{0.905882352941176,0.541176470588235,0.764705882352941}
\definecolor{color4}{rgb}{0.650980392156863,0.847058823529412,0.329411764705882}
\definecolor{color5}{rgb}{1,0.85098039215,0.18431372549}
\definecolor{color6}{rgb}{0.89803921568,0.76862745098,0.58039215686}

\begin{axis}[
legend cell align={left},
legend style={
  fill opacity=0.8,
  draw opacity=1,
  text opacity=1,
  at={(0.97,0.03)},
  anchor=south east,
  draw=white!80!black,
  font=\footnotesize
},
label style={font=\small},
ymajorgrids=true,
xmajorgrids=true,
tick align=outside,
tick pos=left,
x grid style={white!69.0196078431373!black},
xlabel={Average harvested power $P_{\rm PV}$ [W]},
xmin=250, xmax=400,
xtick style={color=black},
xtick={250, 275, 300, 325, 350, 375, 400},
xticklabels={250, 275, 300, 325, 350, 375, 400},
y grid style={white!69.0196078431373!black},
ylabel={Energy efficiency $\eta$},
ymin=0.754414948331404, ymax=1,
ytick style={color=black}
]
\path [draw=color1, fill=color1, opacity=0.2]
(axis cs:250,0.883440605880725)
--(axis cs:250,0.87705457891587)
--(axis cs:265,0.897302566459446)
--(axis cs:280,0.913996365060719)
--(axis cs:295,0.930783656078225)
--(axis cs:310,0.94442485064018)
--(axis cs:325,0.955229456448946)
--(axis cs:340,0.964903019808405)
--(axis cs:355,0.973786478869367)
--(axis cs:370,0.980972021758519)
--(axis cs:385,0.986107966939378)
--(axis cs:400,0.989050567316494)
--(axis cs:400,0.990811789779283)
--(axis cs:400,0.990811789779283)
--(axis cs:385,0.98803180509386)
--(axis cs:370,0.983207685273372)
--(axis cs:355,0.97625662238871)
--(axis cs:340,0.968042414796029)
--(axis cs:325,0.958773702015014)
--(axis cs:310,0.948305070402537)
--(axis cs:295,0.935338274296459)
--(axis cs:280,0.919392104710676)
--(axis cs:265,0.902786566842077)
--(axis cs:250,0.883440605880725)
--cycle;

\path [draw=color3, fill=color3, opacity=0.2]
(axis cs:250,0.778566116174065)
--(axis cs:250,0.77053549298251)
--(axis cs:265,0.794181431664634)
--(axis cs:280,0.817125102300621)
--(axis cs:295,0.84020203086352)
--(axis cs:310,0.863283578430501)
--(axis cs:325,0.885000450469994)
--(axis cs:340,0.905365445960385)
--(axis cs:355,0.924419691782525)
--(axis cs:370,0.940923696528397)
--(axis cs:385,0.951813234272197)
--(axis cs:400,0.958795934034671)
--(axis cs:400,0.962221580511594)
--(axis cs:400,0.962221580511594)
--(axis cs:385,0.955800430906903)
--(axis cs:370,0.945162120697165)
--(axis cs:355,0.928945766675396)
--(axis cs:340,0.910222298476745)
--(axis cs:325,0.89018032066046)
--(axis cs:310,0.868810099636235)
--(axis cs:295,0.846941253533435)
--(axis cs:280,0.824268349286608)
--(axis cs:265,0.801391305614607)
--(axis cs:250,0.778566116174065)
--cycle;

\path [draw=color4, fill=color4, opacity=0.2]
(axis cs:250,0.773539664090862)
--(axis cs:250,0.765668881139673)
--(axis cs:265,0.788052358952434)
--(axis cs:280,0.810334694712491)
--(axis cs:295,0.832322245299243)
--(axis cs:310,0.854051986102745)
--(axis cs:325,0.874781635735532)
--(axis cs:340,0.894292621316162)
--(axis cs:355,0.912763900121138)
--(axis cs:370,0.929009221242106)
--(axis cs:385,0.940695068749369)
--(axis cs:400,0.948813336398845)
--(axis cs:400,0.952607272842067)
--(axis cs:400,0.952607272842067)
--(axis cs:385,0.944550890045527)
--(axis cs:370,0.933690941919064)
--(axis cs:355,0.917632861793774)
--(axis cs:340,0.899420887455161)
--(axis cs:325,0.880211250074749)
--(axis cs:310,0.859963366876876)
--(axis cs:295,0.838894692944342)
--(axis cs:280,0.81713706007104)
--(axis cs:265,0.795515062198819)
--(axis cs:250,0.773539664090862)
--cycle;

\path [draw=color5, fill=color5, opacity=0.2]
(axis cs:250,0.78712430302418)
--(axis cs:250,0.779328820183779)
--(axis cs:265,0.79969398431484)
--(axis cs:280,0.820994225757701)
--(axis cs:295,0.842410821818773)
--(axis cs:310,0.864283578430501)
--(axis cs:325,0.884203743808306)
--(axis cs:340,0.902757256600169)
--(axis cs:355,0.920008217945072)
--(axis cs:370,0.935113869310621)
--(axis cs:385,0.945737670350892)
--(axis cs:400,0.95422772552079)
--(axis cs:400,0.957600619161446)
--(axis cs:400,0.957600619161446)
--(axis cs:385,0.949643171140278)
--(axis cs:370,0.939742449614025)
--(axis cs:355,0.924700378894684)
--(axis cs:340,0.90788236312541)
--(axis cs:325,0.889541055417541)
--(axis cs:310,0.869810099636235)
--(axis cs:295,0.849033789065991)
--(axis cs:280,0.827533919208734)
--(axis cs:265,0.806922755785928)
--(axis cs:250,0.78712430302418)
--cycle;

\path [draw=color6, fill=color6, opacity=0.2]
(axis cs:250,0.773756377500998)
--(axis cs:250,0.766112005855168)
--(axis cs:265,0.788635475079056)
--(axis cs:280,0.811548499064886)
--(axis cs:295,0.833677215333934)
--(axis cs:310,0.855560826486765)
--(axis cs:325,0.876744243128302)
--(axis cs:340,0.896383816516438)
--(axis cs:355,0.91493720743384)
--(axis cs:370,0.931519944017449)
--(axis cs:385,0.942942138691213)
--(axis cs:400,0.950688089743616)
--(axis cs:400,0.954514980298012)
--(axis cs:400,0.954514980298012)
--(axis cs:385,0.947051594686931)
--(axis cs:370,0.935981040594802)
--(axis cs:355,0.919792847734529)
--(axis cs:340,0.901252201877456)
--(axis cs:325,0.88208763197177)
--(axis cs:310,0.861480299545648)
--(axis cs:295,0.839948772776408)
--(axis cs:280,0.818495179065686)
--(axis cs:265,0.7962539083386)
--(axis cs:250,0.773756377500998)
--cycle;

\addplot [thick, color1, mark=square*, mark size=3.5, mark options={solid,draw=white}]
table {%
250 0.880230113118895
265 0.897677071875937
280 0.916872608373147
295 0.933091067483757
310 0.946356505751051
325 0.957058032807868
340 0.966518909792586
355 0.97505130545154
370 0.982025925976004
385 0.987084733079489
400 0.989951330499653
};
\addlegendentry{EASE}
\addplot [thick, color3, mark=*, mark size=3.5, mark options={solid,draw=white}]
table {%
250 0.774727542268609
265 0.797772509722329
280 0.820751130286541
295 0.843535180043535
310 0.865907629464735
325 0.887574917455443
340 0.907775543491566
355 0.926672769321554
370 0.943051646999213
385 0.953763605198777
400 0.96047253278892
};
\addlegendentry{keep}
\addplot [thick, color4, mark=pentagon*, mark size=3, mark options={solid,draw=white}]
table {%
250 0.76959244860688
265 0.791798823885865
280 0.813834821187156
295 0.835607588418042
310 0.856929201420287
325 0.877508299787428
340 0.896811924836829
355 0.915165552129276
370 0.931409079893558
385 0.942644268827936
400 0.950679102412054
};
\addlegendentry{migrate}
\addplot [thick, color5, mark=otimes*, mark size=3, mark options={solid,draw=white}]
table {%
250 0.783147655146574
265 0.803369677315164
280 0.824282056783332
295 0.845639070870107
310 0.866968144847723
325 0.886968144847723
340 0.905283376398178
355 0.922332533851082
370 0.937380267025248
385 0.947777368560665
400 0.955815767270868
};
\addlegendentry{threshold}
\addplot [thick, color6, mark=oplus*, mark size=3, mark options={solid,draw=white}]
table {%
250 0.769862916093228
265 0.792532478554931
280 0.81494270927209
295 0.836785657757564
310 0.858562538383833
325 0.879368327338511
340 0.898891761150869
355 0.917412936833316
370 0.933704585725736
385 0.945011969394971
400 0.952628483521335
};
\addlegendentry{lyapunov}
\end{axis}

\end{tikzpicture}}}
  \label{fig:comparison}
 \setlength\belowcaptionskip{-.4cm}
  \caption{\textbf{Comparison between EASE ($\bf{T=5}$) and other approaches of the literature.} Average processing (\ref{fig:proc_power_comp}) and migration (\ref{fig:migr_power_comp}) power dissipation of the edge servers. Energy efficiency with respect to the generation probability (\ref{fig:eff_prob_comp}) and to the power generated by the \ac{PV} (\ref{fig:eff_power_comp}) cells.}
  \label{fig:efficiency}
\end{figure*}
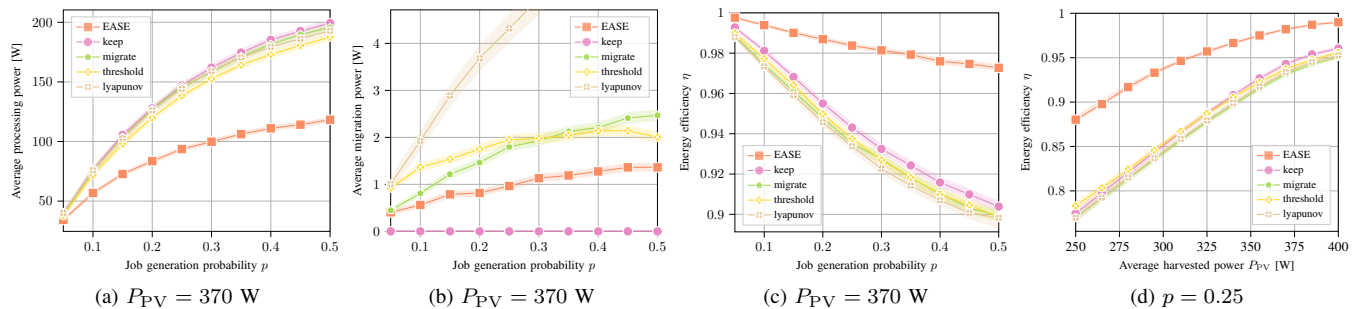

Fig.~\ref{fig:proc_power_comp} shows the processing power, which has an increasing concave trend for all the strategies. As it can be seen, \ac{EASE} allows substantial savings, e.g., as much as $70$~W at $p=0.5$ (a gain of $33\%$) with respect to the benchmarks. The ``threshold'' policy provides a slight improvement over the other heuristics, due to a better organization of the computational resources, as its migration decisions depend on energy considerations. The average power used to migrate the jobs is shown in Fig.~\ref{fig:migr_power_comp}. Since the ``keep'' strategy never migrates tasks, its job migration power is always zero. On the other hand, the strategy with the highest migration power is ``lyapunov'', as it potentially migrates multiple replicas of the service to increase the probability of correctly following the user. The ``migrate'' and ``threshold'' strategies consume consistently more than the optimized EASE, as they migrate services in a blind way, even when the target \ac{MEH} processes them inefficiently. In Fig.~\ref{fig:eff_prob_comp}, the energy efficiency $\eta$ is shown as a function of the job generation probability. All the strategies show an almost linear decrease for increasing $p$. However, the absolute slope of such decrease is larger for the benchmark strategies with respect to \ac{EASE}. At $p=0.5$, EASE allows gaining about $7\%$ in efficiency: the harvested energy can fully support the edge network for at least $97\%$ of the total energy requirement. The energy efficiency is also evaluated by varying the amount of harvested energy (Fig.~\ref{fig:eff_power}), with the \ac{PV} panel generating power in $[P^{\rm PV}_{\rm min}, P^{\rm PV}_{\rm max}]$~W. \ac{EASE} can entirely sustain the edge at least $87.5\%$ of the time when the harvested energy is at its minimum, i.e., $P_{\rm PV}=250$~W, leading to a gain of $10\%$ with respect to the other strategies, thus resulting in a significantly reduced carbon footprint. At $P_{\rm PV}=400$~W the gain is lower, but \ac{EASE} performs very close to complete carbon neutrality (efficiency $\approx 99\%$). Note that $400$~W are just sufficient to \mbox{self-sustain} (on average) the less powerful HP ProLiant server, but not the Nettrix computing unit at full load. As a final consideration, from Figs.~\ref{fig:eff_prob_comp} and~\ref{fig:eff_power_comp}, it can be seen that the largest gain is achieved when either the computing demand is high (large $p$) or the harvested energy is scarce. These are the cases where it is important to use the available resources wisely, and EASE succeeds to do so. 

The results about the jobs drop rate and the fraction of jobs finishing in the \ac{MEH} co-located with the serving \ac{eNB} (dubbed ``minimum latency'') are summarized in Tab.~\ref{tab:results}. In addition to being consistently more energy efficient, \ac{EASE} never discards jobs, while the benchmark strategies drop a significant percentage of the tasks. The ``migrate'' and ``lyapunov'' strategies are the best in following the vehicles' trajectories, i.e., they seek to minimize the latency by transferring the jobs to the closest MEH. EASE takes a different approach, by considering latency deadlines, and seeking to migrate the jobs in a way that minimizes the overall energy that is drained from the power grid, subject to such deadlines. This leads to migration paths where jobs do not necessarily (strictly) follow the users. As a second-order optimization criterion, and only if feasible, \ac{EASE} migrates jobs to the next predicted user location (\ac{eNB}). 

%We observe that \ac{EASE} with $T=20$ achieves a gain over $T=5$ of about $2\%$ on this last aspect, while $T=2$ performs the best, being, however, the least energy efficient.  %Actually, besides the fact that the latter server has a maximum power of $468$~W, the fixed power of keeping the communication circuits switched on is also taken into account, which adds about $70$~W of dissipation. %To this end, the occasional transmission power must be added for the strategies that have the chance to displace the containers.

\begin{table*}[tb]
\centering
\resizebox{0.8\textwidth}{!}{%
\begin{tabular}{c|c|c|c|c|c|c|c}
& \textbf{EASE ($T=2$)} & \textbf{EASE ($T=5$)} & \textbf{EASE ($T=20$)} & \textbf{keep} & \textbf{migrate} & \textbf{threshold} & \textbf{lyapunov} \vspace{1mm}\\
\hline\hline
\textbf{minimum latency jobs} & $33\%$                & $28\%$                & $30\%$                 & --            & $75\%$           & $58\%$             &                   $78\%$ \\
\textbf{drop rate}            & --                    & --                    & --                     & $1.5\%$       & $0.5\%$          & $0.5\%$            & $1.5\%$     \\
\hline             
\end{tabular}%
}
 \setlength\belowcaptionskip{-.3cm}
\caption{Minimum latency executions and drop rates for $p=0.3$ and $P_{\rm PV}=370$~W.}
\label{tab:results}
\end{table*}
\subsection{Convergence of the dual ascent}
\label{sec:convergence}
\begin{figure}
  \centering
  \resizebox{0.8\columnwidth}{!}{\input{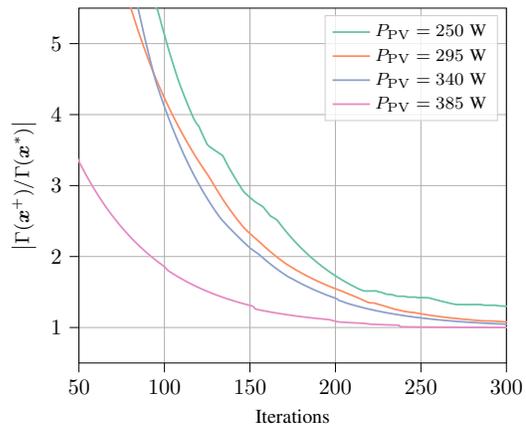}}
 \setlength\belowcaptionskip{-.4cm}
  \caption{Ratio between the value of the cost at iteration $m$ and the optimal cost computed with CVXPY ($90$th percentile). Job generation probability $p=0.25$.}
  \label{fig:convergence_energy}
\end{figure}

In Fig.~\ref{fig:convergence_energy}, the convergence speed of the proposed decentralized solution is evaluated. Specifically, the cost value reached at the current iteration is compared with the optimal solution obtained with CVXPY~\cite{diamond2016cvxpy}, considering the absolute value of their ratio \mbox{$\left|\Gamma(\bm x^+)/\Gamma(\bm x^*)\right|$}. In the plot, the $90$th percentile is shown, discarding hence $10\%$ of outliers. Thus, whenever the ratio settles down to approximately $1$, the nodes have reached the global minimum of the cost function. The results show that the power availability impacts the convergence speed: the more harvested energy $P_{\rm PV}$ is available, the quicker the algorithm reaches the minimum. %This happens because
This descends from the fact that a high energy availability leads to a rare activation of the $\max$ term in function~\eqref{eq:distr_cost}. When the $\max$ term returns $0$ and the constraint~\eqref{eq:distr_proc_constraint} is not active, the optimum is simply given by \mbox{$\bm o_i = \bar{\bm w}_i$}, i.e., the selected action is to follow the vehicle movements. The nodes will be very fast in retrieving this particular solution, as the Lagrange multipliers associated with all the constraints remain null after the first two iterations, leading to accepting the solution.
%The nodes will be very fast in retrieving this particular solution, as, at the second iteration, they will see that the Lagrange multipliers associated with all constraints are still null, and will therefore accept the solution.
Similar reasoning holds for the job generation probability that determines the load of the servers. Here, in the interest of space, we omit the associated plot as it is very similar to Fig.~\ref{fig:convergence_energy}. Specifically, the convergence requires more iterations as $p$ increases. In fact, an increase in the average load experienced by the servers activates the constraint~\eqref{eq:distr_proc_constraint}, modifying the optimal solution or even activating the penalties $\hat{\delta}_i$. As it is known, the dual ascent is slow when being close to constraint boundaries. 
However, as a general result, the number of iterations required to converge even with complex initializations is between $200$ and $500$. The communication overhead can be evaluated considering that two communication rounds (of a few bytes) are required per iteration (see Algorithm~\ref{alg:dual_ascent_server}). Although this may actually appear to be a high number of exchanged messages, we remark that:
%Note that, although this may actually appear to be a high number of exchanged messages, there are three observations to consider.
\begin{inparaenum}[i)]
\item the subsequent step of the proposed pipeline rounds the solution, and, in turn, it is not necessary to retrieve the exact optimum, but it is sufficient to obtain a decent cost value in the continuous domain;
\item we considered slots of $\tau=3$~s, which is the amount of time available to make a migration decision. Longer time slots can be used, leaving more time for the decision process.
%\item similar fully decentralized algorithms can be deployed also in accelerated versions, like what we presented in~\cite{perin2021towards}, at the cost of possibly losing the prospect of finding a nice \mbox{closed-form} solution of the primal problem. 
\end{inparaenum}

\subsection{Rounding algorithm performance}

To test the performance of the rounding Algorithm~\ref{alg:association}, the cost function~\eqref{eq:opt_distr} is evaluated with the obtained rounded solution \mbox{$\bm o^r =\{\bm o_i^r \mid i \in \mathcal{N}\}$}. The comparison is performed with the solution given by each server $i$ simply following the desired $\bar{\bm w}_i$, i.e., the solution corresponding to the ``migrate'' strategy. Specifically, the ratio between the cost values of the ``migrate'' strategy and the rounded solution is computed, considering the cases where it is energetically inefficient to follow the desired migrations. Indeed, in the other case $\bm o_i = \bm o_i^r = \bar{\bm w}_i$, for all servers, i.e., $\bar{\bm w}_i$ is the optimal solution and it is a feasible one in the discrete domain, thus the costs are equal. As an example, with prediction horizon $T=5$, job generation probability $p=0.3$, and $P_{\rm PV}=300$~W, the gain of using the proposed relax and round optimization procedure of \ac{EASE} over the ``migrate" strategy is on average $10$ folds. More in the detail, the gain has a median of $3.8$, the $10$th percentile is $1.3$, meaning that rarely a gain lower than $30\%$ is observed, and the $90$th percentile is $17$. Hence, often, the rounding step of \ac{EASE} induces a high gain over the blind ``migrate" strategy from an energy perspective.

%Hence, although it can be impossible to reach the cost values computed in continuous form by the dual ascent of algorithm~\ref{alg:dual_ascent_server}, adopting the proposed rounding policy leads to a considerable energy gain  

\section{Conclusions}\label{sec:Conclusions}
%!TEX root = ./paper.tex

In this paper, we proposed \ac{EASE}, a novel strategy for online job scheduling in a \ac{MEC}-enabled network co-powered by the grid and renewable energy resources, considering an \ac{IoV} scenario. %Besides the high potential of the new \ac{MEC} paradigm entailing the \mbox{de-location} of computing power at the network edge, several challenges are still open for research. 
%Specifically, in this work we 
\ac{EASE} tackles the problem of ensuring computing service continuity as the users move within the resources-constrained network area. %While for communication service continuity effective handover algorithms are already widely tested and implemented, the handling of the computing services in case the user changes its point of attachment to the network is not yet standardized. 
%To address this point, we devised a complete framework that 
It allows deciding whether to migrate the jobs following the \ac{UE}, or to continue the execution on the \ac{MEC} server where it started. This is achieved through the alternation of a local control optimization phase, to estimate future resources, and a distributed consensus step, to reach the migration agreement. %leveraging backhaul links to provide the user with the computation outcome. 
The primary objective is the minimization of the carbon footprint at the network side, guaranteeing adequate \ac{QoS} to the moving users. 
Using \ac{EASE} leads to energy efficiency improvements of up to $10\%$ over heuristic strategies, getting close to carbon neutrality in a wide range of contexts.

%% use section* for acknowledgment
%\ifCLASSOPTIONcompsoc
%  % The Computer Society usually uses the plural form
%  \section*{Acknowledgments}
%\else
%  % regular IEEE prefers the singular form
%  \section*{Acknowledgment}
%\fi
%\ifCLASSOPTIONcaptionsoff
%  \newpage
%\fi

\appendices
\section{Proofs and convergence rate}\label{sec:appendix}
%!TEX root = ./paper.tex
\subsection{Proof of Proposition \ref{prop:3cases}}
\begin{IEEEproof}
i) and~ii) correspond to the cases where the $\max\{\cdot\}$ operator in~\eqref{eq:distr_cost_trans} is replaced by $0$ or $h_i(\bm x_i)$, respectively. Once the optimum is computed, the feasibility check must be done: if the minimum lies in the feasible region, the solution is accepted. However, it can also be that these two optima are both infeasible: in this case, the optimal solution must lie on the plane \mbox{$h_i(\bm x_i) = 0$}, and a constrained problem has to be solved (case~iii)).
\end{IEEEproof}
\begin{remark}
It is impossible that both solutions~i) and~ii) are feasible, otherwise the convex function~\eqref{eq:distr_cost_trans} would have two minima, which is absurd due to its convexity.
\end{remark}
\vspace{-0.3cm}

\subsection{Proof of Proposition \ref{prop:primal_sol}}
\begin{IEEEproof}
The proof is straightforward for cases~i) and~ii): it is sufficient to set the gradient of the function to zero. In the third case, it is necessary to solve the constrained minimization of $u(\bm x_i)$ subject to \mbox{$h(\bm x_i)=0$}. The Lagrange multipliers method can be used, where the Lagrangian of case~iii) is \mbox{$\mathcal{L}'(\bm x_i, \eta_i) = u(\bm x_i) + \eta_i\, h(\bm x_i)$}, and its primal solution is
\begin{equation}
\inf_{\bm x_i} \; \sup_{\eta_i} \; \Vert \bm x_i - \bm b_i \Vert^2_{Q_i}+\bm \nu_i^{\scriptscriptstyle{T}}A_i\bm x_i + \eta_i \left(\bm q_i^{\scriptscriptstyle{T}}\bm x_i - \hat{P}^{\rm H}_{i}\right).
\end{equation}
The partial derivatives with respect to $\bm x_i$, and $\eta_i$ are
\begin{equation}
\begin{aligned}
\frac{\partial \mathcal{L}'(\bm x_i, \eta_i)}{\partial \bm x_i} &= 2 \,Q_i \left(\bm x_{i} - \bm b_{i}\right) + A_{i}^{\scriptscriptstyle{T}} \bm \nu_{i} + \eta_i\, \bm q_{i}, \\
\frac{\partial \mathcal{L}'(\bm x_i, \eta_i)}{\partial \eta_i} &= \bm q_i^{\scriptscriptstyle{T}} \bm x_i - \hat{P}^{\rm H}_i.
\end{aligned}
\end{equation}
Setting them to zero, we obtain
%\begin{align}
%\label{eq:x_first_espr}
%\bm x_i &= \bm b_i - \frac{1}{\rho} \left(A_i^{\scriptscriptstyle{T}}\bm y_i + \eta\, \bm q_i \right), \quad \text{and}\\
%0 &= \bm q_i^{\scriptscriptstyle{T}}\left[\bm b_i - \frac{1}{\rho} \left(A_i^{\scriptscriptstyle{T}} \bm y_i + \eta\, \bm q_i \right)\right] - \hat{P}_{{\rm H}, i}\nonumber \\
%\label{eq:lambda_star}
%\implies \eta^* &= \frac{1}{\Vert \bm q_i \Vert ^2} \left[ \rho \left(\bm q_i^{\scriptscriptstyle{T}} \bm b_i - \hat{P}_{{\rm H}, i} \right) - \bm q_i^{\scriptscriptstyle{T}}A_i^{\scriptscriptstyle{T}} \bm y_i \right].
%\end{align}
\begin{equation}
\label{eq:x_first_espr}
\bm x_i = \bm b_i - \frac{1}{2} \left[Q_i^{-1}\left(A_i^{\scriptscriptstyle{T}}\bm \nu_i + \eta_i\, \bm q_i \right)\right]=\frac{\hat{P}^{\rm H}_{i}}{\Vert \bm q_i \Vert^2} \bm q_i,
\end{equation}
from which it is possible to derive the optimal value for the Lagrange multiplier
\begin{equation}
\label{eq:lambda_star}
\eta_i^* = \frac{\bm q_i^{\scriptscriptstyle{T}} \left[ 2\,Q_i \left(\bm b_i - \frac{\hat{P}^{\rm H}_{i}}{\Vert \bm q_i \Vert ^2}\bm q_i\right) - A_i^{\scriptscriptstyle{T}} \bm\nu_i \right]}{\Vert \bm q_i \Vert ^2} .
\end{equation}
Now, plugging \eqref{eq:lambda_star} into \eqref{eq:x_first_espr} returns the optimal value $\bm x_i^+$ for case~iii).
\end{IEEEproof}
\vspace{-0.3cm}

\subsection{Convergence rate of the dual ascent}
\begin{remark}
For quadratic programs, it is possible to find a condition on the step size $\alpha$ for which the algorithm is ensured to converge. This only depends on the constraint matrices $A_1$ and $A_2$, and on the quadratic cost matrix $Q$ defining the curvature. Since these values do not change among the three different primal optimization cases, a common condition can be obtained, i.e.,
\begin{equation}
\label{eq:alpha_condition}
\alpha \le \frac{2}{\left\Vert 
\left[
\begin{matrix}
A_1 \\
A_2
\end{matrix} \right]
Q^{-1}
\left[
\begin{matrix}
A_1 \\
A_2
\end{matrix} \right]^{\scriptscriptstyle{T}} \right\Vert}.
\end{equation} 
\end{remark}
\begin{IEEEproof}
This result can be derived using proposition 2.3.2 of~\cite{bertsekas1999nonlinear}.
% Do we need to say more on the proof? Now I should modify $\alpha$ on the algorithms to make it fixed for all multipliers, and it can become a scalar?}
\end{IEEEproof}
\vspace{-0.3cm}

\bibliographystyle{IEEEtran}
\bibliography{IEEEabrv,biblio}

\begin{IEEEbiography}[{\includegraphics[width=1in,height=1.25in,clip,keepaspectratio]{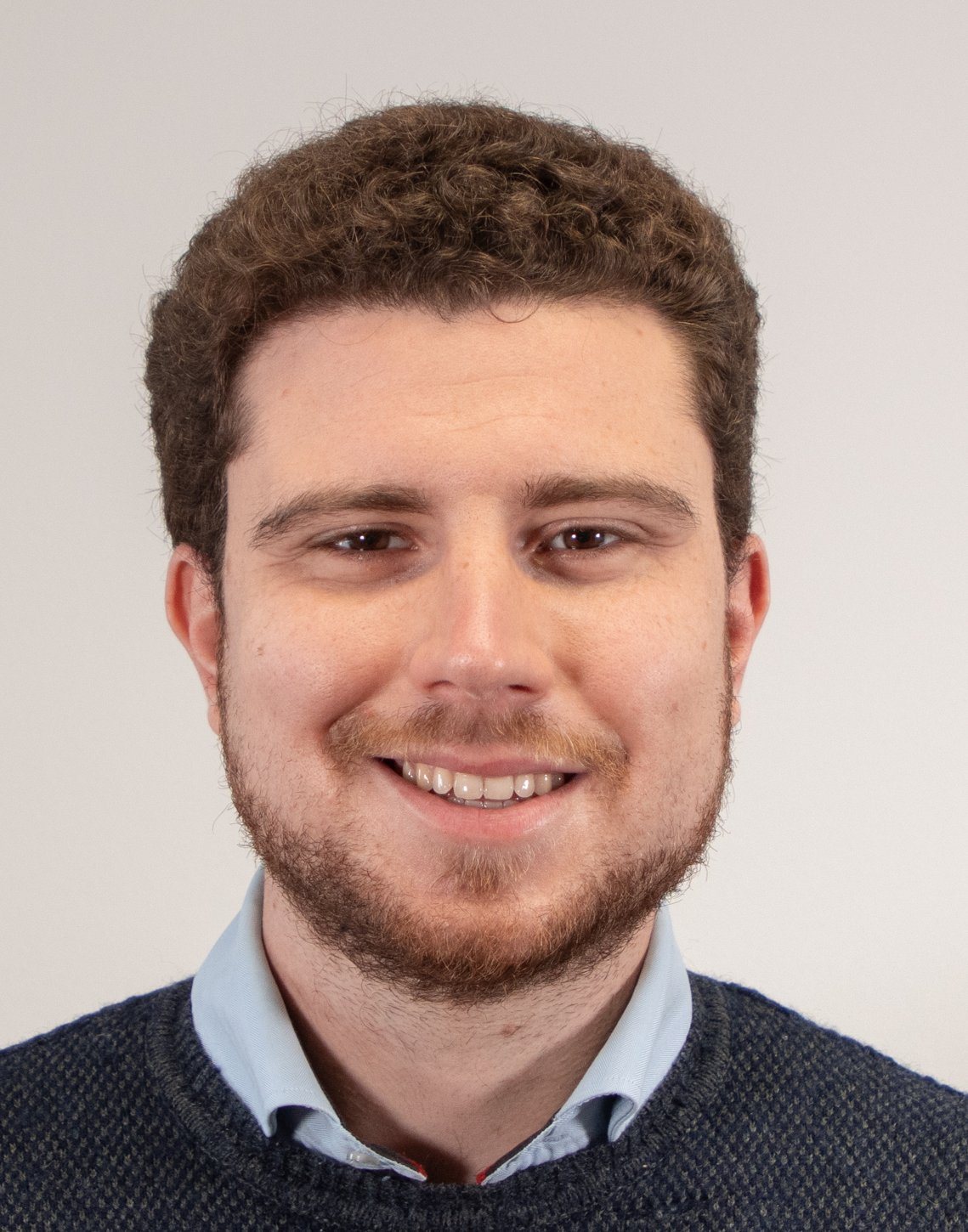}}]{Giovanni Perin} (Graduate Student Member, IEEE) received the B.Sc. degree in Information Engineering and the M.Sc. degree in ICT for Internet and Multimedia (\emph{summa cum laude}) from the University of Padova, Italy, in 2017 and 2019, respectively, where he is currently pursuing the Ph.D. degree in Information Engineering, joining the PRIN project no. 2017NS9FEY tackling the real-time control of 5G wireless networks. In 2019, he spent six months as a visiting student with the Deutsche Telekom Chair of Communication Networks, Technical University of Dresden, Germany, working on broadcast routing, while in 2022 he was visiting scholar at the University of California, Irvine, USA, conducting research on vehicular communications and edge computing. His research focuses on sustainable edge computing, distributed optimization and processing, and federated learning.
\end{IEEEbiography}

\vspace{-1cm}

\begin{IEEEbiography}[{\includegraphics[width=1in,height=1.25in,clip,keepaspectratio]{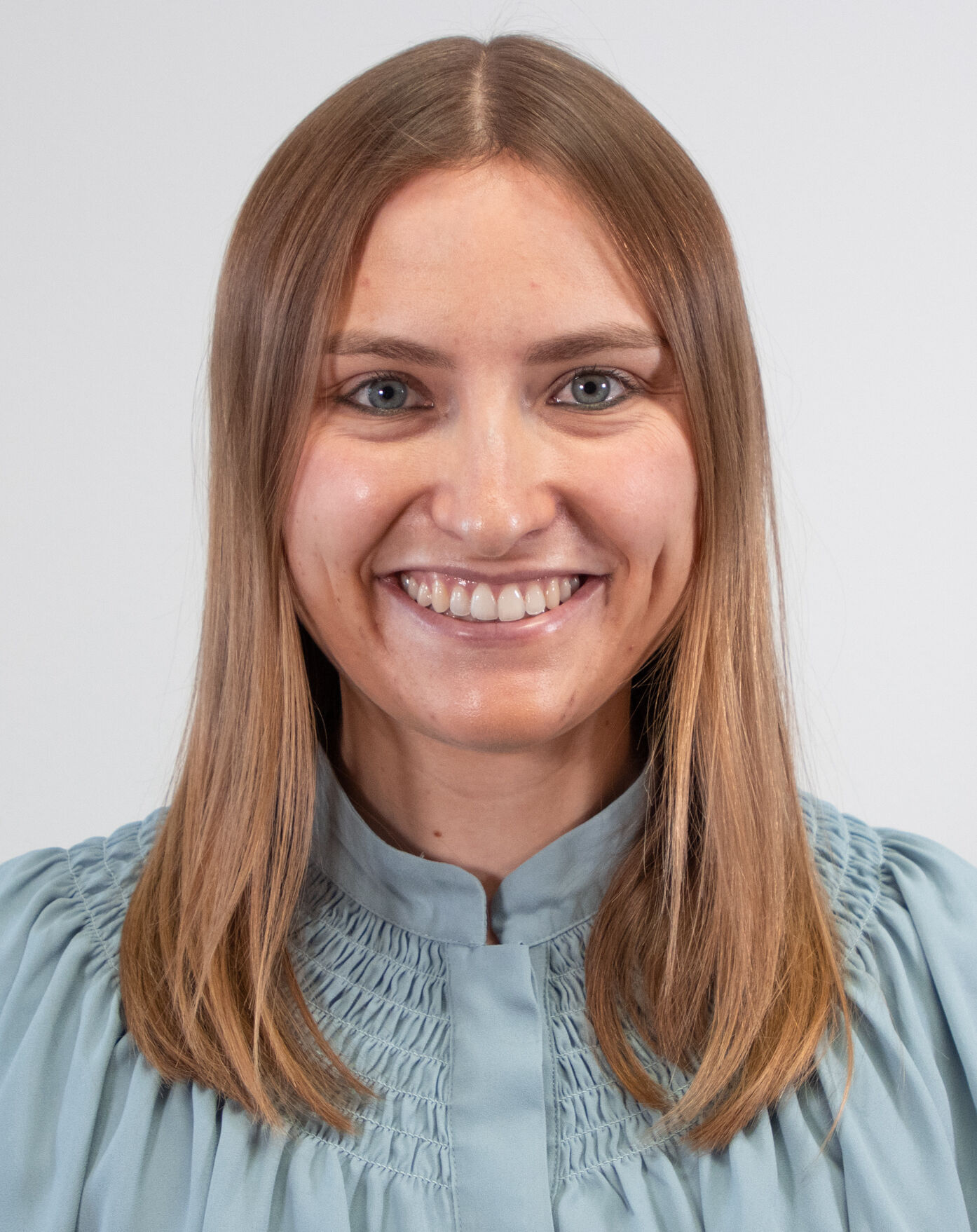}}]{Francesca Meneghello} (Member, IEEE) received the Ph.D. degree in Information Engineering in 2022 from the University of Padova and is currently a postdoctoral researcher with the Department of Information Engineering at the same university. Her research interests include \mbox{deep-learning} architectures and signal processing with application to remote radio frequency sensing and wireless networks. She was a recipient of the Best Student Paper Award at WUWNet 2016, the Best Student Presentation Award at the IEEE Italy Section SSIE 2019 and received an honorary mention in the 2019 IEEE ComSoc Student Competition. She was awarded with a Fulbright-Schuman visiting scholar fellowship for the a.y. 2022-2023.
\end{IEEEbiography}

\vspace{-1cm}

\begin{IEEEbiography}[{\includegraphics[width=1in,height=1.25in,clip,keepaspectratio]{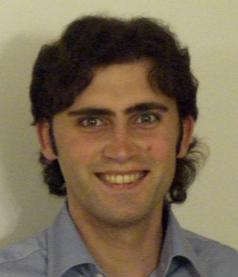}}]{Ruggero Carli} (Member, IEEE) received the Laurea degree in computer engineering and the Ph.D. degree in information engineering from the University of Padova, Padova, Italy, in 2004 and 2007, respectively. From 2008 to 2010, he was a Postdoctoral Fellow with the Department of Mechanical Engineering, University of  California, Santa Barbara, CA, USA. He is currently an Associate
Professor with the Department of Information Engineering, University of Padova. His research interests include multiagent robotics, distributed optimization, estimation and control, and nonparametric estimation.
\end{IEEEbiography}

\vspace{-1cm}

\begin{IEEEbiography}[{\includegraphics[width=1in,height=1.25in,clip,keepaspectratio]{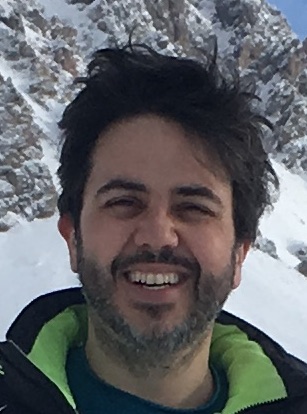}}]{Luca Schenato} (Fellow, IEEE) received the Dr. Eng. degree in electrical engineering from the University of Padova in 1999 and the Ph.D. degree in Electrical Engineering and Computer Sciences from the U.C. Berkeley, in 2003. He held a post-doctoral position in 2004 and a visiting professor position in 2013-2014 at U.C. Berkeley. Currently he is Full Professor at the Information Engineering Department at the University of Padova. His interests include networked control systems, multi-agent systems, wireless sensor networks, smart grids and cooperative robotics. Luca Schenato has been awarded the 2004 Researchers Mobility Fellowship by the Italian Ministry of Education, University and Research (MIUR), the 2006 Eli Jury Award in U.C. Berkeley and the EUCA European Control Award in 2014, and IEEE Fellow in 2017. He served as Associate Editor for IEEE Trans. on Automatic Control from 2010 to 2014 and he is he is currently Senior Editor for IEEE Trans. on Control of Network Systems and Associate Editor for Automatica.
\end{IEEEbiography}

\vspace{-1cm}

\begin{IEEEbiography}[{\includegraphics[width=1in,height=1.25in,clip,keepaspectratio]{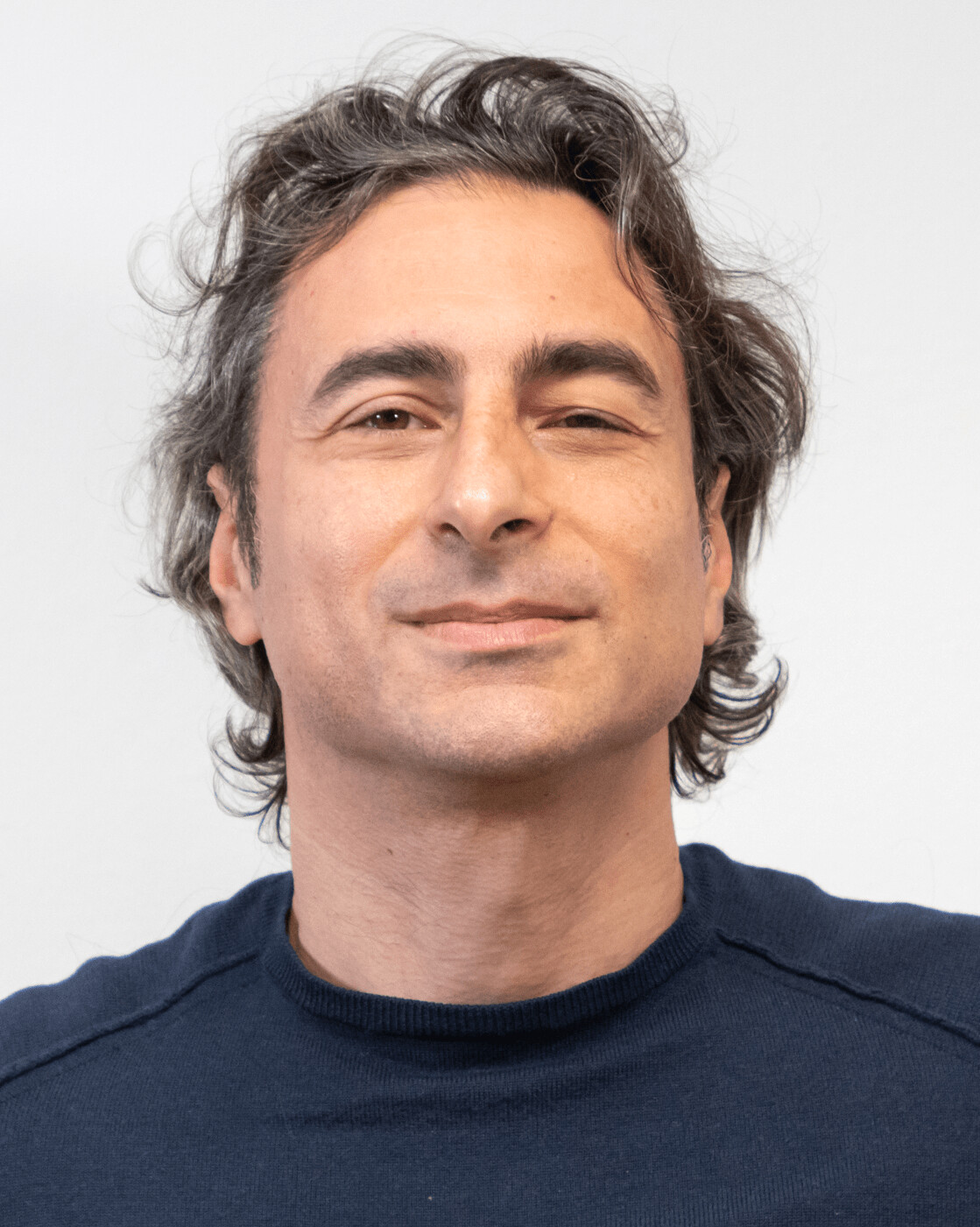}}]{Michele Rossi} (Senior Member, IEEE) is Full Professor in the Department of Information Engineering (DEI) at the University of Padova (UNIPD), Italy, where is the head of the Master's Degree in ICT for internet and Multimedia (\url{http://mime.dei.unipd.it/}). Since 2017, he has been the Director of the DEI/IEEE Summer School of Information Engineering (\url{http://ssie.dei.unipd.it/}). His research interests lie broadly in wireless sensing systems -- with current focus on radar networks and joint communications and sensing, green mobile networks, edge and wearable computing. Over the years, he has been involved in several EU projects on wireless sensing and IoT, and has collaborated with major companies such as Ericsson, DOCOMO, Samsung and INTEL. His research is currently supported by the European Commission through the H2020 projects MINTS (grant no. 861222) on ``mmWave networking and sensing"; and GREENEDGE (grant no. 953775) on ``green edge computing for mobile networks"; (project coordinator). Dr. Rossi has been the recipient of seven best paper awards from the IEEE and currently serves on the Editorial Boards of the IEEE Transactions on Mobile Computing, and of the Open Journal of the Communications Society. For further information on current activities, see: \url{http://www.dei.unipd.it/~rossi/}
\end{IEEEbiography}

\end{document}